\documentclass[seceq]{ptptex}

\usepackage{wrapft}
\usepackage{subfigure}
\usepackage{ulem}

\usepackage[dvips]{graphicx,psfrag,color}
\usepackage{comment}
\usepackage{amsmath,amssymb}
\usepackage{type1cm}
\usepackage{bm}

\newcommand\beq{\begin{equation}}
\newcommand\eeq{\end{equation}}
\newcommand\beqa{\begin{eqnarray}}
\newcommand\eeqa{\end{eqnarray}}
\renewcommand{\d}{\partial}

\newcommand\E{\epsilon}

\title{A Model Analysis of Triaxial Deformation Dynamics 
in Oblate-Prolate Shape Coexistence Phenomena}
\author{
Koichi \textsc{Sato}$^{1,2}$, 
Nobuo \textsc{Hinohara}$^{2}$, 
Takashi \textsc{Nakatsukasa}$^{2}$, \\
Masayuki \textsc{Matsuo}$^{3}$ 
and Kenichi \textsc{Matsuyanagi}$^{2,4}$
}

\inst{
$^{1}$Department of Physics, Graduate School of Science, 
Kyoto University, \\
Kyoto 606-8502, Japan \\
$^{2}$Theoretical  Nuclear Physics Laboratory, 
RIKEN Nishina Center, \\ 
Wako 351-0198, Japan \\
$^{3}$Department of Physics, Faculty of Science, \\
Niigata University, Niigata 950-2181, Japan\\
$^{4}$Yukawa Institute for Theoretical Physics, Kyoto University, \\
Kyoto 606-8502, Japan
}

\abst{
From a viewpoint of oblate-prolate symmetry and its breaking, 
we adopt the quadrupole collective Hamiltonian to study dynamics of triaxial deformation in shape coexistence phenomena.
It accommodates 
the axially symmetric rotor model, the $\gamma$-unstable model, 
the rigid triaxial rotor model and an ideal situation for 
the oblate-prolate shape coexistence as particular cases. 
Numerical solutions of this model yield a number of interesting suggestions. 
(1) The relative energy of the excited $0^+$ state can be a signature
of the potential shape along the $\gamma$ direction. (2) Specific $E2$
transition probabilities are sensitive to the breaking of the
oblate-prolate symmetry. (3) Nuclear rotation may induce the localization
of collective wave functions in the $(\beta,\gamma)$ deformation space. 
}

\begin{document}

\maketitle

\section{Introduction}

In recent years, experimental data suggesting coexistence of 
the ground-state rotational band with the oblate shape 
and the excited band with the prolate shape have been obtained 
in proton-rich unstable nuclei.
\cite{fis00,fis03,lju08} 
Together with a variety of shape coexistence phenomena 
observed in various regions of nuclear chart,
\cite{woo92,and00,dug03,rod04} 
some of which involve the spherical shape also, 
these discoveries stimulate development of 
nuclear structure theory capable of describing this new class of phenomena.
\cite{mat00,kob05,nik09,li09,ben08,gir09}
Recently, Hinohara et al. \cite{hin08,hin09}
carried out detailed microscopic calculations for 
the oblate-prolate shape mixing 
by means of the adiabatic self-consistent collective coordinate (ASCC) 
method.\cite{mat00}
They suggest that the excitation spectrum of $^{68}$Se may be regarded 
as a case corresponding to an intermediate situation between 
the well-developed oblate-prolate shape coexistence limit, 
where the shapes of the two coexisting rotational bands are 
well localized in the $(\beta, \gamma)$ deformation space, 
and the $\gamma$-unstable limit, 
where a large-amplitude shape fluctuation takes place in the 
$\gamma$ degree of freedom. 
Here, $\beta$ and $\gamma$ are well-known dynamical variables 
denoting the magnitude of the quadrupole deformation and 
the degree of axial symmetry breaking, respectively. 
These calculations indicate  the importance of
the coupled motion of the large-amplitude shape fluctuation in the 
$\gamma$ degree of freedom and 
the three-dimensional rotation associated with the triaxial shape.    

In order to discuss the oblate-prolate shape coexistence phenomena 
in a wider perspective including their relations to other classes of 
low-energy spectra of nuclei, we introduce 
in this paper a simple phenomenological model capable of describing 
the coupled motion of the large-amplitude $\gamma$-vibrational 
motions and the three-dimensional rotational motions.     
We call it {\it (1+3)D model }  in order to explicitly indicate 
the numbers of vibrational ($\gamma$) and rotational (three Euler angles) 
degrees of freedom.  
This model is able to describe 
several interesting limits in a unified perspective. 
It includes the axially symmetric rotor model,
the $\gamma$-unstable model, \cite{wil56}
the triaxial rigid rotor model \cite{dav58} 
and an ideal situation of the oblate-prolate shape coexistence.  
It also enables us 
to describe intermediate situations between these different limits 
by varying a few parameters.
This investigation will provide a new insight concerning connections 
between microscopic descriptions of oblate-prolate shape mixing and various 
macroscopic pictures on low-energy spectra in terms of phenomenological models. 
It is intended to be complementary to the microscopic approach 
we are developing on the basis of the ASCC method.

The (1+3)D model is introduced on the basis of 
the well-known five-dimensional (5D) quadrupole collective Hamiltonian
\cite{boh75}
by fixing the axial deformation parameter $\beta$. 
Simple functional forms are assumed for the collective potential 
and the collective mass (inertial function) 
with respect to the triaxial deformation $\gamma$.  
We analyze properties of excitation spectra, 
quadrupole moments and transition probabilities 
from a viewpoint of {\it oblate-prolate symmetry and its breaking} 
varying a few parameters characterizing the collective potential 
and the collective mass. 
Specifically, we investigate the sensitivity of these properties to 
1) the barrier-height between the oblate and prolate local minima 
in the collective potential,
2) the asymmetry parameter that controls the degree of 
oblate-prolate symmetry breaking in the collective potential, 
and 3) the mass-asymmetry parameter introducing the 
oblate-prolate asymmetry in the vibrational and rotational collective mass 
functions.
Dynamical mechanism determining localization 
and delocalization of collective wave functions in the deformation space 
is investigated. 
We find a number of interesting features 
which have received little attention until now: 
1) the unique behavior of the excited $0^+$ state as a function of 
the barrier-height parameter, 
2) specific $E2$ transition probabilities sensitive
to the degree of oblate-prolate symmetry breaking,
3) the rotation-assisted localization of collective wave functions 
in the deformation space.   
We also examine the validity of  the (1+3)D model 
by taking into account the $\beta$ degree of freedom, 
i.e., by solving the collective Schr\"odinger equation 
for a 5D quadrupole collective Hamiltonian 
that simulates the situation under consideration.

In the present paper, therefore, we intend to clarify the role of 
the $\beta$-$\gamma$ dependence of the collective mass.
Since the original papers by Bohr and Mottelson, 
\cite{boh52,boh53}
the collective Schr\"odinger equation with 
the 5D quadrupole collective Hamiltonian\cite{boh75}
has been widely used as the basic framework 
to investigate low-frequency quadrupole modes of excitation in nuclei. 
The collective potential and the collective masses appearing 
in the Hamiltonian have been introduced either phenomenologically
\cite{gne71,hes80,lib82,tro92}
or through microscopic calculations.  
\cite{mar62,bel65,kum67a,kum67b,bar65,roh77,pro99,lib99,pro04,sre06}
In recent years, powerful algebraic methods of solving the 
collective Schr\"odinger equation have been developed 
\cite{row04,row09}
and analytical solutions have been found 
\cite{iac00,cap07,cap09,bae06,bon07}
for some special forms of the collective potential 
(see the recent review \cite{for05} for an extensive list of references).
 However, all collective masses are assumed to be equal 
and a constant in these papers\cite{iac00,cap07,cap09,bae06,bon07}. 
It should be emphasized that this approximation is justified only for 
harmonic vibrations about the spherical shape. 
The collective masses express inertia of vibrational and 
rotational motions, so that they play crucial roles 
in determining the collective dynamics.  
In general, they are coordinate-dependent, i.e.,  
functions of $\beta$ and $\gamma$.  
In fact, various microscopic calculations for the collective masses 
indicate their significant variations 
as functions of $\beta$ and $\gamma$. 
\cite{kum67b,bar65,roh77,pro99,lib99,pro04,sre06,lie75,tam79,wee80,sak95,yam93}
In phenomenological analysis of experimental data, for instance,  
Jolos and Brentano \cite{jol09} 
have shown that it is necessary to use   
different collective masses for the rotational and 
$\beta$- and $\gamma$-vibrational modes in order to describe 
interband $E2$ transitions in prolately deformed nuclei.

This paper is organized as follows. 
After recapitulating the 5D quadrupole collective Hamiltonian and 
the collective Schr\"odinger equation in \S2, 
we introduce in \S3 the (1+3)D model of triaxial deformation dynamics. 
In \S4, using the (1+3)D model, 
we first discuss properties of excitation spectra for 
the collective potentials possessing oblate-prolate symmetry. 
We then investigate how they change when this symmetry is broken 
in the collective potential and/or in the collective mass. 
In \S5 we examine the validity of the (1+3)D model by 
introducing $\beta$-$\gamma$ coupling effects on the basis of
the 5D quadrupole collective Hamiltonian.  
Concluding remarks are given in \S6.

\section{Five-dimensional quadrupole collective Hamiltonian}

We start with the 5D quadrupole collective Hamiltonian 
involving five collective coordinates, i.e., 
two deformation variables ($\beta, \gamma$) and three Euler angles:
\begin{align}
H &= T_{\rm vib} + T_{\rm rot} +  V(\beta,\gamma), 
\label{eq:classical1}\\
T_{\rm vib} &= \frac{1}{2}D_{\beta\beta}(\beta,\gamma){\dot \beta}^2
 +D_{\beta\gamma}(\beta,\gamma){\dot \beta} {\dot \gamma}
 +\frac{1}{2}D_{\gamma\gamma}(\beta,\gamma){\dot \gamma}^2, 
 \label{eq:classical2}\\
T_{\rm rot} &= \sum_{k=1}^3 \frac{1}{2}\mathcal{J}_k(\beta,\gamma)\omega_k^2. 
\label{eq:classical3}
\end{align}
Here, the collective potential $V(\beta,\gamma)$ is a function of 
two deformation coordinates, $\beta$ and $\gamma$, which represent 
the magnitudes of quadrupole deformation and triaxiality, respectively.
It must be a scalar under rotation, 
so that it can be written as a function of
$\beta^2$ and $\beta^3\cos 3\gamma$\cite{boh75}.
As is well known, one can restrict the range of $\gamma$ to be 
$0^\circ \le \gamma \le 60^\circ$ by virtue of the transformation 
properties between different choices of the principal axes. 
The first term $T_{\rm vib}$ in Eq.~(\ref{eq:classical1}) 
represents the kinetic energies of shape vibrations; 
it is a function of $\beta$ and $\gamma$ as well as their time derivatives 
$\dot \beta$ and $\dot \gamma$.  
The second term $T_{\rm rot}$ represents the rotational energy 
written in terms of three angular velocities $\omega_k$ 
which are related to the time derivatives of the Euler angles.  
The three moments of inertia can be written as  
\begin{equation}
\mathcal{J}_k(\beta,\gamma)=4\beta^2D_k(\beta,\gamma)\sin^2\gamma_k,
\label{eq:inertia}
\end{equation}
with respect to the principal axes ($k=1-3$), 
where $\gamma_k=\gamma-(2\pi k)/3$.
In this paper, we adopt Bohr-Mottelson's notation \cite{boh75} 
for the six collective mass functions, $D_{\beta\beta}, D_{\beta\gamma}, 
D_{\gamma\gamma}, D_1, D_2$ and $D_3$.
They must fulfill the following conditions, 
\begin{align}
D_1(\beta,\gamma=0^\circ )&=D_2(\beta,\gamma=0^\circ ) \label{eq:condition1}, \\
D_3(\beta,\gamma=0^\circ )&=D_{\gamma\gamma}(\beta,\gamma=0^\circ)\beta^{-2}  
\label{eq:condition2}, \\
D_1(\beta,\gamma=60^\circ )&=D_3(\beta,\gamma=60^\circ )  
\label{eq:condition3} ,\\
D_2(\beta,\gamma=60^\circ )&=D_{\gamma\gamma}(\beta,\gamma=60^\circ)
\beta^{-2}, \label{eq:condition4}
\end{align}
in the prolate ($\gamma=0^\circ$) and the oblate ($\gamma=60^\circ$) 
axially symmetric limits.\cite{kum67a}

The classical collective Hamiltonian (\ref{eq:classical1}) is 
quantized according to the Pauli prescription. 
Then, the explicit expressions for the vibrational and rotational kinetic 
energies are given by \cite{bel65}
\begin{align}
\hat T_{\rm vib}=\frac{-\hbar^2}{2\sqrt{WR}}\left\{ \frac{1}{\beta^4} 
\left[\d_\beta \left(\beta^2\sqrt{\frac{R}{W}}D_{\gamma\gamma}\d_\beta\right)
-\d_\beta \left(\beta^2\sqrt{\frac{R}{W}}D_{\beta\gamma}\d_\gamma\right)
\right] \right.\notag\\
\left.+\frac{1}{\beta^2\sin 3\gamma}\left[-\d_\gamma \left(\sqrt{\frac{R}{W}}
\sin 3\gamma D_{\beta\gamma}\d_\beta\right)
+\d_\gamma \left(\sqrt{\frac{R}{W}}\sin 3\gamma D_{\beta\beta}\d_\gamma\right) 
\right]
\right\}
\label{eq:Tvib}
\end{align}
and
\begin{equation}
\hat T_{\rm rot}=\sum_{k=1}^3\frac{\hat I_k^2}{2\mathcal{J}_k(\beta,\gamma)},
\label{eq:Trot}
\end{equation}
respectively, 
where $W$ and $R$ are the abbreviations of 
\begin{align}
W(\beta,\gamma)&=\beta^{-2}\left[D_{\beta\beta}(\beta,\gamma)D_{\gamma\gamma}(\beta,\gamma)
-D_{\beta\gamma}^2(\beta,\gamma)\right], \\ 
R(\beta,\gamma)&=D_1(\beta,\gamma)D_2(\beta,\gamma)D_3(\beta,\gamma), 
\end{align}
and ${\hat I_k}$ are the angular momentum operators with respect to 
the principal-axis frame associated with a rotating deformed nucleus (the body-fixed PA frame). 

The collective Schr\"odinger equation is written as
\begin{align}
[\hat T_{\rm vib} + \hat T_{\rm rot} + V(\beta,\gamma) ]
\Psi_{IM\alpha}(\beta,\gamma,\Omega)
=E_{I,\alpha}\Psi_{IM\alpha}(\beta,\gamma,\Omega),
\label{eq:Hcoll}
\end{align}
where the collective wave function $\Psi_{IM\alpha}(\beta,\gamma,\Omega)$
is specified by the total angular momentum $I$, 
its projection $M$ onto the $z$-axis in the laboratory frame and
$\alpha$ distinguishing eigenstates possessing the same values of $I$ and $M$. 
In Eq.~(\ref{eq:Hcoll}), $\Omega$ denotes a set of the three Euler angles, 
which are here dynamical variables describing 
the directions of the body-fixed PA frame with respect to the laboratory frame. 
Using the rotational wave functions $\mathcal{D}^I_{MK}(\Omega)$, 
the orthonormalized collective wave functions 
in the laboratory frame can be written as  
\begin{equation}
\Psi_{IM\alpha}(\beta,\gamma,\Omega)=\sum_K 
\Phi_{IK\alpha}(\beta,\gamma)\langle \Omega|IMK\rangle ,
\label{eq:collectivewavefunction}
\end{equation}
where
\begin{equation}
\langle \Omega |IMK \rangle =\sqrt{ \frac{2I+1} {16\pi^2 (1+\delta_{K0})} } 
\left( \mathcal{D}^I_{MK}(\Omega)+(-)^I\mathcal{D}^I_{M-K}(\Omega) \right). 
\end{equation}
The functions $\Phi_{IK\alpha}(\beta,\gamma)$ are vibrational wave functions 
orthonormalized by 
\begin{align}
 \int d\tau' \sum_K \Phi^\ast_{IK\alpha}(\beta,\gamma) 
 \Phi_{IK\alpha'}(\beta,\gamma) = \delta_{\alpha\alpha'} 
\label{eq:normalization}
\end{align}
with the intrinsic volume element  
\begin{align}
 d\tau' = 2\beta^4 \sqrt{W(\beta,\gamma)R(\beta,\gamma)} 
          \sin{3\gamma} d\beta d\gamma. 
\label{eq:metric}
\end{align}
In Eqs.~(\ref{eq:collectivewavefunction}) and (\ref{eq:normalization}), 
the sum is taken over even values of $K$ from 0 to $I$ for even $I$
(from 2 to $I-1$ for odd $I$). 
Detailed discussions on the symmetries and boundary conditions 
in the vibrational wave functions $\Phi_{IK\alpha}(\beta,\gamma)$ are given, 
e.g., in Ref. 27).

\section{Reduction to the (1+3)-dimensional collective Hamiltonian}

For the reason mentioned in \S1, 
we are particularly interested in triaxial deformation dynamics. 
In order to concentrate on the $\gamma$ degree of freedom,
we introduce a simple (1+3)D model 
involving only one vibrational coordinate $\gamma$ and three rotational 
coordinates. This is done by freezing  the $\beta$ degree of freedom 
in the 5D quadrupole collective Hamiltonian (2.1) as explained below.

The collective potential of this model takes a very simple form: 
\begin{equation}
V(\gamma)=V_0\sin^2 3\gamma+V_1 \cos3\gamma.
\label{eq:1Dpotential}
\end{equation}
This form is readily obtained by retaining up to the second order 
with respect to $\beta^3\cos 3\gamma$ and fixing $\beta$ 
at a constant value in an expansion of $V(\beta,\gamma)$ 
in powers of $\beta^2$ and $\beta^3\cos 3\gamma$.\cite{oni86}  
When $V_1=0$, the collective potential is symmetric about $\gamma=30^\circ$ 
with respect to the transformation $\gamma \to 60^\circ-\gamma$. 
For brevity, let us call this symmetry and transformation 
{\it OP (oblate-prolate) symmetry} and {\it OP inversion}, respectively.  
For positive $V_0$, two degenerate minima appear 
at the oblate ($\gamma=60^\circ$) and the prolate ($\gamma=0^\circ$) shapes,  
and they are separated by a barrier 
that takes the maximum at $\gamma=30^\circ$. 
We therefore call $V_0$ {\it barrier-height parameter}. 
On the other hand, the maximally triaxial shape at $\gamma=30^\circ$ 
becomes the minimum for negative $V_0$, and 
it becomes deeper as $|V_0|$ increases.
When $V_1 \ne 0$, the OP symmetry is broken, and 
the oblate (prolate) shape becomes the minimum for 
a combination of positive $V_0$ and positive (negative) $V_1$. 
We call $V_1$ {\it asymmetry parameter} after its controlling the magnitude of the OP symmetry breaking.
{
Thus, by varying the two parameters $V_0$ and $V_1$, we can see  how the excitation spectrum depends on the barrier height and symmetry breaking between the two local minima in the collective potential. 
}

We present in Fig.~1 some examples of the collective potential $V(\gamma)$ 
that seem to be relevant to a variety of oblate-prolate shape coexistence 
phenomena. In this figure, we can clearly see 
how the asymmetry between the oblate and prolate minima 
grows as a function of $V_1$ 
and how the barrier height measured from the second minimum 
sensitively depends on the ratio of $V_1$ to $V_0$.  

\begin{figure}[h]
\begin{center}
\includegraphics[width=0.5\textwidth,clip]{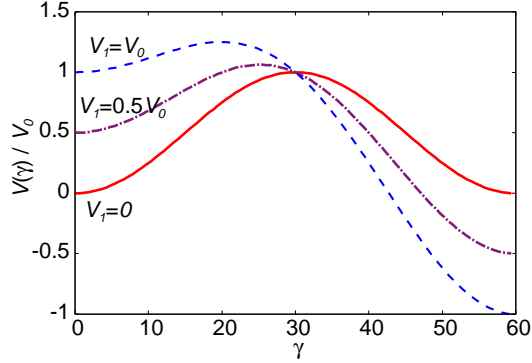}
\caption{The collective potentials $V(\gamma)$ defined by Eq.~(2.4) are plotted 
  for the $V_0>0$ case with solid, dash-dotted and dashed lines 
  as functions of $\gamma$ for $V_1/V_0=0.0, 0.5$ and $1.0$, respectively. }
\label{fig:potential}
\end{center}
\end{figure}


The vibrational kinetic energy term reduces, in the (1+3)D model Hamiltonian, 
to the following form: 
\begin{align}
\hat T_{\rm vib}=\frac{-\hbar^2}{2\sqrt{D_{\gamma\gamma}(\beta_0,\gamma)R(\beta_0,\gamma)}}
\frac{1}{\sin 3\gamma}\d_\gamma \left(\sqrt{\frac{R(\beta_0,\gamma)}{D_{\gamma\gamma}(\beta_0,\gamma)}}\sin 3\gamma \d_\gamma\right).
\label{eq:Tvib2}
\end{align} 
We parametrize the collective mass functions as 
\begin{align}
D_{\gamma\gamma}( \gamma ) &\equiv D_{\gamma\gamma}(\beta_0,\gamma ) =D\beta_0^{2}(1+\E\cos3\gamma), 
\label{eq:mass-function1} \\
D_{k}(\gamma ) &\equiv D_k(\beta_0,\gamma) \hspace{0.5em} =D(1+\E\cos\gamma_k).
\label{eq:mass-function2}
\end{align}
These are the most simple forms involving  
only one parameter $\E$ that controls the degree of 
OP symmetry breaking in these four mass functions 
under the requirement that they should fulfill  
the conditions (\ref{eq:condition1})--(\ref{eq:condition4}). 
We call $\E$ {\it mass-asymmetry parameter}.
These functional forms are obtainable also by 
taking the lowest-order term that brings about the OP symmetry breaking 
in the expressions of the collective mass functions microscopically 
derived  by Yamada\cite{yam93} using the SCC method. \cite{mar80}
In this paper, we use $\beta_0^2=0.1$ and $D=50$ MeV$^{-1}$, 
which roughly simulate the values obtained in the microscopic 
ASCC calculation\cite{hin08,hin09}.

We solve the collective Schr\"odinger equation (\ref{eq:Hcoll}) 
replacing $V(\beta,\gamma)$, $\hat T_{\rm vib}$ and $\mathcal{J}_k(\beta,\gamma)$ in $\hat T_{\rm rot}$ 
by Eqs.~(\ref{eq:1Dpotential}), (\ref{eq:Tvib2}) and $\mathcal{J}_k(\beta_0,\gamma)$, respectively, 
with Eqs. ~(\ref{eq:mass-function1}) and ~(\ref{eq:mass-function2}).
Accordingly, the collective wave function is denoted 
$\Psi_{IM\alpha}(\gamma,\Omega)$. 
Note that the sign change $V_1 \to -V_1$ corresponds to 
the OP inversion.  
One can then easily confirm that the OP inversion is equivalent to the simultaneous sign change of 
the parameters, $(\E, \pm V_1) \to (-\E, \mp V_1)$, in the (1+3)D model Hamiltonian. 
Therefore, it is enough to study only the case of positive $\E$.

\begin{figure}[h]
\begin{tabular}{cc}
\subfigure[Mass Parameters]
{\includegraphics[width=0.5\textwidth]{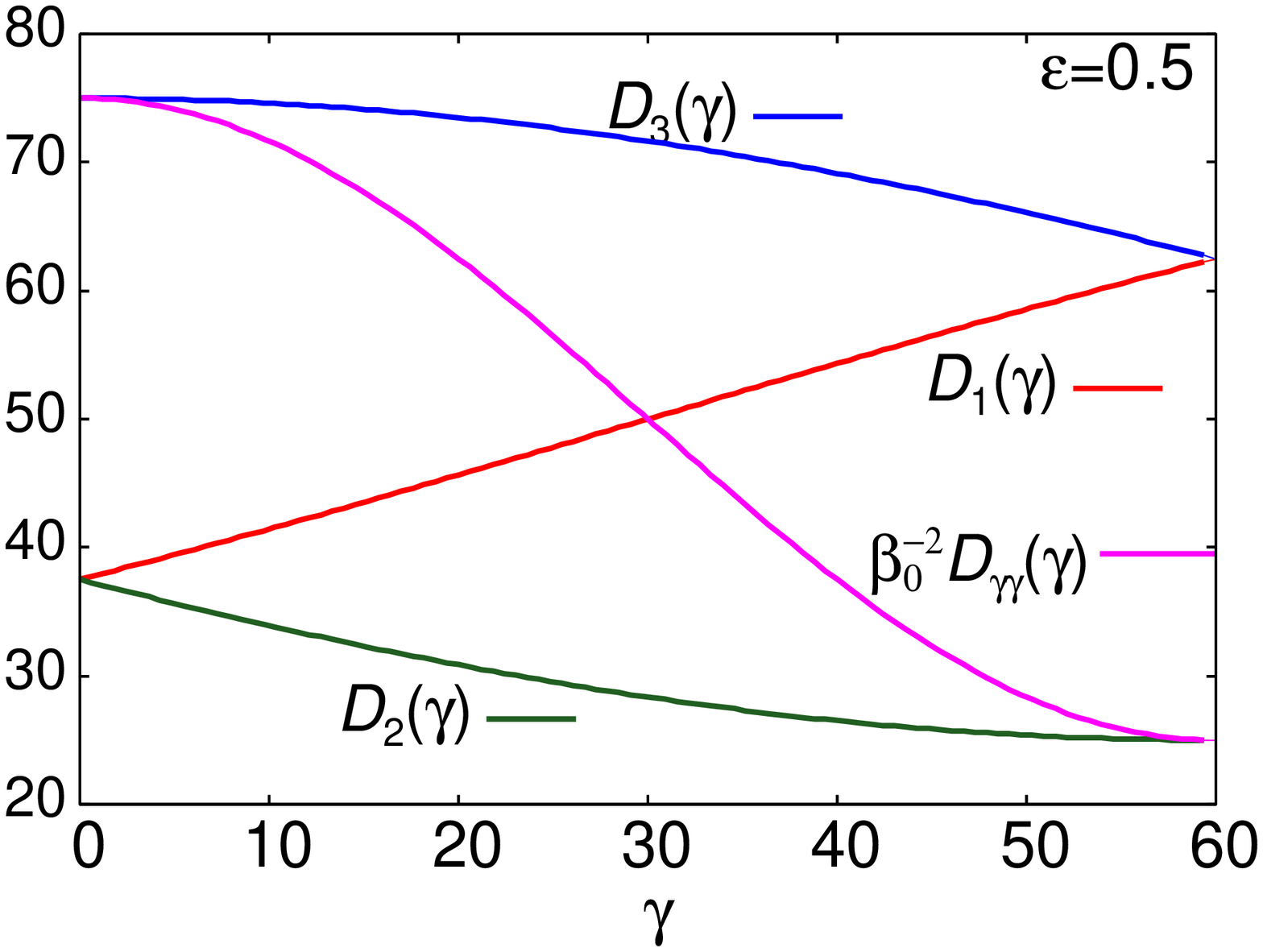} }
\subfigure[Moments of Inertia]
{\includegraphics[width=0.5\textwidth]{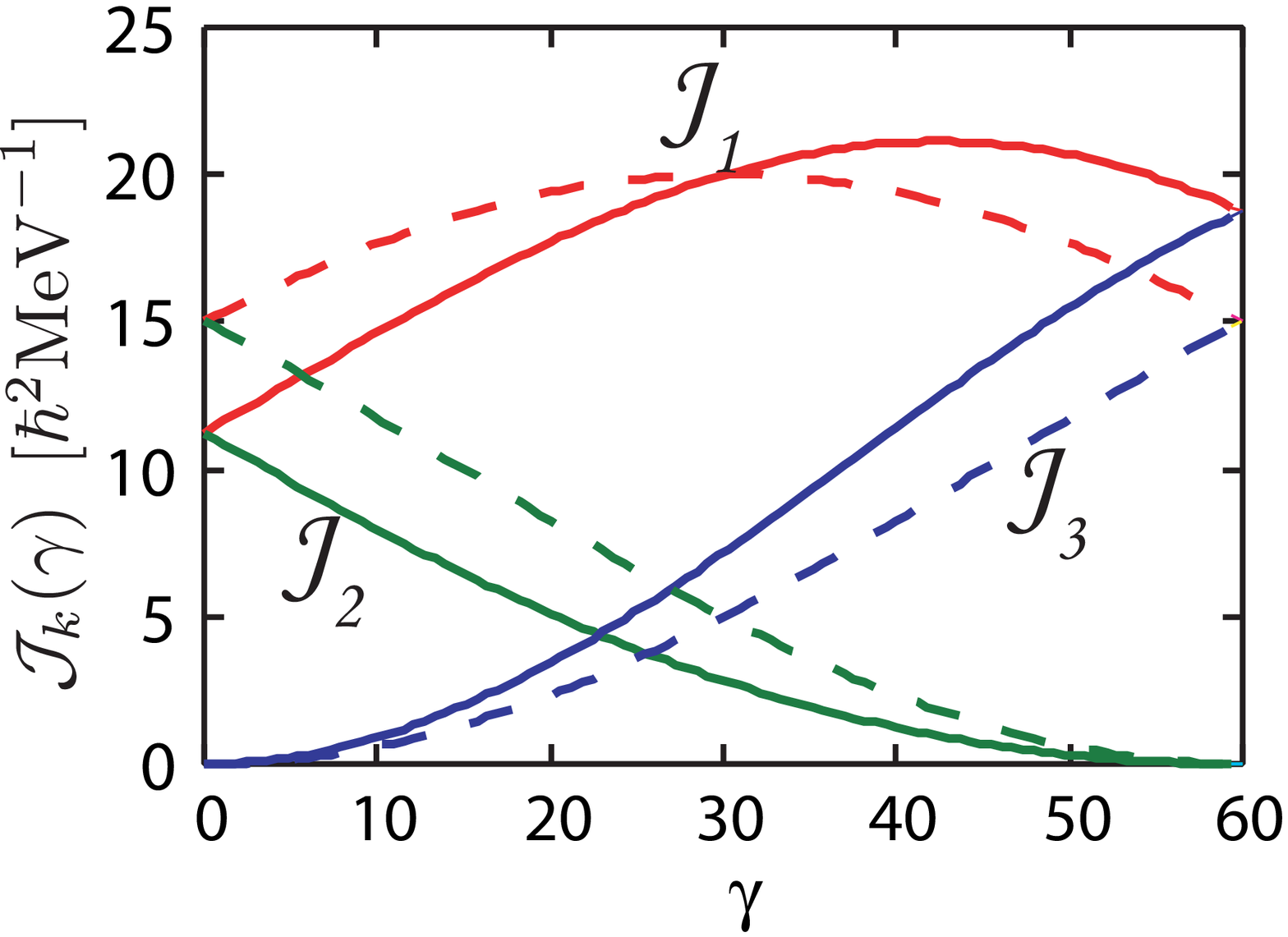}}
\end{tabular}
 \caption{(a) The vibrational and rotational collective mass functions 
    $D_{\gamma\gamma}(\gamma)$ and $D_k(\gamma)$, 
    defined by Eqs.~(2.2) and (2.3), 
    are plotted as functions of $\gamma$ 
    for the mass-asymmetry parameter $\E =0.5$. 
(b) Three rotational moments of inertia $\mathcal{J}_k(\beta,\gamma)$ 
    with $k=$1, 2 and 3, defined by Eqs.~(1.2) and (2.3), are plotted 
    with solid lines as functions of $\gamma$ 
    for the mass-asymmetry parameter $\E =0.5$. 
    The magnitude of the quadrupole deformation $\beta$ is fixed 
    at $\beta^2=0.1$. 
    For comparison, the moments of inertia for the $\E=0$ case are also 
    plotted with dashed lines.
}
\label{figs:DandMoI}
\end{figure}


Figure~2(a) shows, for an example of $\E=0.5$, 
how the collective mass functions $D_{\gamma\gamma}$ and $D_{k}$ 
behave as functions of $\gamma$. 
Figure~2(b) indicates the degree of oblate-prolate asymmetry 
in the moments of inertia $\mathcal{J}_k(\gamma)$ 
brought about by the terms involving $\E$.   
From these figures we can anticipate that, 
for positive (negative) $\E$ and under the parameterizations of 
Eqs.~(\ref{eq:mass-function1}) and (\ref{eq:mass-function2}), 
the rotational energy favors the oblate (prolate) shape  
while the vibrational energy prefers the prolate (oblate) shape.

\section{Triaxial deformation dynamics}

We have built a new computer code to solve the collective 
Schr\"odinger equation (\ref{eq:Hcoll}) for general 5D 
quadrupole collective Hamiltonian as well as its reduced version for the 
(1+3)D model. Numerical algorithm similar to that of Kumar and Baranger 
\cite{kum67a}  
is adopted in this code, except that we discretize the ($\beta,\gamma$) plane 
using meshes in the $\beta$ and $\gamma$ directions   
in place of their triangular mesh. 
Numerical accuracy and convergence were checked 
by comparing our numerical results with analytical solutions in the spherical 
harmonic vibration limit, in the so-called square-well type $\beta$ 
dependence limit and in the $\gamma$-unstable model. \cite{wil56} 
In the following, we discuss { how  the solutions of 
the collective Schr\"odinger equation 
depend} on the barrier-height parameter $V_0$, 
the asymmetry parameter $V_1$ and the mass-asymmetry parameter $\E$. 
We use the adjectives {\it yrast} and {\it yrare}  
for the lowest and the second-lowest states for a given angular momentum $I$, 
respectively, and distinguish the two by suffices as $I_1$ and $I_2$.

\subsection{Excitation spectra in the presence of OP symmetry}

Let us first discuss the situation where $V_1=0$ and $\E=0$.
In this case, both the collective potential and the collective mass functions 
are symmetric about $\gamma=30^\circ$, so that the (1+3)D model Hamiltonian 
possesses OP symmetry. 
Furthermore, the collective mass parameter $D$ and $\beta_0$ 
enter the collective Sch\"odinger equation only in the form of 
the overall factor $(2D\beta_0^2)^{-1}$ in the kinetic energy terms.   
Hence, only the ratio of the barrier-height parameter $V_0$ to this factor 
is important to determine the collective dynamics.
In the particular case of $V_0=0$, i.e., $V(\gamma)=0$, 
which is well known as the Wilets-Jean $\gamma$-unstable model,\cite{wil56}    
the excitation spectra are completely scaled by this factor.

\begin{figure}[b]
\begin{center}
  \includegraphics[width=0.8\textwidth]{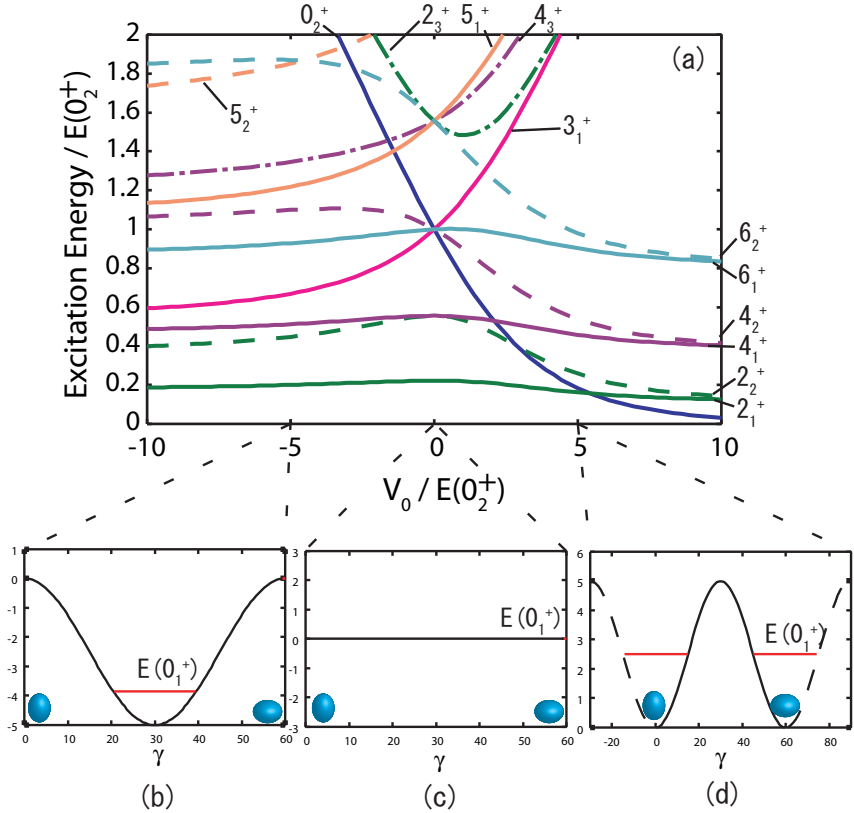}
  \caption{In the upper panel (a), dependence of excitation spectrum on the parameter 
  $V_0$ is displayed. Excitation energies as well as $V_0$ values are 
  normalized by the excitation energy $E(0_2^+)$ 
  of the second (the first excited) $0^+$ state for $V(\gamma)=0$.
  In the lower panels (b), (c) and (d), the potentials $V(\gamma)$ and 
  the ground $0^+$ state energies $E(0_1^+)$ are illustrated 
  for three different values of $V_0/E(0_2^+) = -5.0, 0.0$ and 5.0, 
  respectively.  Note that $E(0_1^+)=0$ for $V_0=0$. 
  Because the collective potential $V(\gamma)$ is a periodic function of 
  $60^\circ$ in $\gamma$, only the region $0^\circ \le \gamma \le 60^\circ$ is 
  drawn with a solid line in (d).  }
\end{center}
\end{figure}


In Fig.~3, we show the excitation spectra for the case of $V_1=0$ and $\E=0$ 
as functions of $V_0$.  Here, the excitation energies are normalized 
by the excitation energy of the first excited $0^+$ state 
(denoted $0_2^+$) at $V_0=0$, $E(0_2^+)$  
(which is 1.8 MeV for $\beta_0^2=0.1$ and $D=50$ MeV$^{-1}$ adopted in the 
present calculation). Accordingly, this figure is valid for any value of 
$(2D\beta_0^2)^{-1}$ by virtue of the scaling property mentioned above.  
In the lower panels of this figure, the collective potentials $V(\gamma)$ 
and the ground $0^+$ state energies $E(0_1^+)$ are illustrated 
for three typical situations: 1) a triaxially deformed case 
where a deep minimum appears at the triaxial shape with $\gamma=30^\circ$, 
2) the $\gamma$-unstable case, where the collective potential is flat 
with respect to $\gamma$, and 3) an extreme case of shape coexistence 
where the oblate and prolate minima are  exactly degenerate in energy.
(Strictly, shape coexistence does not appear in a case where the two minima are {\it  exactly} degenerate as we shall see in Fig.~9.)
Note that the collective potential $V(\gamma)$ is a periodic function of 
$60^\circ$ in $\gamma$. 
Therefore, it is displayed by the solid line  
only in the region $0^\circ \le \gamma \le 60^\circ$.  

In the positive-$V_0$ side of this figure, 
it is clearly seen that a doublet structure emerges 
when the barrier-height parameter $V_0$ becomes very large. 
In other words, approximately degenerate pairs of eigenstates appear 
for every angular momentum 
when $V_0/E(0_2^+) \gg 1$. 
This is nothing but the doublet pattern known well in 
the problems of double-well potential. \cite{lan58} 
In the present case, this doublet structure is associated with the OP symmetry.  
Furthermore, one immediately notices a very unique behavior
of the $0_2^+$ state. When the barrier-height parameter $V_0$ decreases 
(from the limiting situation mentioned above), 
its energy rises more rapidly than  
those of the yrare $2_2^+$, $4_2^+$ and $6_2^+$ states. 
Thus, a level crossing of the $0_2^+$ and $2_2^+$ states takes place 
at $V_0/E(0_2^+) \simeq 3$. When $V_0$ further decreases and approaches zero, 
the excitation energy of the $0_2^+$ state  approaches 
those of the $4_2^+$ and $6_1^+$ states. 
In the $\gamma$-unstable limit of $V_0=0$, they are degenerate. 

In the negative-$V_0$ side, the excitation energies of 
the $3_1^+$ and $5_1^+$ states drastically decrease as $V_0$ decreases, and 
the excitation spectrum characteristic to 
the Davydov-Filippov rigid triaxial rotor model \cite{dav58} appears 
when the triaxial minimum becomes very deep, 
i.e., when $V_0/E(0_2^+) \ll -1$.

\begin{figure}[h]
\begin{center}
  \includegraphics[width=0.7\textwidth]{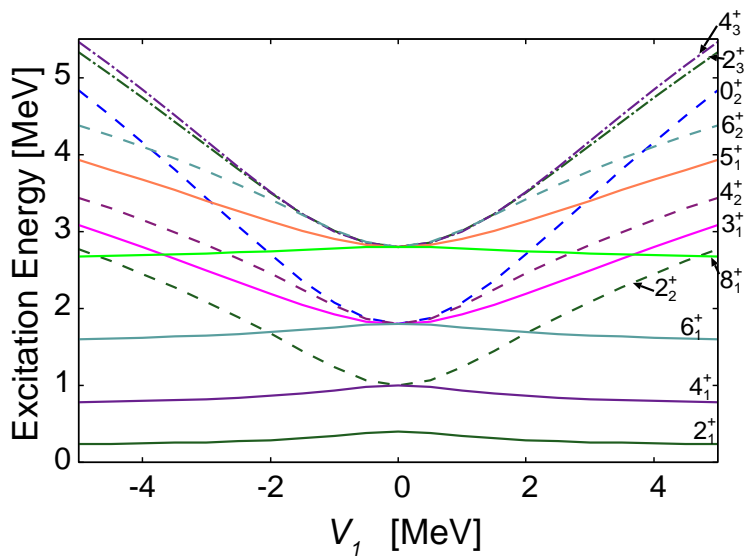}
  \caption{  Dependence of excitation spectrum on the asymmetry parameter $V_1$ 
  is displayed for the case of $V_0=0$.}
\end{center}
\label{fig:Ex_V0+0_V1}
\end{figure}


\subsection{Breaking of the OP symmetry in the collective potential}

Next, let us investigate effects of the OP-symmetry-breaking term 
$V_1 \cos3\gamma$ in the collective potential $V(\gamma)$.  
The effects are manifestly seen for the case of $\E=0$ and $V_0=0$.  
We present in Fig.~4 the excitation spectrum for this case 
as a function of the asymmetry parameter $V_1$.
For $V_1=0$, the spectrum exhibits the degeneracy characteristic 
to the $\gamma$-unstable model: \cite{wil56} e.g., 
the $0_2^+, 3_1^+, 4_2^+$ and $6_1^+$ states are degenerate. 
With increasing the magnitude of $V_1$, such degeneracies are lifted and 
the yrare $0_2^+, 2_2^+, 4_2^+$ and $6_2^+$ states 
as well as the odd angular momentum $3_1^+$ and $5_1^+$ states rise in energy.  
As a consequence, the well-known ground-state rotational band spectrum 
appears for sufficiently large values of $|V_1|$.  
For instance, we can see in Fig.~4 that, as $V_1$ increases, the ratio of the yrast $2_1^+$ and $4_1^+$ energies, $E(4_1^+)/E(2_1^+)$, increases from 2.5 at $V_1=0$, which is the value peculiar to the $\gamma$-unstable model, to be 3.3 for sufficiently large $V_1$.
This figure beautifully demonstrates the fact that the breaking of 
spherical symmetry is not sufficient for the appearance of 
regular rotational spectra even if the magnitude of quadrupole deformation 
is considerably large: we also need appreciable amount 
of the OP symmetry breaking.   
We also note that the spectrum does not depend on the sign of $V_1$.
This is because, for $\E=0$, both the vibrational and rotational 
kinetic energy terms in the (1+3)D collective Hamiltonian possess 
the symmetry under the OP inversion. 
In short, the inversion $V_1 \to -V_1$ merely interchanges 
the roles of the oblate and the prolate shapes.    

\begin{figure}[h]
\begin{tabular}{c}
\subfigure[$V_0=-2.0$MeV, $V_1=1.0$MeV]{
   \includegraphics[width=0.45\textwidth]{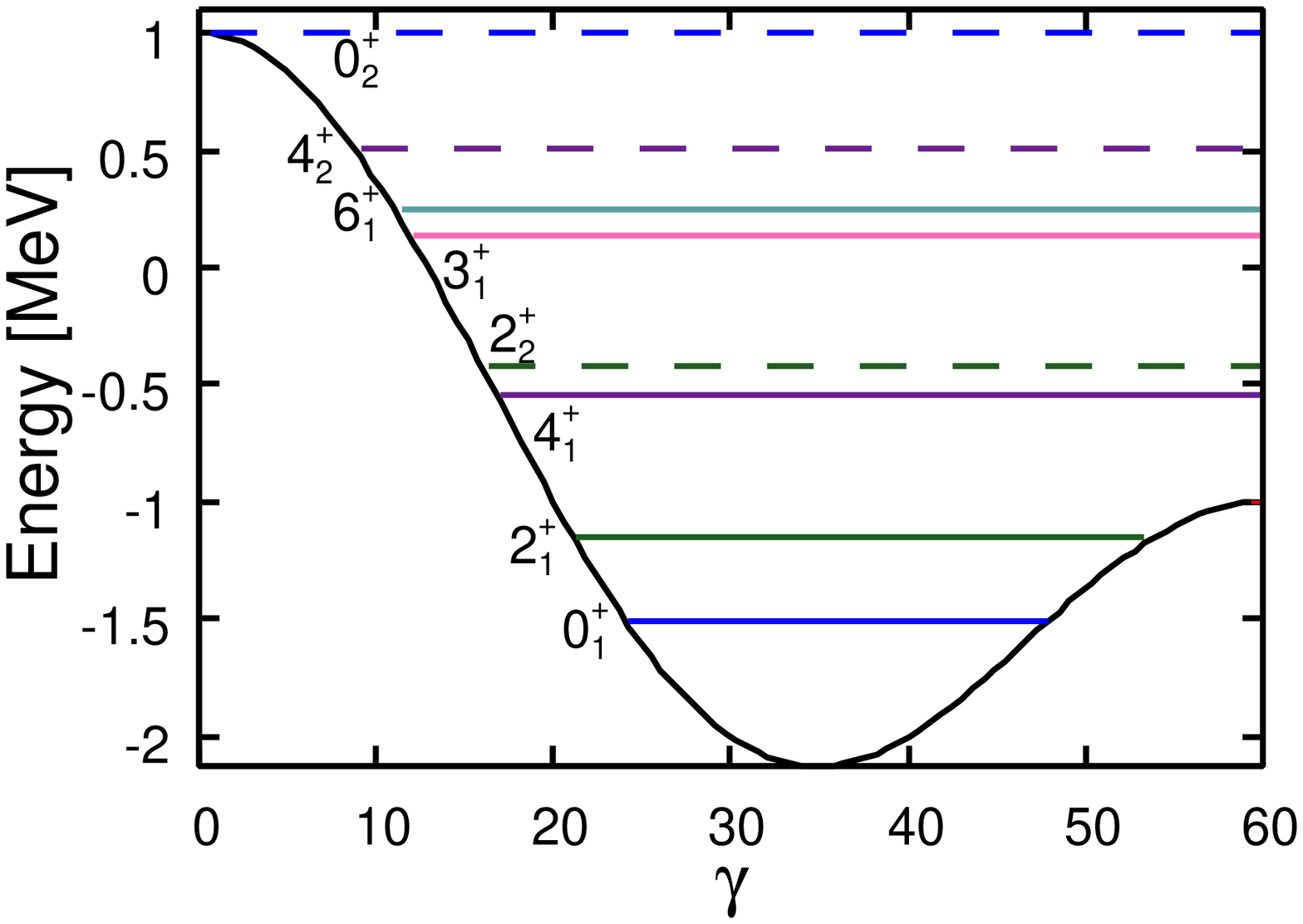} 
} 
\setcounter{subfigure}{2}
\subfigure[$V_0=10.0$MeV, $V_1=1.0$MeV]{ 
  \includegraphics[width=0.45\textwidth]{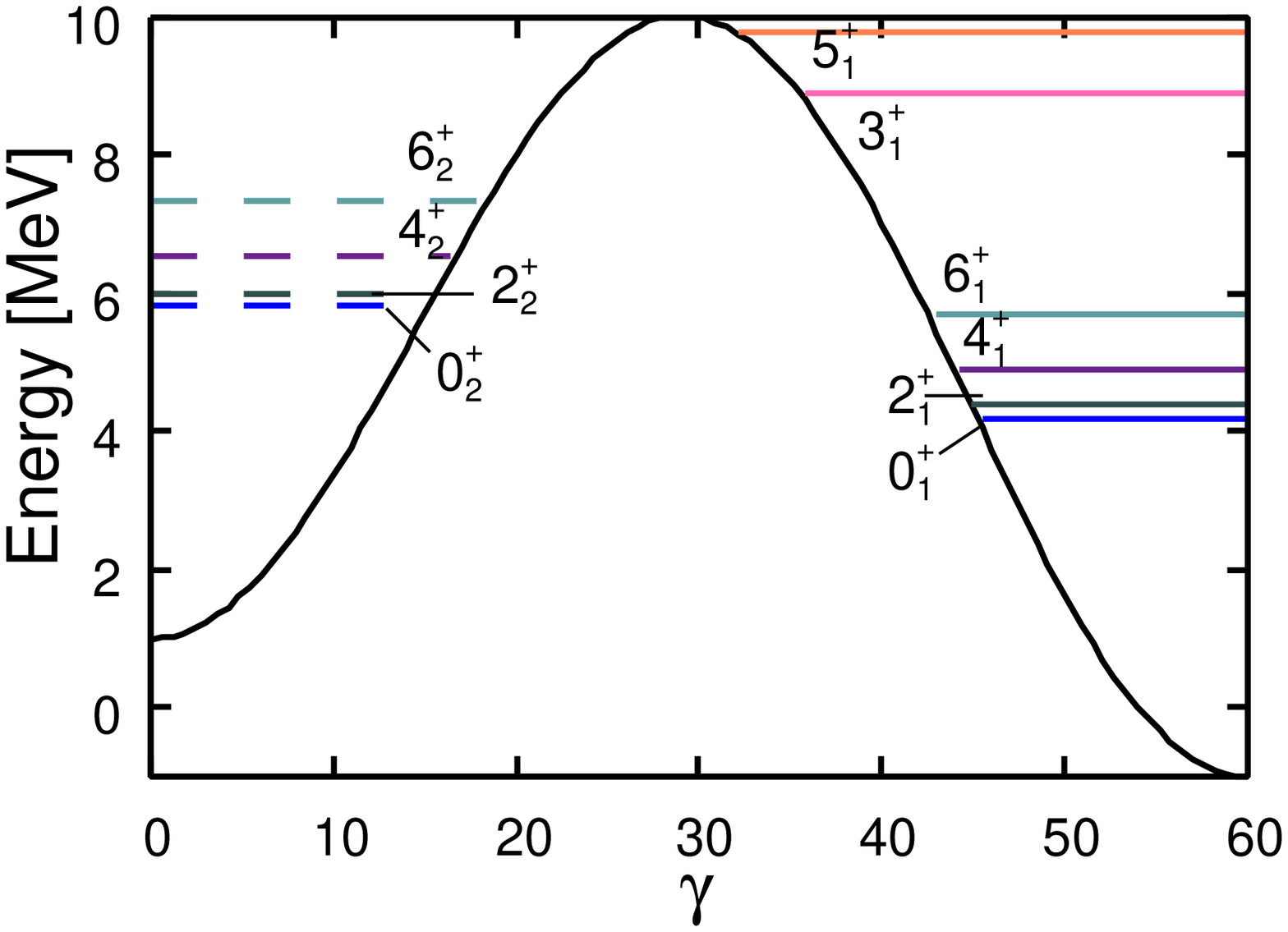} 
}
\\
\vspace{-1em}
\setcounter{subfigure}{1}
\subfigure[$V_0=1.0$MeV, $V_1=0.5$MeV]{
   \includegraphics[width=0.45\textwidth]{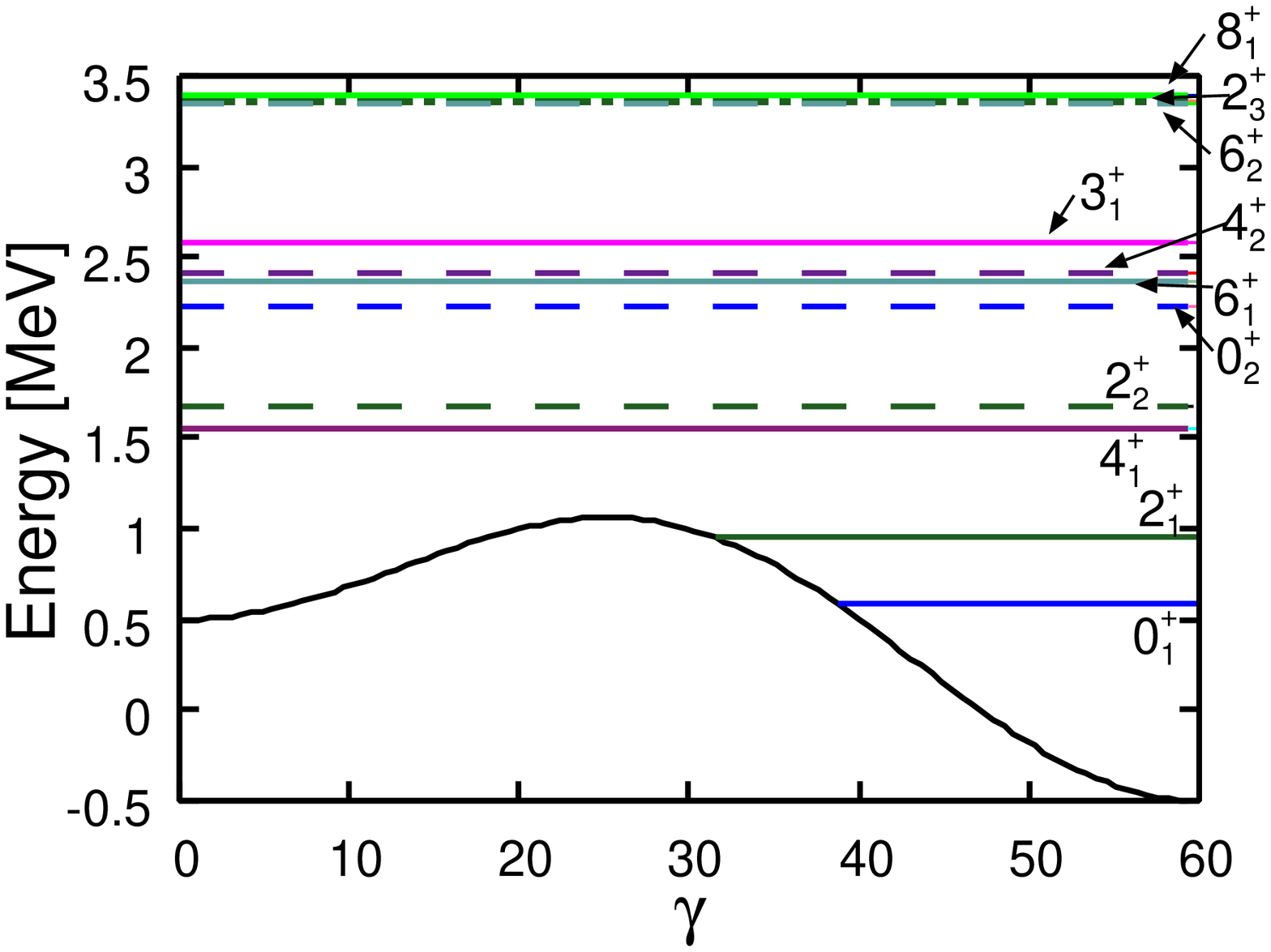} 
} 
\end{tabular}
\caption{The collective potentials $V(\gamma)$ and eigen-energies are displayed 
  in panels (a), (b) and (c) for $(V_0, V_1)=(-2.0, -1.0),(1.0, 0.5)$ 
  and (10.0, 1.0) MeV, respectively.}
\label{fig:3potentials}
\end{figure}


\clearpage
Let us proceed to more general situations 
where $\E=0$ but both $V_0$ and $V_1$ are nonzero.
Three typical situations are illustrated in Fig.~5.
In this figure,  the collective potentials and energy spectra
are drawn for three different sets of parameters, 
$(V_0,V_1) = (-2, 1), (1, 0.5)$ and (10, 1) MeV. 
Panel (a) simulates a situation where the minimum of the collective 
potential occurs at a triaxial shape but it is rather shallow, 
so that the potential pocket accommodates only the ground $0^+$ state 
and the first excited $2^+$ state.  
Panel (b) simulates the situation encountered in the microscopic ASCC 
calculation \cite{hin08,hin09}, where two local minima 
appear both at the oblate and prolate shapes but the barrier between them 
is so low that strong shape mixing may take place. 
Panel (c) illustrates an ideal situation for shape coexistence, where 
the barrier between the oblate and prolate minima is so high 
that two rotational bands associated with them retain their identities.

\begin{figure}[b]
  \begin{tabular}{ccc}
\subfigure[$V_0=-2.0$MeV]
{\includegraphics[width=0.5\textwidth]{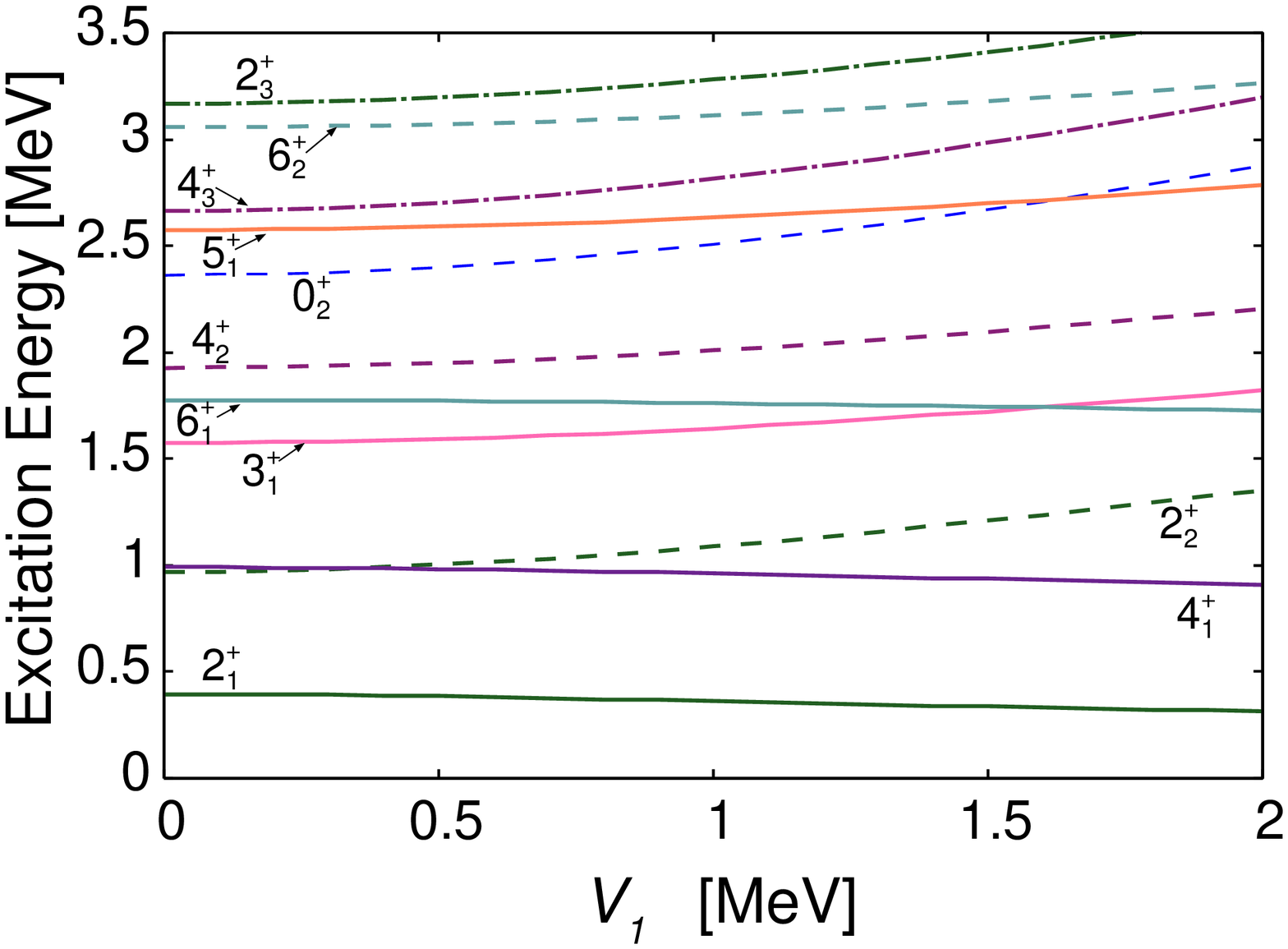}}
\setcounter{subfigure}{2}
\subfigure[$V_0=10.0$MeV]
{\includegraphics[width=0.5\textwidth]{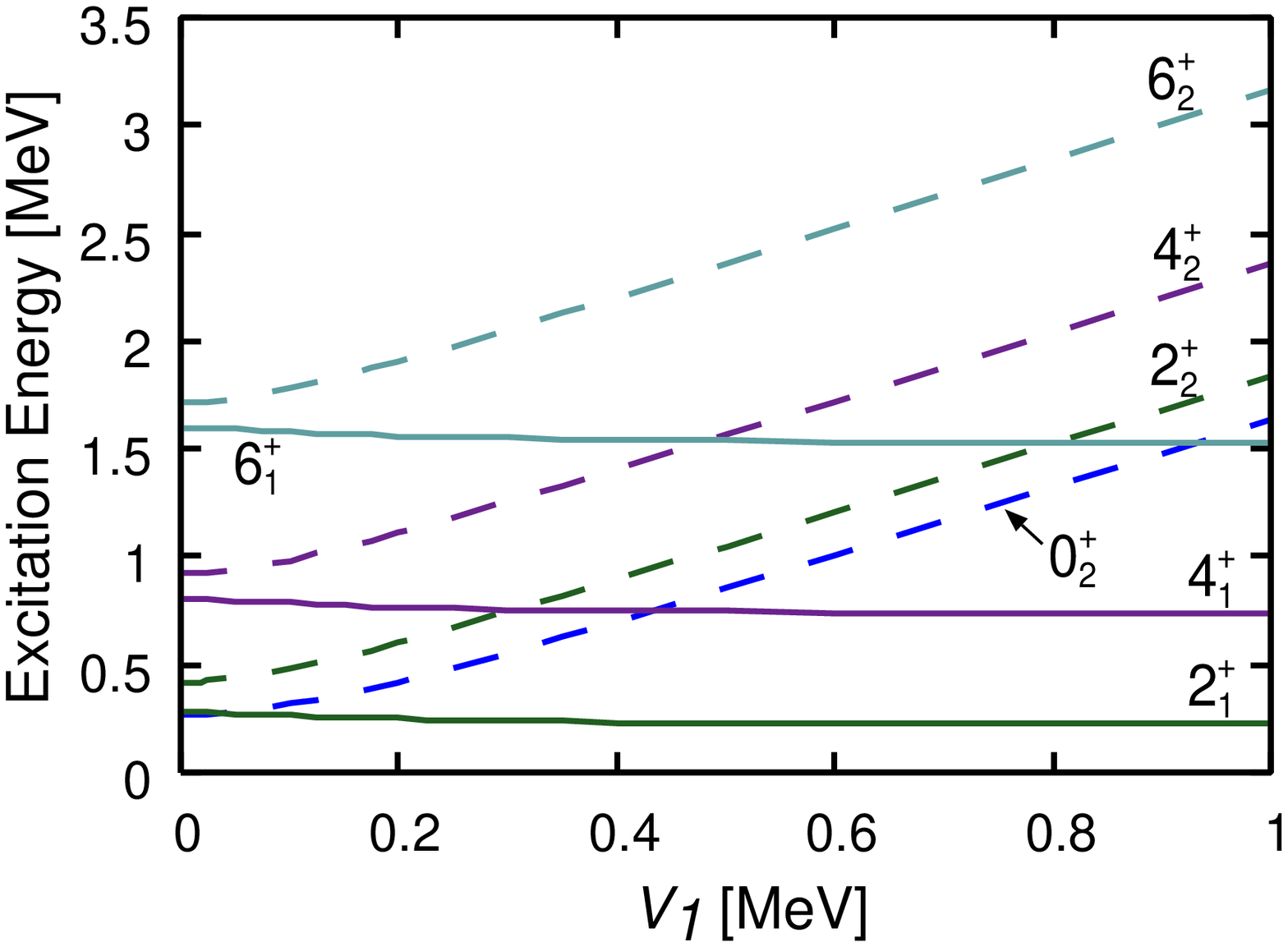}} \\
\setcounter{subfigure}{1}
\subfigure[$V_0=1.0$MeV]
{\includegraphics[width=0.5\textwidth]{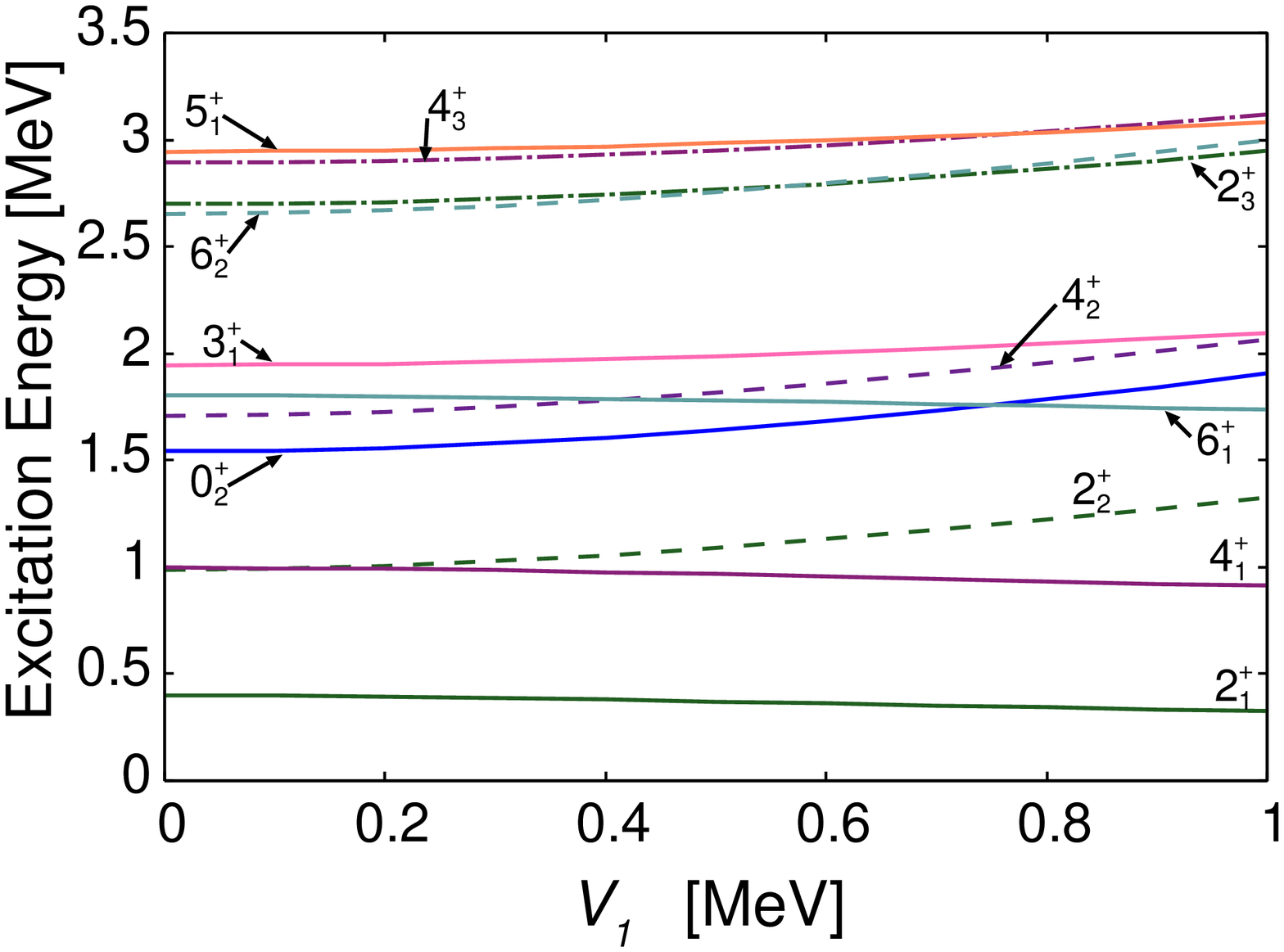} }
\end{tabular}
\caption{  Dependence of excitation spectrum on the asymmetry parameter $V_1$ is displayed 
  for $V_0=-2.0$ (panel a), 1.0 (panel b) and 10.0 MeV (panel c). }
\label{figs:Ex_V1}
\end{figure}


For three typical situations illustrated in Fig.~5, we examine  
the sensitivity of excitation spectra to the asymmetry parameter $V_1$.   
This is done in Fig.~6, (Panels (a), (b) and (c) display the excitation 
spectra as functions of $V_1$ for $V_0= -2, 1$ and 10 MeV, respectively.) 
We see that the dependence on $V_1$ is rather weak in the cases (a) and (b).
In contrast, the effect of the $V_1$ term is extremely strong in the case (c): 
the approximate doublet structure at $V_0=0$ is quickly broken as soon as 
the  $V_1$ term is switched on, and the
excitation energies of the yrare partners $(0_2^+, 2_2^+, 4_2^+, 6_2^+)$ 
remarkably increase as $V_1$ increases.
Quantitatively, the energy splittings between the yrast and yrare states 
with the same angular momenta are proportional to $V_1^2$ for 
$V_1 \lesssim 0.2$ MeV and then increase almost linearly in the region 
of $V_1 \gtrsim 0.2$ MeV. 
The quadratic dependence on $V_1$ in the small-$V_1$ region 
can be understood as the second-order perturbation effects 
in the double-well problem. 
In this case, what is important is not the absolute 
magnitude of $V_1$ but its ratio to the energy splitting 
due to the quantum tunneling through the potential barrier 
between the two minima. 
The structure of the collective wave functions is drastically 
changed by the small perturbation. 
Indeed, for $V_1 \gtrsim 0.2$ MeV, they are already well localized 
in one of the potential pockets, 
as we shall discuss below in connection with Fig.~9. 
Once the collective wave functions are well localized, 
the energy shifts are mainly determined by the diagonal matrix 
elements of the $V_1$ term (with respect to the localized wave functions),  
leading to the almost linear dependence on $V_1$.

\begin{figure}[b]
\begin{center}
  \includegraphics[width=0.6\textwidth]{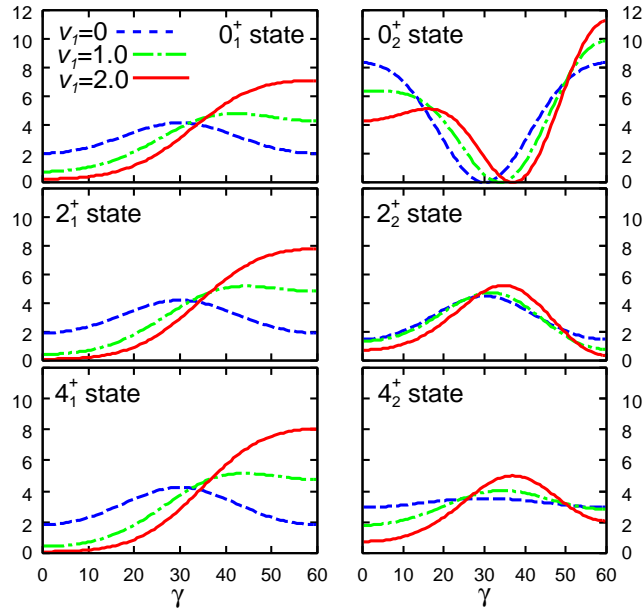}
   \caption{Collective wave functions squared $\sum_K |\Phi_{IK\alpha}(\gamma)|^2$ 
   of the yrast and the yrare states for the collective potentials 
   $V(\gamma)$ with $V_1=0.0, 1.0$ and $2.0$ MeV. 
   The barrier-height parameter $V_0$ is fixed at $-2.0$ MeV.}
   \label{fig:wfs_V0-2000_V1}
\end{center}
\end{figure}

\begin{figure}[t]
\begin{center}
  \includegraphics[width=0.6\textwidth]{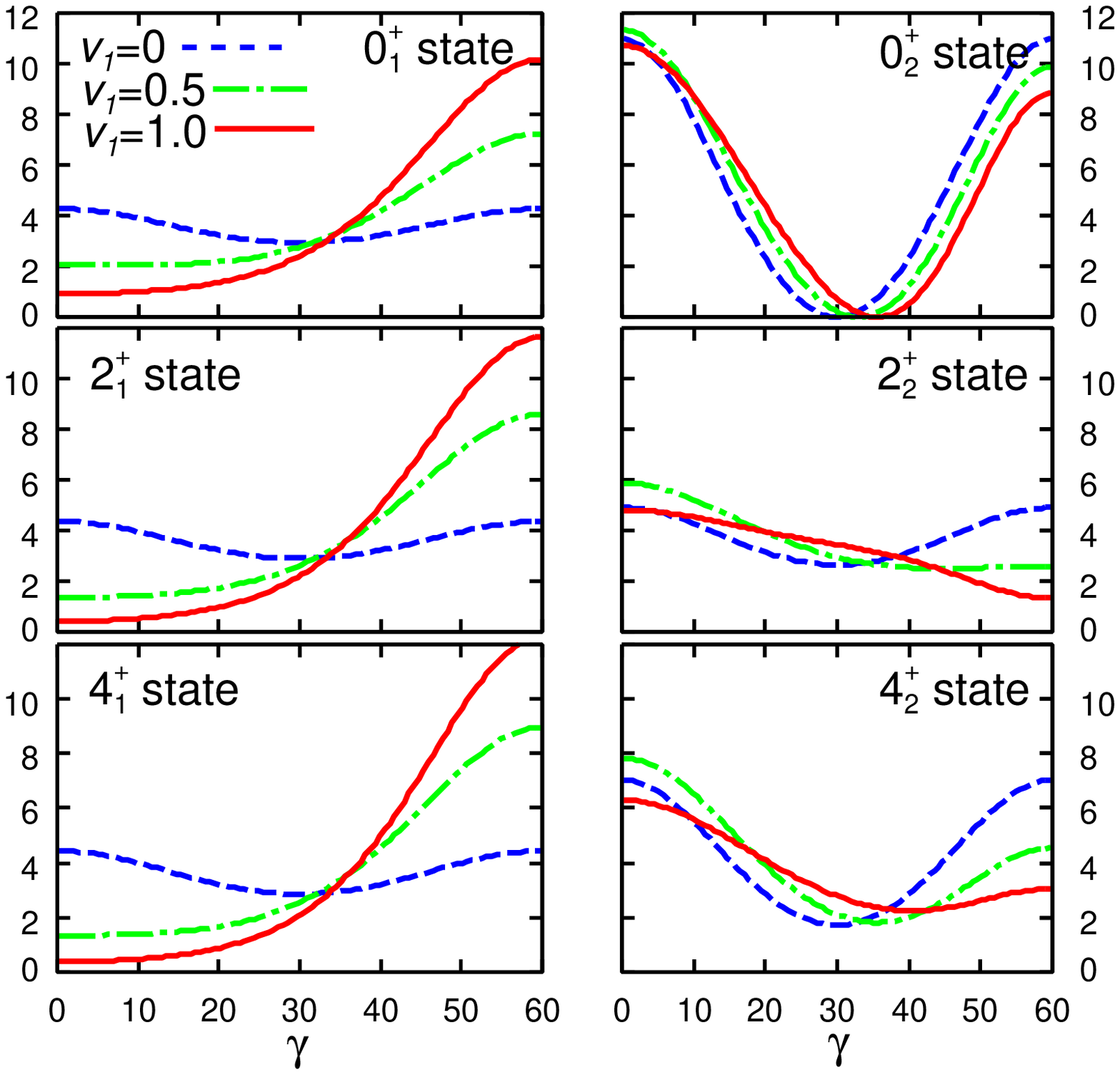}
   \caption{The same as Fig.~7 but for $V_0=1.0$ and $V_1=$0.0, 0.5 and 1.0 MeV.}
   \label{fig:wfs_V0+1000_V1}
\end{center}
\end{figure}


\begin{figure}[h]
\begin{center}
  \includegraphics[width=0.6\textwidth]{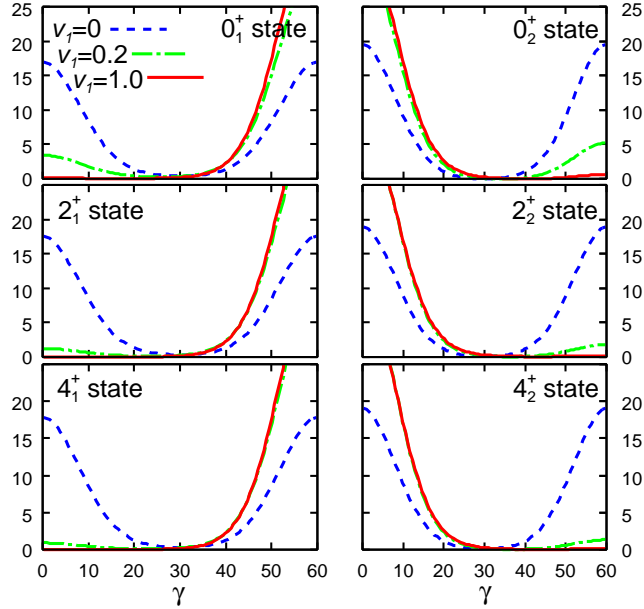}
   \caption{The same as Fig.~7 but for $V_0=10.0$ and $V_1=$0.0, 0.2 and 1.0 MeV.}
   \label{fig:wfs_V0+10000_V1}
\end{center}
\end{figure}
  

The effects of the $V_1$ term are seen more clearly in the collective wave functions.  
Figures~7, 8 and 9 show the collective wave functions squared 
$\sum_K |\Phi_{IK\alpha}(\gamma)|^2$ of the yrast and yrare states 
for $V_0 = -2, 1$ and 10 MeV, 
which respectively correspond to the potentials in panels (a), (b) and (c) of Fig.~5.  
In Figs.~7 and 8, one sees that the localization 
(with respect to the $\gamma$ coordinate) of the collective wave 
functions of the yrast states grows as $V_1$ increases, 
while that of the yrare states are insensitive to $V_1$.
This can be easily understood  by considering
the kinetic-energy effects tend to dominate over the potential-energy effects 
with increase in the excitation energies. 
It is interesting to note that in Fig.~8, 
although the yrast $4_1^+$ state is situated above the potential barrier, 
its wave function is well localized around the oblate minimum. 
Concerning the yrare $4_2^+$ state, although its wave function has the maximum 
at the prolate shape due to the orthogonality to the $4_1^+$ state, 
it is considerably extended over the entire region of $\gamma$. 
For the reason mentioned above, the effects of the $V_1$ term on localization properties of the yrare states are 
much weaker than those of the yrast states.

In Fig.~9, we see drastic effects of the $V_1$ term.
At $V_1=0$, the wave functions squared are symmetric 
about $\gamma=30^\circ$. 
This symmetry is immediately broken after 
the $V_1$ term is switched on:  
the wave functions of the yrast states rapidly localize 
around the potential minimum 
even if the energy difference between the two local minima is very small.  
In striking contrast to the situation presented in Fig.~8, 
the wave functions of the yrare states also localize 
about the second minimum of the potential. 
We confirmed that $V_1$ of 0.2 MeV is sufficient to bring about 
such strong localization.  
As pointed out above in connection with the double-well problem, 
this value of $V_1$ is small but comparable to the energy splittings. 
Thus, one can regard Fig.~9 as a very good example demonstrating 
that even small symmetry breaking in the collective potential 
is able to cause a dramatic change in the properties of 
the collective wave function, 
provided that the barrier is sufficiently high.

\begin{figure}[b]
\begin{center}
  \includegraphics[width=0.5\textwidth]{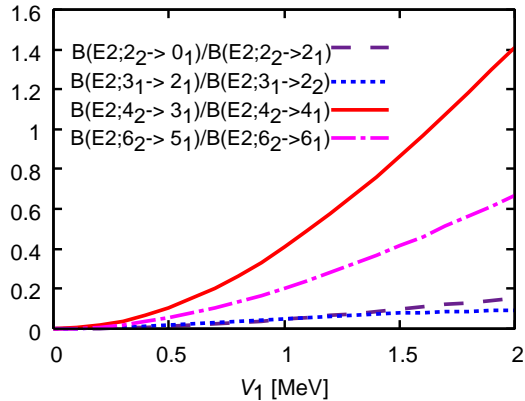}
  \caption{Dependence on the asymmetry parameter $V_1$ of selected $E2$ branching ratios 
  that vanish in the oblate-prolate symmetric limit $V_1=0$. 
  The barrier-height parameter $V_0$ is fixed at $-2.0$ MeV.}
  \label{fig:BE2_V0-2}
\end{center}
\end{figure}


In the remaining part of this subsection, 
we discuss how the $V_1$ term affects properties of 
electric quadrupole ($E2$) transitions and moments. 
In Fig.~10,  the $B(E2)$ ratios 
that vanish in the $V_1=0$ limit are plotted as functions of $V_1$ 
for the $V_0=-2.0$ MeV case corresponding to Figs.~5(a), 6(a) and 7. 
Vanishing of these ratios is well known as one of the signatures of 
the triaxial shape with $\gamma=30^\circ$ 
in the rigid triaxial rotor model. \cite{dav58}
It should be emphasized, however, 
that this is in fact a consequence of OP symmetry: 
these $E2$ transitions vanish due to { the exact cancellation between 
 the contribution from the prolate side ($0^\circ \le \gamma < 30^\circ$) 
and that from the oblate side ($30^\circ < \gamma \le 60^\circ$).}   
Therefore, the localization around $\gamma=30^\circ$ is 
not a necessary condition. 
In fact, these ratios vanish also in the $\gamma$-unstable model. 
\cite{wil56}
As anticipated, we see in Fig.~10 that these ratios increase 
as the collective potential becomes more asymmetric with respect to 
the oblate and prolate shapes. 
In particular, the significant rises of the ratios, 
$B(E2;4_2^+ \rightarrow 3_1^+)/B(E2;4_2^+ \rightarrow 4_1^+)$ and 
$B(E2;6_2^+ \rightarrow 5_1^+)/B(E2;6_2^+ \rightarrow 6_1^+)$, 
are remarkable. 
This may be interpreted as an incipient trend that the sequence of 
states $(2_2^+, 3_1^+, 4_2^+, 5_1^+$ and $6_2^+)$ forms 
a $\gamma$-vibrational bandlike structure 
about the axially symmetric shape (oblate in the present case) 
when $V_1$ becomes very large.

\begin{figure}[t]
\begin{tabular}{cc}
\subfigure[$V_0=1.0$MeV]
{\includegraphics[width=0.5\textwidth]{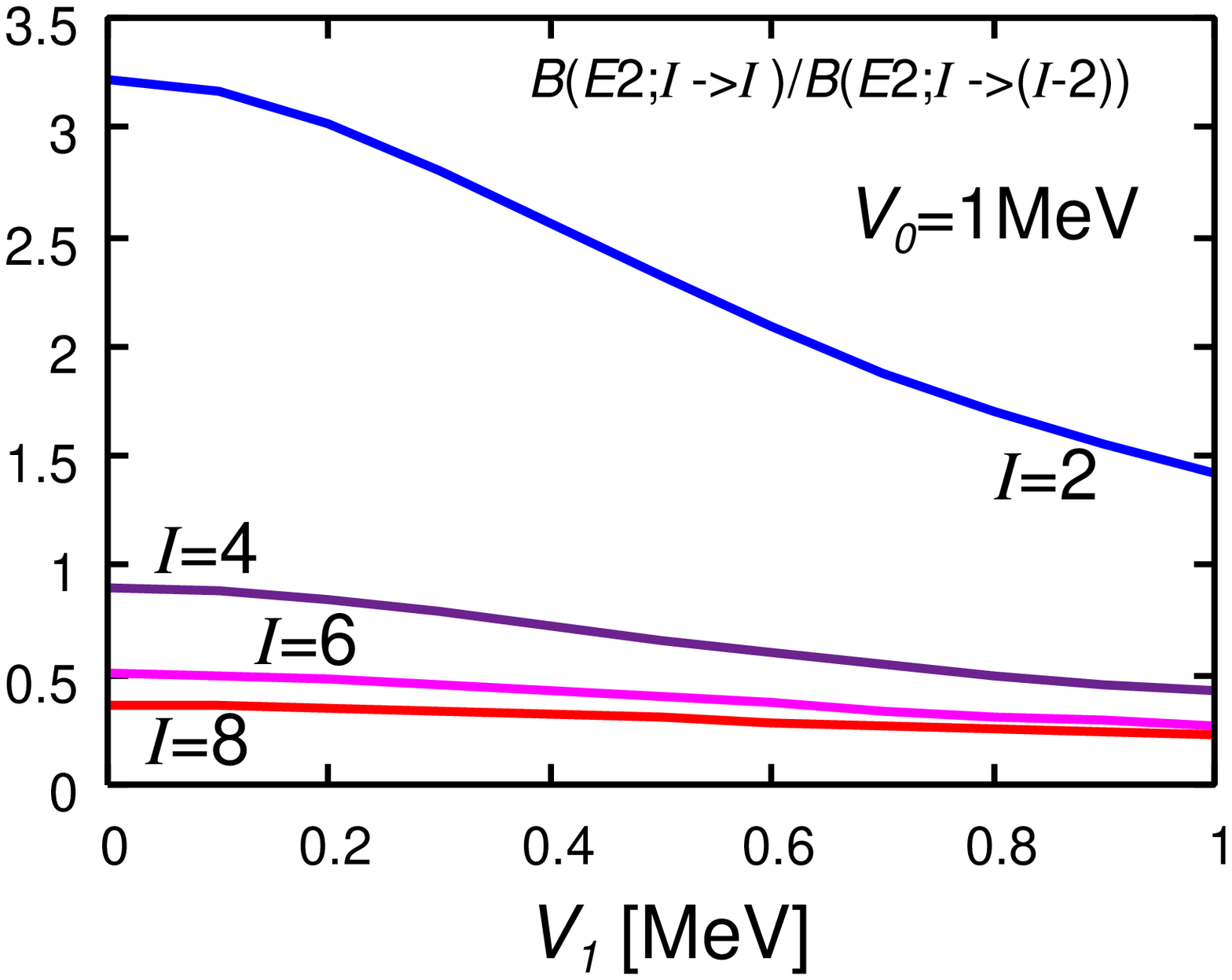} }
\subfigure[$V_0=10.0$MeV]
{  \includegraphics[width=0.5\textwidth]{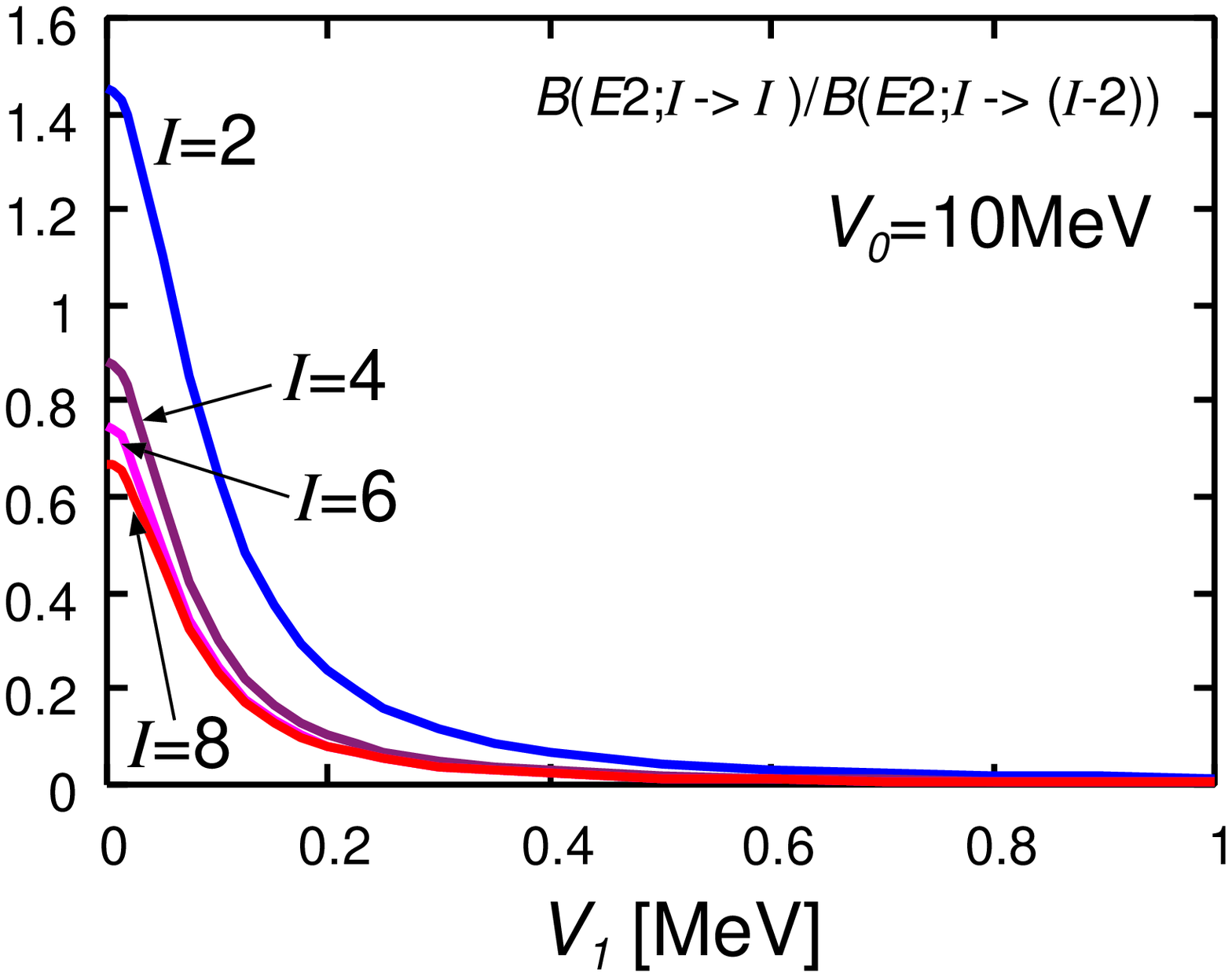}}
\end{tabular}
\caption{  Dependence on the asymmetry parameter $V_1$ of 
  the branching ratios $B(E2;I_{\rm yrare} \rightarrow I_{\rm yrast})$/ 
  $B(E2;I_{\rm yrare} \rightarrow (I-2)_{\rm yrare})$. 
  Here, $I_{\rm yrast}$ and $I_{\rm yrare}$ denote the yrast and the yrare states 
  having the same values of angular momenta $I$, while 
  $(I-2)_{\rm yrare}$ the yrare states with angular momentum $I-2$. 
  The left and right panels display the results of calculation 
  for $V_0=1.0$ and 10.0 MeV, respectively.}
\label{figs:BE2_V1}
\end{figure}


In Fig.~11, the ratios 
$B(E2;I_{\rm yrare} \rightarrow I_{\rm yrast})$/ 
$B(E2;I_{\rm yrare} \rightarrow (I-2)_{\rm yrare})$ 
are plotted as functions of $V_1$ for the two cases of $V_0=1.0$ and 10.0 MeV.  
The $V_0=1.0$ case corresponds to Figs.~5(b), 6(b) and 8, while 
the $V_0=10.0$ case corresponds to Figs.~5(c), 6(c) and 9. 
Here, the numerator denotes $B(E2)$ values for $\Delta I=0,~E2$ transitions 
from the yrare to the yrast states with the same angular momenta $I$, 
while the denominator indicates $\Delta I=-2,~E2$ transitions 
between the yrare states. 
One immediately notices a sharp contrast between the two cases: 
when the barrier between the oblate and prolate local minima is very low 
($V_0=1.0$ MeV), these values are sizable even at $V_1=1.0$MeV, although they gradually 
decrease with increase of $V_1$. 
In contrast, when the barrier is very high ($V_0=10.0$ MeV), 
they quickly decrease once the OP-symmetry-breaking term is turned on. 
The reason why the interband $E2$ transitions almost vanish is apparent 
from Fig.~9; the collective wave functions of the yrast and yrare states are 
well localized around the oblate and prolate shapes, respectively.

\begin{figure}[h]
 \begin{tabular}{ccc}
\subfigure[$V_0=1.0$MeV]
{\includegraphics[width=0.5\textwidth]{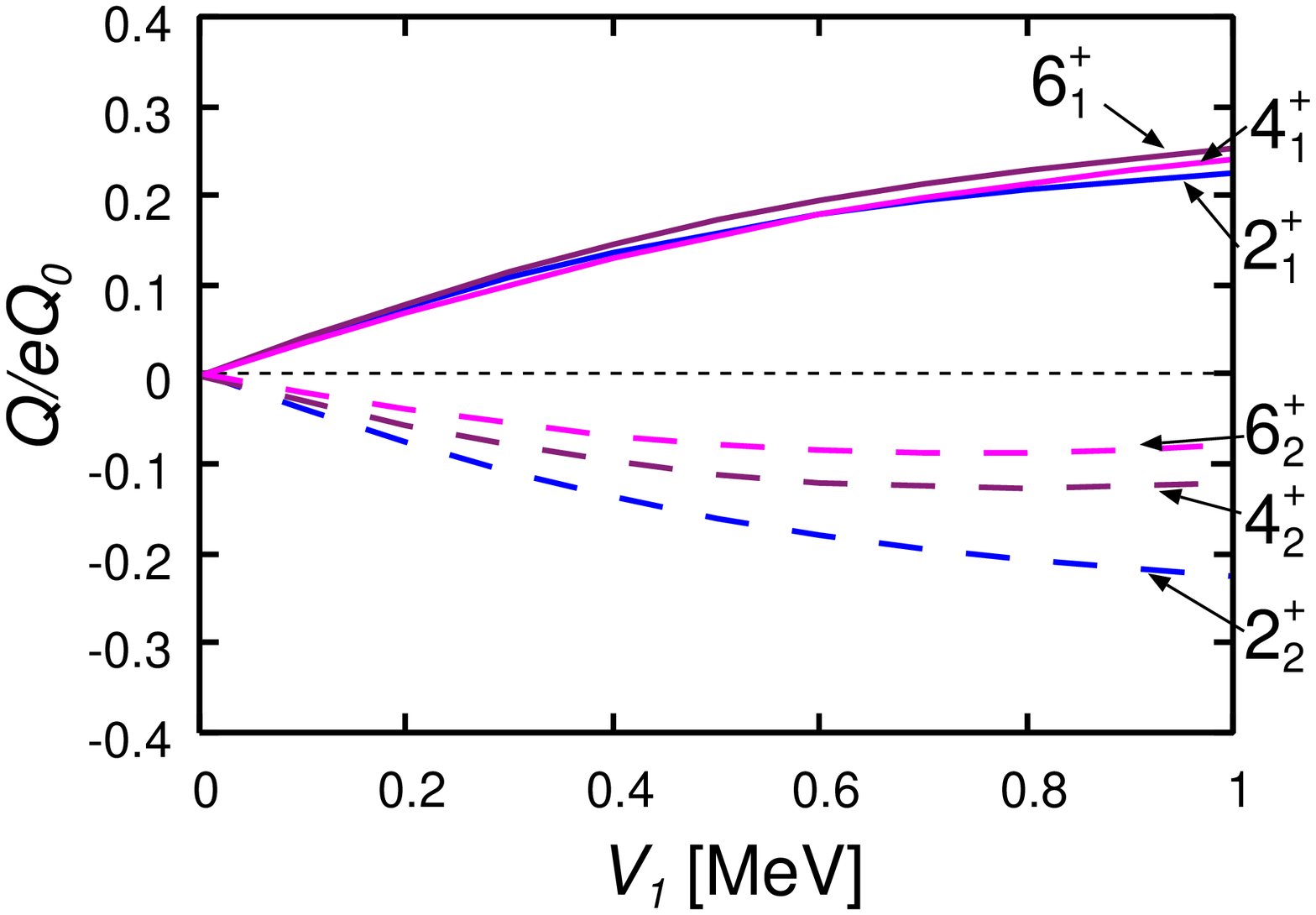}}
\subfigure[$V_0=10.0$MeV]
{\includegraphics[width=0.5\textwidth]{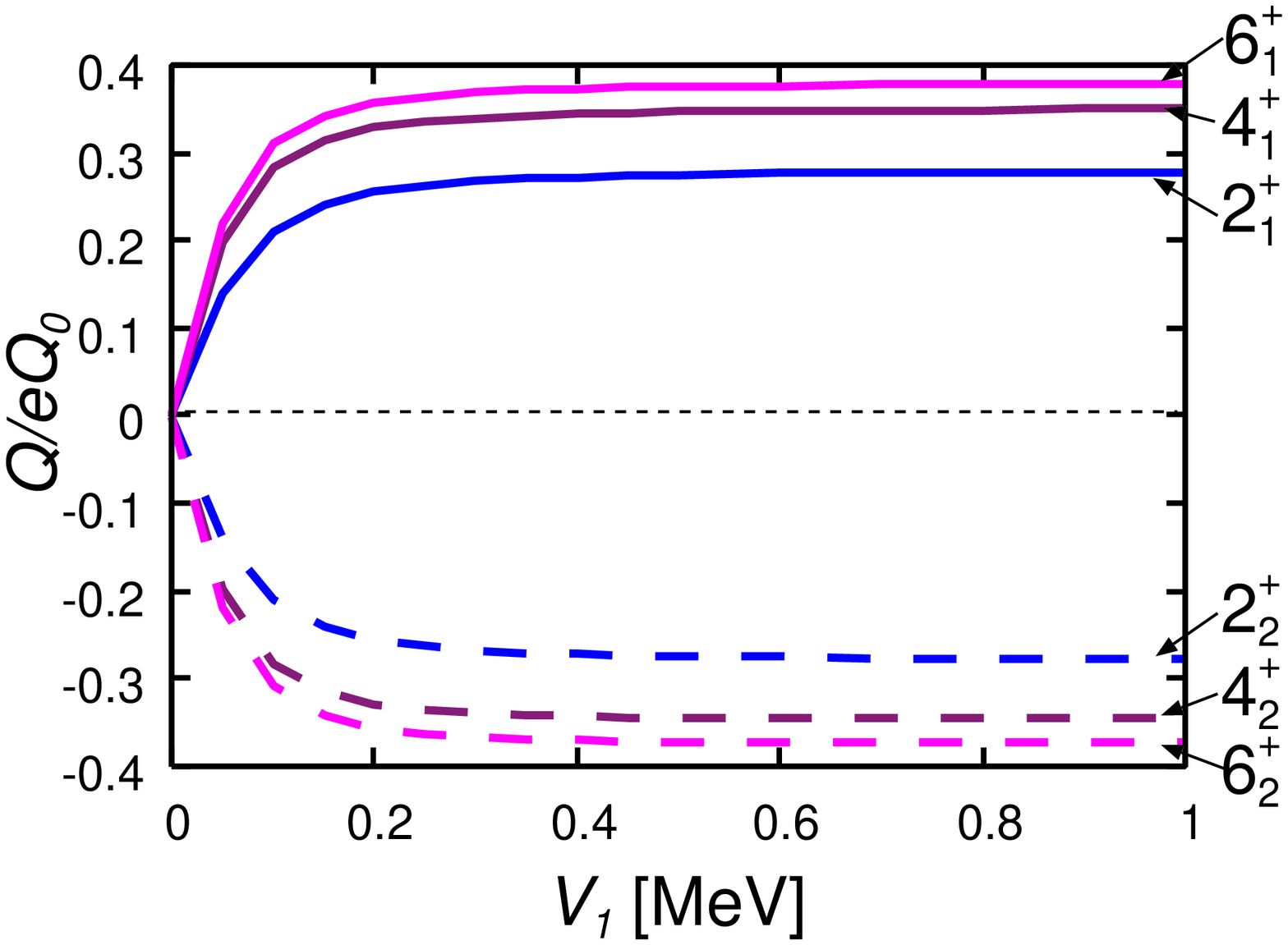}}  
 \end{tabular}
     \caption{Dependence of the spectroscopic quadrupole moments 
  on the asymmetry parameter $V_1$, 
  calculated for $V_0=1.0$ MeV (left panel) and $V_0=10.0$ MeV (right panel). 
  Their values are plotted in units of the intrinsic quadrupole moment, $eQ_0=3e/\sqrt{5\pi} ZR_0^2\beta_0$.}
     \label{figs:Q_V1}
\end{figure}
  

One can further confirm the same point by looking at the spectroscopic quadrupole 
moments displayed in Fig.~12 in a way parallel to Fig.~11. 
They vanish in the presence of the OP symmetry due to the exact cancellation 
between the contributions from the oblate side and from the prolate side.  
In Fig.~12(a) for the low barrier case ($V_0=1.0$ MeV), 
the quadrupole moments of both the yrast and yrare states first increase 
after the $V_1$ term is switched on, 
but those for the yrare states eventually saturate. 
This trend is obvious especially for the $4_2^+$ and $6_2^+$ states. 
These results are easily understandable from the properties of their wave 
functions displayed in Fig.~8. 
That is, while the localization in the yrast states develops with $V_1$ increasing, the wave functions of the yrare states widely 
extend over the entire region of $\gamma$, 
and the effects of the $V_1$ term  are rather weak. 
In contrast, Fig.~12(b) for the high barrier case ($V_0=10.0$ MeV) 
demonstrates that both the yrast and yrare states quickly acquire 
quadrupole moments as soon as the $V_1$ term is switched on. 
This is a direct consequence of the wave function localization 
displayed in Fig.~9. 
Note again that the sign change, $V_1 \leftrightarrow -V_1$, 
corresponds to the OP inversion.

When $V_0$ is large and $V_1$ is small, just as in the above case, 
the yrast and the yrare states can be grouped, in a very good approximation, 
into two rotational bands: 
one associated with the oblate shape and the other with the prolate shape.
This is an ideal situation for the emergence of an oblate-prolate shape coexistence phenomenon. 
According to the realistic HFB calculations for the collective potential 
\cite{yam01}, however, it seems hard to obtain such a large value of $V_0$. 
Therefore, for the shape coexistence phenomena 
we need to take into account dynamical effects going beyond the consideration 
on the collective potential energies. 
We shall discuss this point in the succeeding subsection.

\subsection{Breaking of OP symmetry in the collective mass}

We  examine dynamical effects on the localization properties of the collective wave functions. 
As mentioned in \S1, we are particularly interested in 
understanding the nature of the shape coexistence phenomena 
observed in nuclei 
for which approximately degenerate oblate and prolate 
local minima and a rather low barrier between them 
are suggested from the microscopic potential energy calculations. 
\cite{yam01,hin08,hin09} 
In the followings, we therefore concentrate our discussion on 
the case of the collective potential with a low barrier ($V_0=1.0$) and 
weak OP asymmetry ($V_1=0.5$) represented in Fig.~5(b). 
We shall investigate how the results 
discussed in the previous subsections 
for the $\E=0$ case, where the collective mass 
$D_{\gamma\gamma}(\gamma), D_{k}(\gamma)$ and 
the rotational moments of inertia $\mathcal{J}_k(\gamma)$ are 
symmetric functions about $\gamma=30^\circ$, 
are modified when  the mass-asymmetry parameter $\E$ 
becomes nonzero.

\begin{figure}[t]
\begin{center}
  \begin{tabular}{ccc}
    \begin{minipage}[b]{0.5\textwidth}
      \includegraphics[width=\textwidth, trim=0 0 0 0,clip]{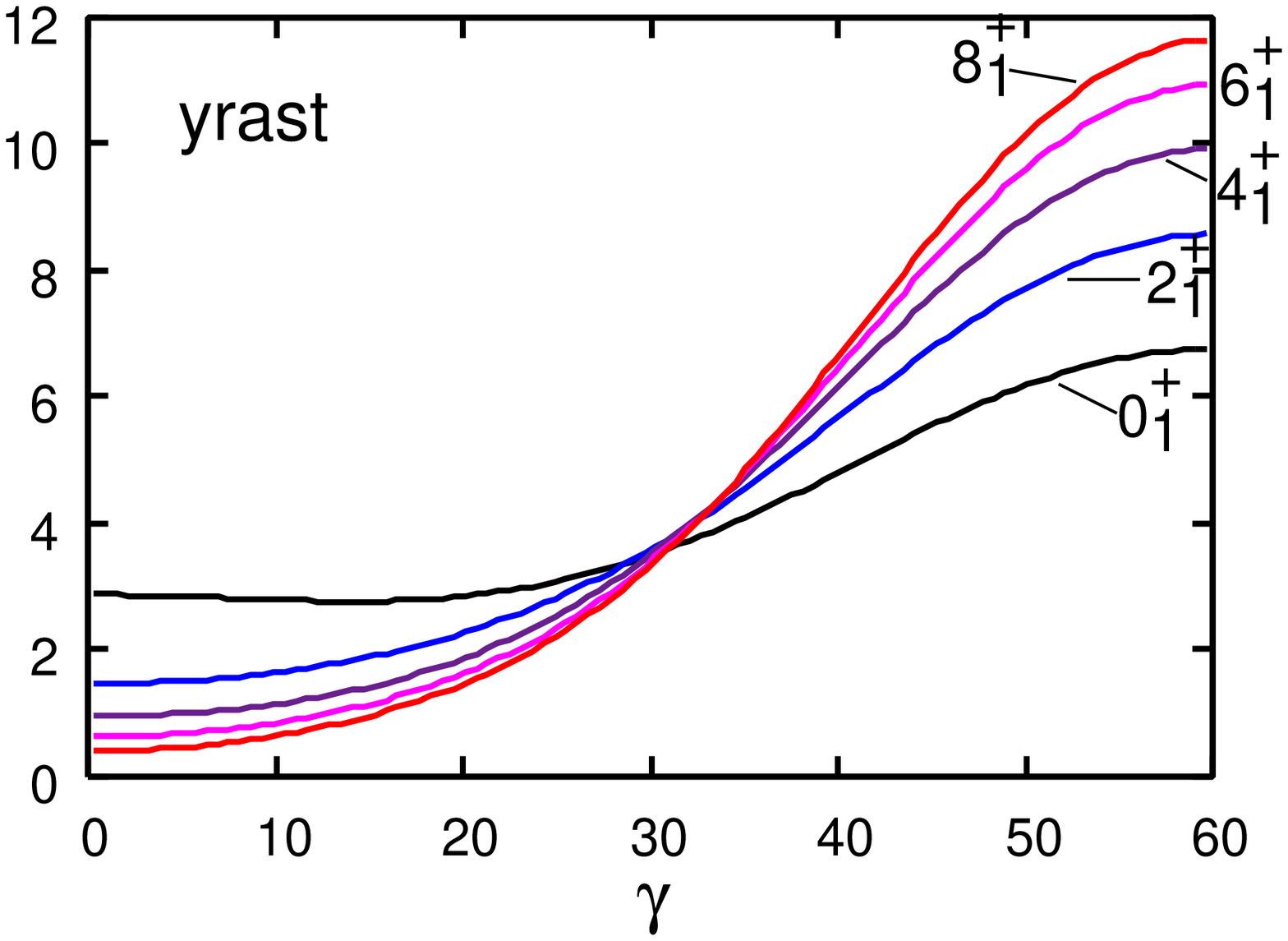}
    \end{minipage}
    \begin{minipage}[b]{0.5\textwidth}
      \includegraphics[width=\textwidth, trim=0 0 0 0,clip]{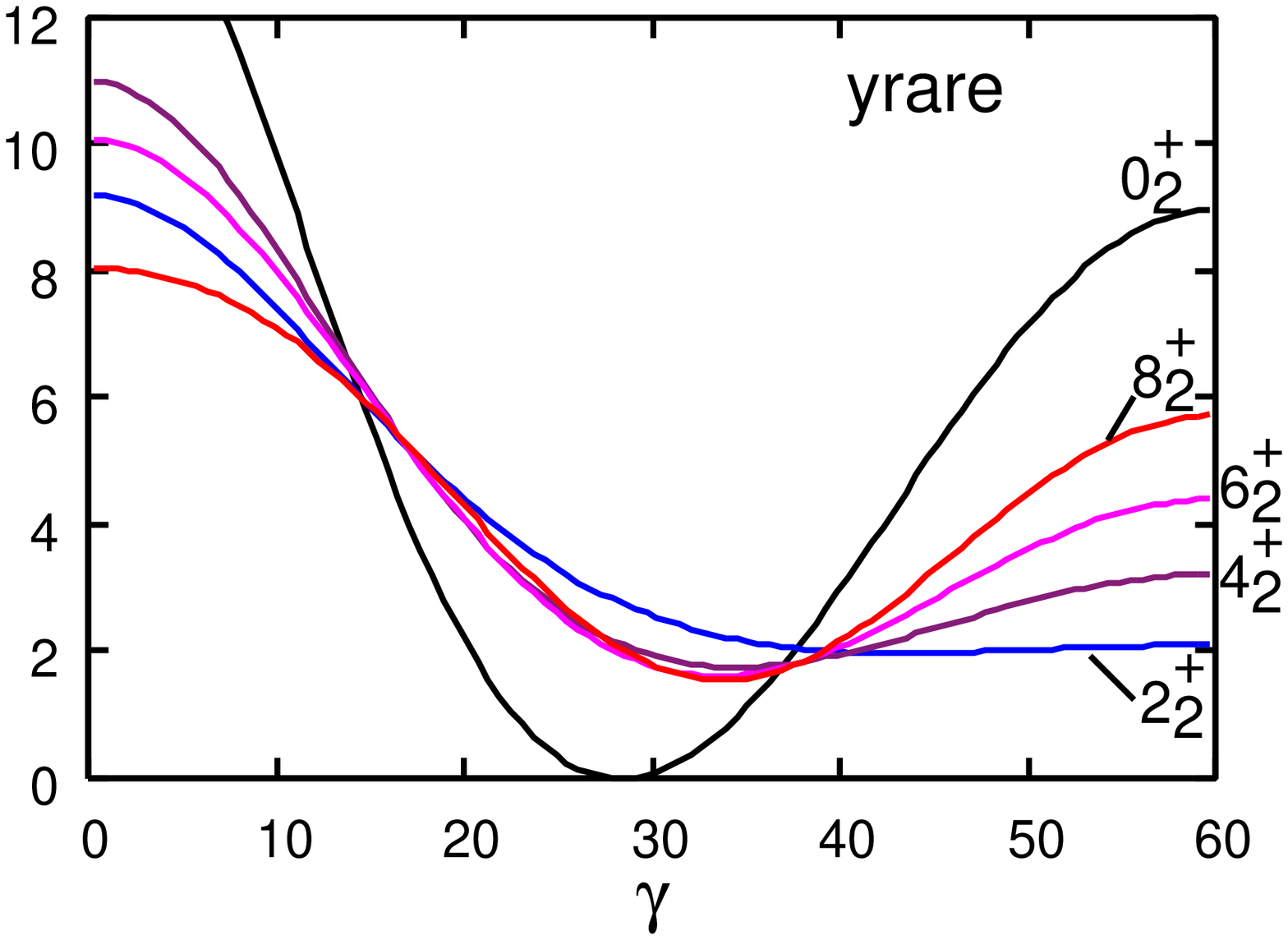}
    \end{minipage}
    \end{tabular}
    \caption{  Dependence on angular momentum of the collective wave functions 
  calculated for the mass-asymmetry parameter $\E=0.5$ and  
  the collective potential $V(\gamma)$ with $V_0=1.0$ and $V_1=0.5$ MeV. 
  The left and the right panels display the results of calculation 
  for the yrast and the yrare states, respectively.}
    \label{fig:wfs_yrast_V0+1_V1+0.5_e+0.50}
\end{center}
\end{figure} 


In Fig.~13, the collective wave functions squared 
$\sum_K|\Phi_{IK\alpha}(\gamma)|^2$ calculated for $\E=0.5$ are displayed. 
It is clearly seen that, for the yrast states, 
the localization around the oblate shape ($\gamma=60^\circ$) 
develops with increase in the angular momentum. 
The reason is easily understood: for positive $\E$, 
the rotational moments of inertia 
perpendicular to the oblate symmetry axis (2nd axis), 
$\mathcal{J}_1(\gamma=60^\circ)$ and $\mathcal{J}_3(\gamma=60^\circ)$, 
are larger than those perpendicular to the prolate symmetry axis (3rd axis),
$\mathcal{J}_1(\gamma=0^\circ)$ and $\mathcal{J}_2(\gamma=0^\circ)$, 
as shown in Fig.~2(b). Therefore, the rotational energy for a given angular momentum 
decreases by increasing the probability of existence around the oblate shape. 
Since the rotational energy dominates 
over the vibrational and potential energies,  
the localization is enhanced for higher angular momentum states.
We call this kind of dynamical effect 
{\it rotation-assisted localization.} 
On the other hand, the wave functions of the yrare states 
exhibit  two-peak structure: the first peak at the prolate shape ($\gamma=0^\circ$) 
and the second at the oblate shape($\gamma=60^\circ$)  
except in the $2_2^+$ state.
One might naively expect that the yrare states would localize about 
the prolate shape because of the orthogonality requirement to the yrast states.
However, 
the second peak is formed around the oblate shape 
in order to save the rotational energy  keeping the orthogonality condition.

\begin{figure}[h]
\begin{center}
  \includegraphics[width=0.6\textwidth]{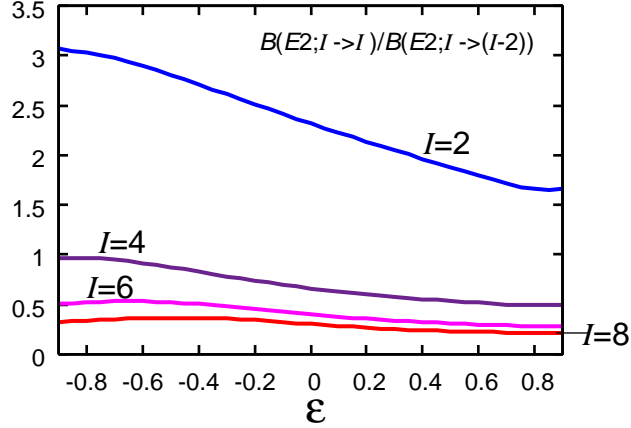}
  \caption{Dependence on the mass-asymmetry parameter $\E$ of the ratios  
  $B(E2;I_{\rm yrare} \rightarrow I_{\rm yrast})/
  B(E2;I_{\rm yrare} \rightarrow (I-2)_{\rm yrare})$, 
  calculated for $V_0=1.0$ and $V_1=0.5$ MeV. 
  Here, $I_{\rm yrast}$ and $I_{\rm yrare}$ denote the yrast and the yrare states 
  having the same values of angular momenta $I$, while 
  $(I-2)_{\rm yrare}$ the yrare states with angular momentum $I-2$.}
  \label{fig:BE2_V0+1_V1+0.5_e}
\end{center}
\end{figure}


Due to the two peak structure of the yrare wave functions mentioned above, 
the ratios of $B(E2)$ values from an yrare state to an yrast state 
to those between the yrare states remain rather large for a wide region of 
the mass-asymmetry parameter $\E$. This is shown in Fig.~14.

\begin{figure}[h]
\begin{center}
  \begin{tabular}{ccc}
    \begin{minipage}[b]{0.5\textwidth}
      \includegraphics[width=\textwidth, trim=0 0 0 0,clip]{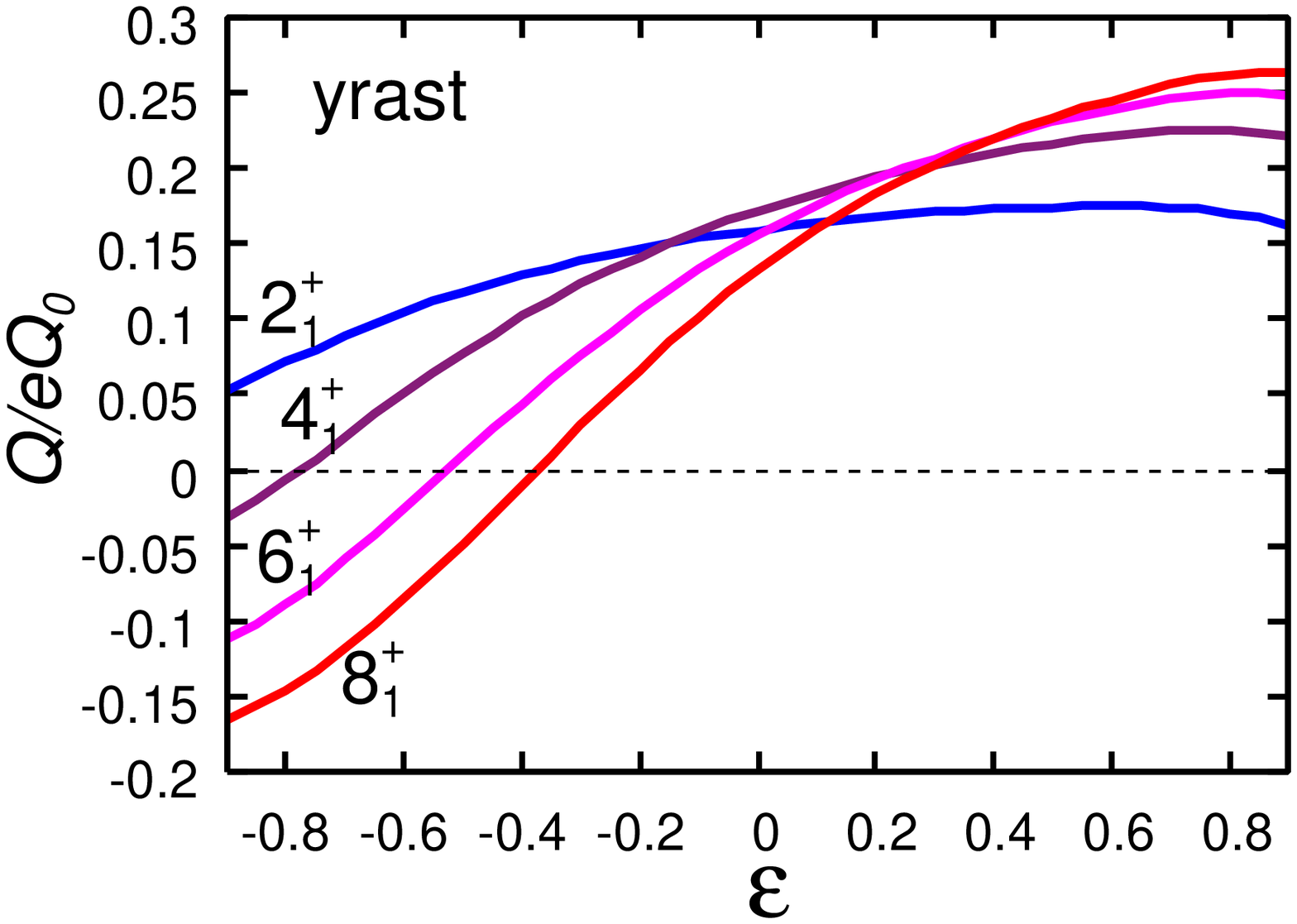}
    \end{minipage}
    \begin{minipage}[b]{0.475\textwidth}
      \includegraphics[width=\textwidth, trim=0 0 0 0,clip]{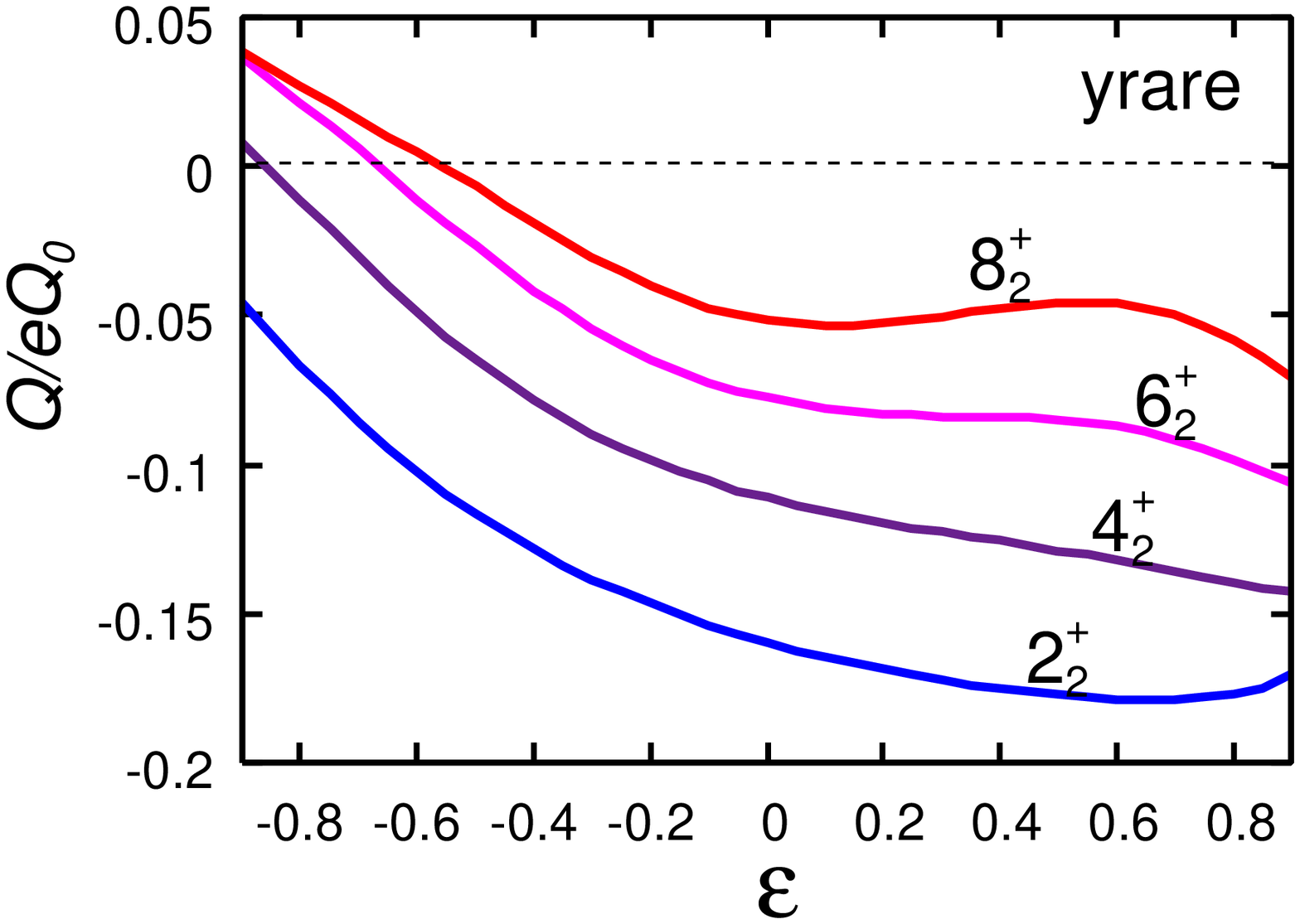}
    \end{minipage}
    \end{tabular}
    \caption{Dependence of the quadrupole moments on the mass-asymmetry parameter $\E$, 
  calculated for $V_0=1.0$ and $V_0=0.5$ MeV. 
  The left and the right panels display the results of calculation 
  for the yrast and the yrare states, respectively.}
    \label{Q_V0+1_V1+0.5_e}
\end{center}
\end{figure}


In spite of the two-peak structure of the yrare wave functions, 
we can find a feature of shape coexistence 
in the spectroscopic quadrupole moments $Q$, 
which are shown in Fig.~15 as functions of $\E$. 
Let us first concentrate on the $\E>0$ part of this figure. 
We see that the $Q$ values of the yrast states 
are positive, indicating their oblatelike character, 
while the yrare states have negative $Q$,  
indicating their prolatelike character.
This can be regarded as a feature of the shape coexistence.
It is furthermore seen that the absolute magnitude of $Q$ increases 
as $\E$ increases, except the $8_2^+$ state. 
As discussed above in connection with Fig.~13, 
for positive $\E$, the oblate shape is favored to lower 
the rotational energy. Consequently, with $\E$ increasing, the collective wave functions in the yrast band tend to 
localize around the oblate shape more and more  
and those of the yrare states localize around the prolate shape because of the orthogonality if the angular momentum is not so high.
 
 
It is also noticeable that the absolute magnitude of $Q$ 
in the yrare band decreases with increase in the angular momentum.   
This is because the cancellation mechanism between the contributions from 
the oblatelike and prolatelike regions of the collective wave function 
works more strongly as the two-peak structure grows.   

Next, let us discuss on the $\E<0$ part of Fig.~15.
For negative $\E$, the prolate shape is favored to lower 
the rotational energy. 
On the other hand, the collective potential under consideration 
($V_1=0.5$ MeV) is lower for the oblate shape. 
Hence, the rotational energy and the potential energy compete 
to localize the collective wave function into the opposite directions. 
It is seen in Fig.~15 that the spectroscopic quadrupole moments of the 
yrast states decrease with $\E$ decreasing and that 
this trend is stronger for higher angular momentum states. 
As a consequence, the $Q$ values of the $6_1^+$ and $8_1^+$ states become 
negative for large negative $\E$, which implies that the rotational effect 
dominates there.

\begin{figure}[b]
  \begin{tabular}{ccc}
\subfigure[$2_1^+$]
{      \includegraphics[width=.5\textwidth, trim=0 0 0 0,clip]{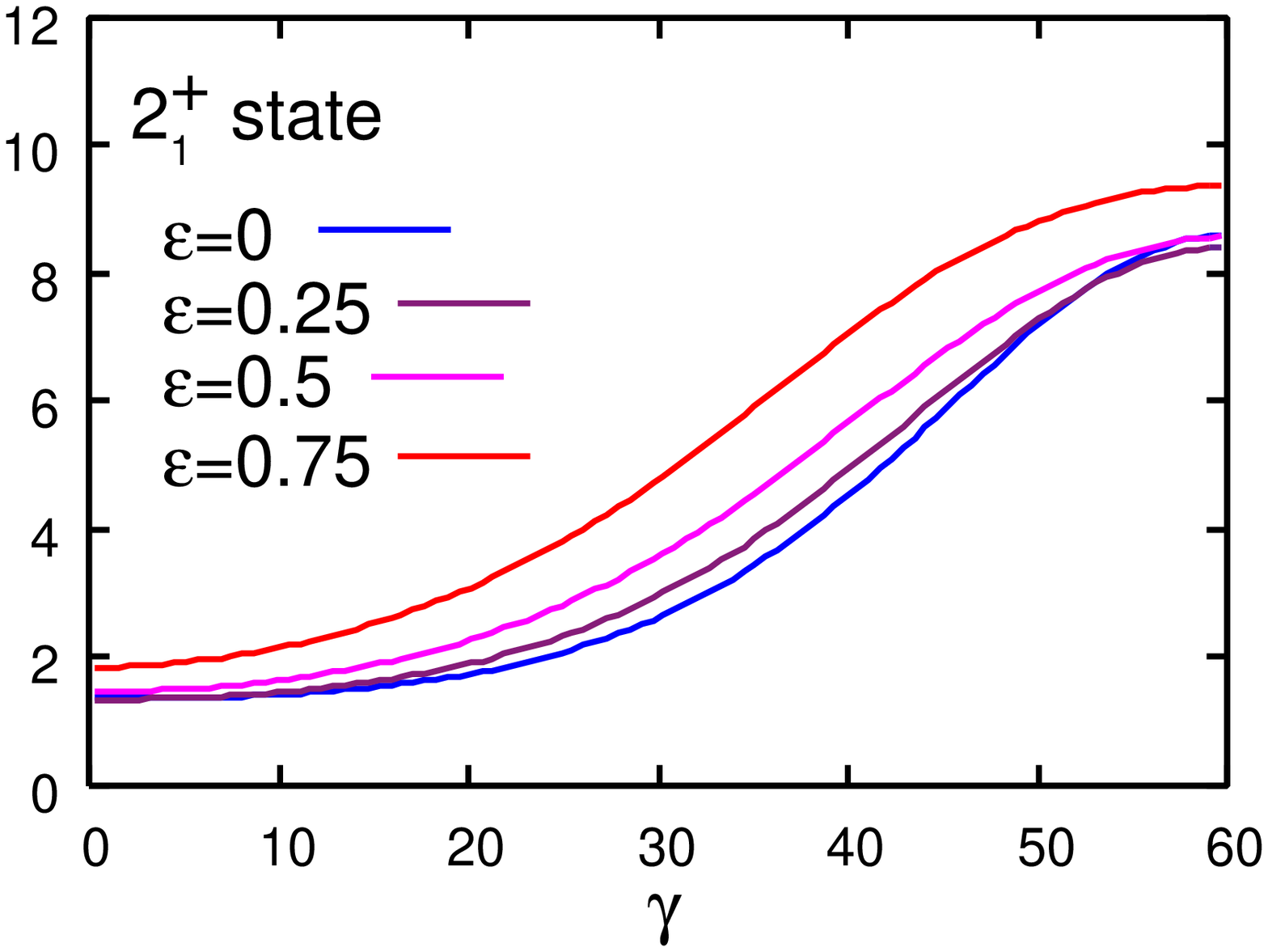}}
\subfigure[$2_2^+$]
{      \includegraphics[width=.5\textwidth, trim=0 0 0 0,clip]{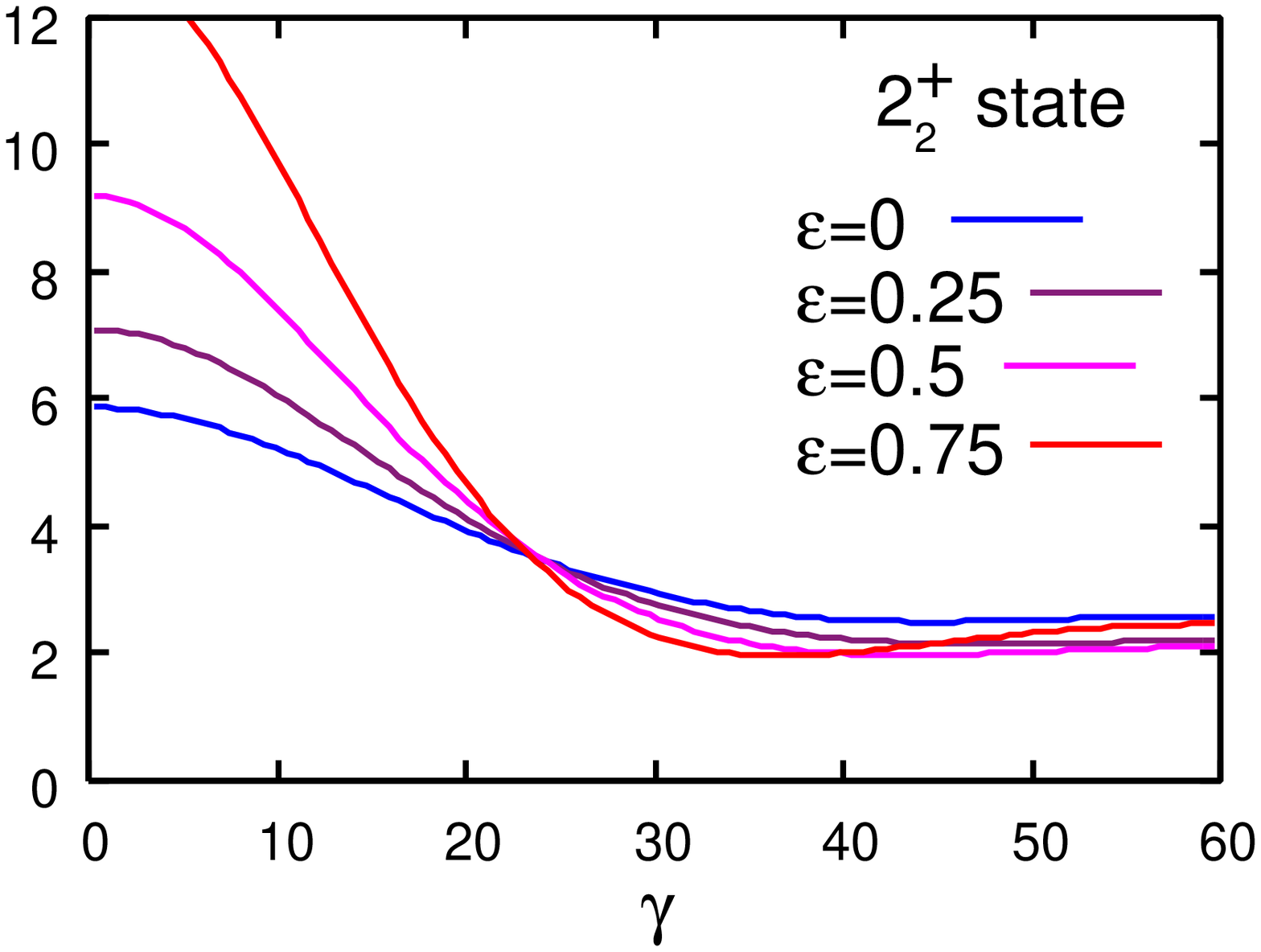}} \\
\subfigure[$8_1^+$]
{      \includegraphics[width=.5\textwidth, trim=0 0 0 0,clip]{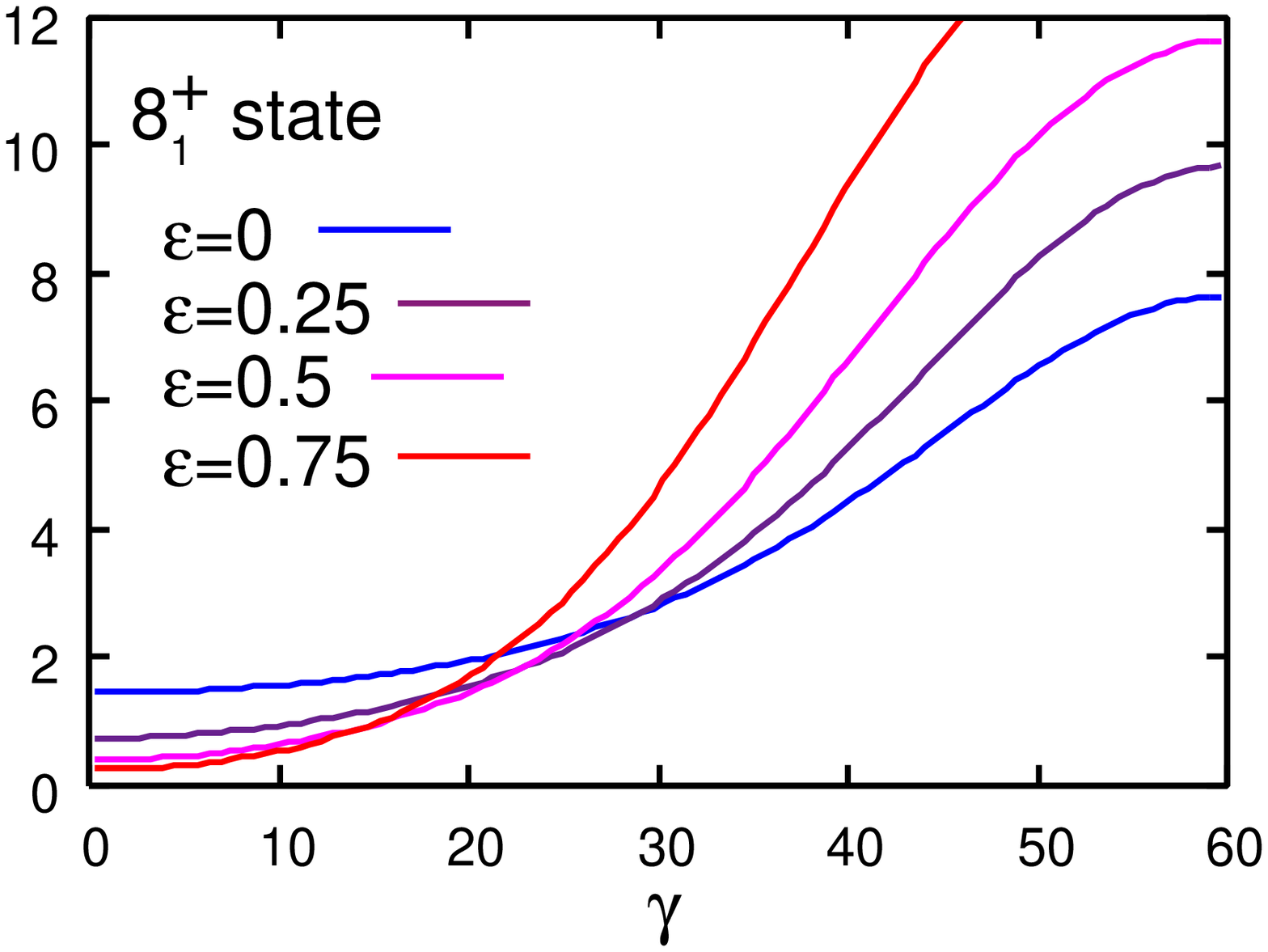}}
\subfigure[$8_2^+$]
{      \includegraphics[width=.5\textwidth, trim=0 0 0 0,clip]{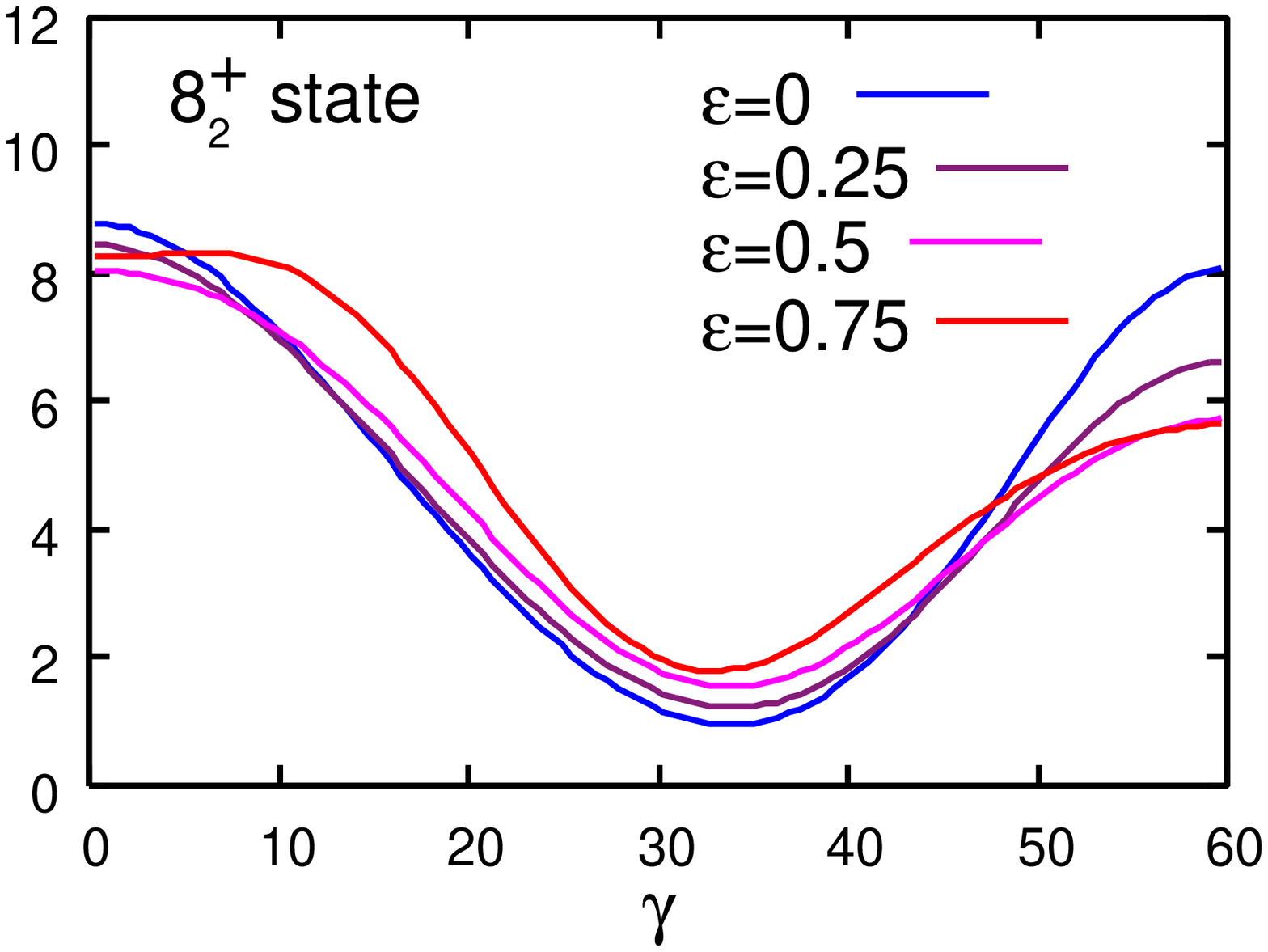} }
\end{tabular}
\caption{Dependence of the collective wave functions 
  on the mass-asymmetry parameter $\E$, 
  calculated for $V_0=1.0$ and $V_1=0.5$ MeV. 
  The upper-left and the upper-right panels display the results of 
  calculation for the yrast and the yrare states with $I^\pi=2^+$, 
  while the lower-left and the lower-right panels show those for 
  the yrast and the yrare states with $I^\pi=8^+$.}
\label{figs:wfs_2and8_V0+1000_V1+500_e} 
\end{figure}


Finally, we show in Fig.~16 how the localization properties of 
the collective wave functions depend on the mass-asymmetry parameter $\E$, 
taking the $2^+$ and $8^+$ states as representatives of 
low and high angular momentum states.
It is clearly seen that, for the yrast states, while the 
the localization of the $2_1^+$ state around the oblate shape is rather insensitive to $\E$, that of $8_1^+$ state 
remarkably develops with increasing $\E$.
For the yrare states,  
the localization of the $2_2^+$ state around the prolate shape grows with 
increasing $\E$, while the $8_2^+$ state retains 
the two-peak structure discussed above for every value of $\E$. 
The different effects of the mass-asymmetry parameter $\E$ 
on the $2^+$ and $8^+$ states are comprehensible from the 
consideration of the relative importance of the rotational and 
vibrational energies shown in Table I. 
We see that the rotational energies dominate in the $8^+$ states,
while the vibrational energies are comparable in magnitude to 
the rotational energies in the $2^+$ states.  
Thus, the effect of the mass-asymmetry parameter $\E$ on 
the localization properties of the $8^+$ states can be 
easily accounted  by the rotational energy. 
On the other hand, 
in the situation characterized by the parameters 
$V_0=1.0$ and $V_1=0.5$ MeV, the properties of the $2^+$ states 
are determined by a delicate competition 
between the rotational and vibrational kinetic energies 
as well as the potential energy. 
The growth of the prolate peak with increasing $\E$ 
in the $2_2^+$ state { turns out due mainly to the increase of the vibrational 
mass $D_{\gamma\gamma}$ at the prolate shape}.

\begin{table} 
\caption{Expectation values of the vibrational and rotational energies 
in units of MeV for the $2_1^+,2_2^+,8_1^+$and $8_2^+$ states.
The results for $\E=0.0$ and $0.5$ are shown.} 
\begin{center}
\begin{tabular}{cccccc} \hline \hline
\multicolumn{2}{c}{ }                              & $2_1^+$ & $2_2^+$  & $8_1^+$ & $8_2^+$  \\ \hline
 $\E=0.0$ & $\langle \hat T_{\rm vib} \rangle $ &  0.23   & 0.42     & 0.35    & 1.09     \\ 
          & $\langle \hat T_{\rm rot} \rangle $ &  0.29   & 0.55     & 2.56    & 2.87     \\ \hline
 $\E=0.5$ & $\langle \hat T_{\rm vib} \rangle $ &  0.20   & 0.51     & 0.49    & 1.03     \\ 
          & $\langle \hat T_{\rm rot} \rangle $ &  0.24   & 0.60     & 2.40    & 3.18     \\ \hline
\end{tabular}
\end{center}
\end{table}

Summarizing this subsection, we have found that 
the OP symmetry breaking in the collective mass plays an important role 
in developing the localization of the collective wave functions 
of the yrast states.
On the other hand, the asymmetry of the collective mass tends to enhance 
the two peak structure of the yrare states.   

\section{Role of $\beta$-$\gamma$ couplings}

In this section, we examine whether or not 
the results obtained in the previous section using the (1+3)D model 
remain valid when we take into account the $\beta$ degree of freedom.  
Because this is a vast subject, we here concentrate on 
the situation in which we are most interested: 
namely, the case with $V_0=1.0, V_1=0.5$ MeV and $\E=0.5$. 

\subsection{A simple (2+3)-dimensional model}

We come back to the collective Schr\"odinger equation (\ref{eq:Hcoll}) 
for the general 5D quadrupole collective Hamiltonian and 
set up the collective potential in the following form:  
\begin{equation}
V(\beta,\gamma)=\frac{1}{2}C(\beta^2-\beta_0^2)^2 -v_0\beta^6\cos^2 3\gamma 
+ v_1 \beta^3\cos 3 \gamma +C_6\beta^6, 
\label{eq:2Dpotential}
\end{equation}
where  
$v_0=V_0/\beta_0^6$ and $v_1=V_1/\beta_0^3$.  
The first term ensures that the collective wave functions localize 
around $\beta \simeq \beta_0$.  
The second and third terms are reduced to the collective potential $V(\gamma)$  
in the (1+3)D model when the collective coordinate $\beta$ is frozen at 
 $\beta = \beta_0$. 
The fourth term guarantees that the potential satisfies the boundary condition, 
$V(\beta,\gamma) \rightarrow \infty$ as $\beta \rightarrow \infty$. 
Obviously, the collective potential (\ref{eq:2Dpotential}) fulfills 
the requirement that it should be a function of 
$\beta^2$ and $\beta^3\cos 3 \gamma$.  
We note that various parameterizations of the collective potential similar 
to Eq.(\ref{eq:2Dpotential}) have been used by many authors.
\cite{kum67a, row09,cap09}
It is certainly interesting and possible to derive 
the coefficients $C, v_0, v_1, C_6$ and $\beta_0$ 
using microscopic theories of nuclear collective motion. 
In this paper, however, 
we simply treat these coefficients as phenomenological parameters 
and determine these values so that 
the resulting collective potential $V(\beta,\gamma)$ qualitatively simulates 
that obtained using the microscopic HFB calculation \cite{yam01} for $^{68}$Se . 
They are $C=800,~V_0=1.0,~V_1=0.5,~C_6=1000$ MeV and  $\beta_0^2=0.1$. 
In Fig.~17, the collective potential with these coefficients 
is drawn in the $(\beta, \gamma$) plane. 
One may immediately notice the following characteristic features of 
this collective potential.   
1) There are two local minima, one at the oblate shape and 
the other at the prolate shape. They are approximately degenerate in energy 
but the oblate minimum is slightly lower. 
2) There is a valley along the $\beta=\beta_0$ line 
connecting the two local minima. 
3) The spherical shape is a local maximum, which is approximately 4 MeV 
higher than the oblate minimum.

\begin{figure}[t]
\begin{center}
  \includegraphics[width=0.5\textwidth]{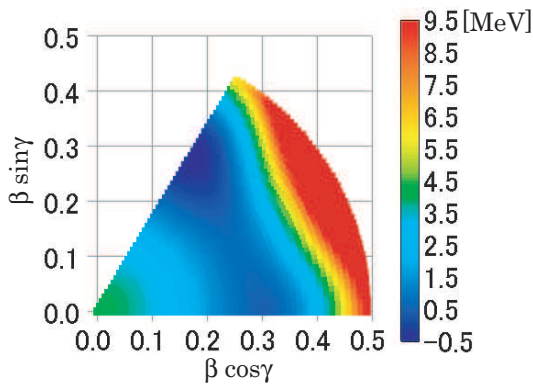}
  \caption{ Map of the two-dimensional collective potential $V(\beta,\gamma)$ 
  defined by Eq.~(\ref{eq:2Dpotential}) 
  for $C=800.0,~V_0=1.0,~V_1=0.5,~C_6=1000.0$ MeV and $\beta_0^2=0.1$.}
  \label{fig:potentialmap_C800}
\end{center}
\end{figure}


We set up the collective mass functions appearing in the collective kinetic 
energy terms as follows:  
\begin{align}
D_{\beta\beta}(\beta,\gamma)&=D(1-\E'\beta\cos3\gamma), \label{eq:2Dmass0}\\
D_{\gamma\gamma}(\beta,\gamma) &=D\beta^2(1+\E'\beta\cos3\gamma), \\
D_{\beta\gamma}(\beta,\gamma)&=D\E'\beta\sin3\gamma, \\
D_k(\beta,\gamma ) &=D(1+\E'\beta\cos\gamma_k), 
\label{eq:2Dmass}
\end{align}
where $\E'=\E/\beta_0$. 
These expressions are adopted to take into account the lowest-order 
$(\beta,\gamma)$ dependence of the collective mass functions 
derived by Yamada \cite {yam93} by means of the SCC method. 
\cite{mar80} 
{
We use the same values for $D$ and $\E$ as in Fig.~13, namely $D$=50MeV$^{-1}$ and $\E=0.5$.
}

\subsection{Comparison of the (1+3)D and the (2+3)D model calculations}

We have solved the collective Schr\"odinger equation (\ref{eq:Hcoll}) 
using the collective potential (\ref{eq:2Dpotential}) and 
the collective masses (\ref{eq:2Dmass0})-(\ref{eq:2Dmass}). 
Below, the results of the numerical calculation are presented 
and compared with those of the (1+3)D model.  

{ 
Figures~18 and 19 display on the ($\beta,\gamma$) plane 
the two-dimensional collective wave functions squared, 
$\sum_K |\Phi_{IK\alpha}(\beta,\gamma)|^2$, and 
the $\beta^4$-weighted ones, 
$\beta^4\sum_K |\Phi_{IK\alpha}(\beta,\gamma)|^2$, 
respectively, 
of the yrast and yrare states with even angular momenta $I=0 - 8$.  
The $\beta^4$ factor carries the major $\beta$ dependence 
of the intrinsic volume element $d\tau'$ given by Eq.~(\ref{eq:metric}).  
We see in Fig.~18 that, except the $0^+$ states, the yrast (yrare) wave 
functions are well localized around the oblate (prolate) shape.
While the localization of the yrast wave functions grows as the angular momentum increases,
the yrare wave functions gradually develop the second peaks 
around the oblate shape. 
These behaviors are qualitatively the same as those we have seen  
for the collective wave functions in the (1+3)D model in Fig.~13.     
For the $0^+$ states, the wave functions squared in Fig.~18 
appear to be spread over a rather wide region around the spherical shape, 
but the $\beta^4$-weighted ones in Fig.~19 take the maxima 
(as functions of $\beta$) near the constant-$\beta$ line 
with $\beta=\beta_0$. Thus, we can see in Fig.~19 
a very good correspondence with the (1+3)D wave functions 
including the $0^+$ states also. 
That is, the behaviors along the constant-$\beta$ line 
of the collective wave functions in the (2+3)D model 
exhibit qualitatively the same features as those of the (1+3)D model.    
(A minor difference is seen only in the relative heights of 
the oblate and prolate peaks of the $0_2^+$ wave function.) 
We note that the localization properties of the collective wave function 
are seen better in the $\beta^4$-weighted wave functions squared 
than those multiplied by the total intrinsic volume element, 
$d\tau'\sum_K |\Phi_{IK\alpha}(\beta,\gamma)|^2$, 
which always vanish at the oblate and prolate shapes 
due to the $\sin3\gamma$ factor contained in $d\tau'$. 
}

\begin{figure}[htb]
\begin{center}
\subfigure[$0_1$ state]{\hspace{-2.em}\includegraphics[height=0.2\textwidth,keepaspectratio,trim= 10 6 160 20,clip]{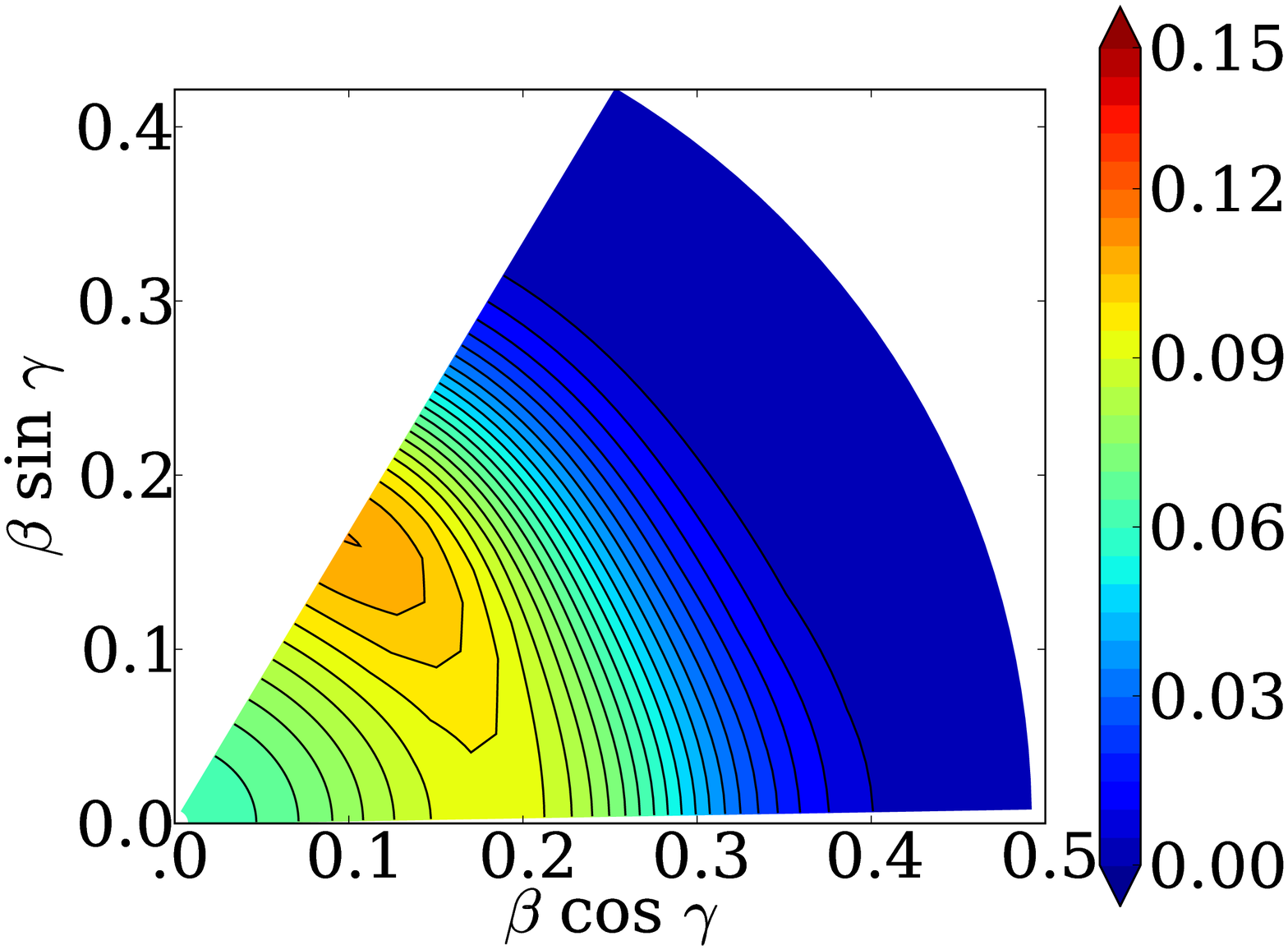}} 
\subfigure[$2_1$ state]{\includegraphics[height=0.2\textwidth,keepaspectratio,trim= 32 6 160 20,clip]{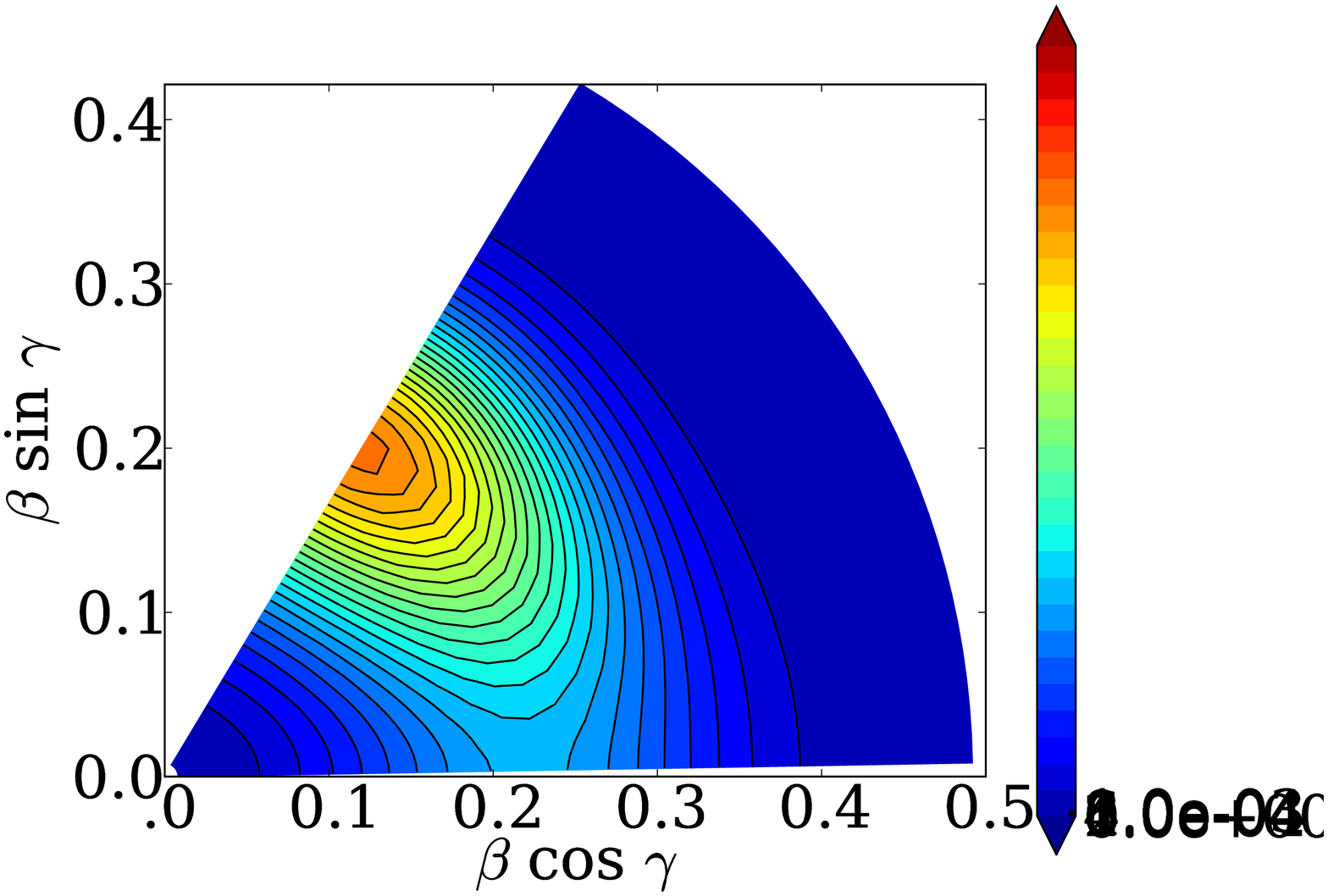}} 
\subfigure[$4_1$ state]{\includegraphics[height=0.2\textwidth,keepaspectratio,trim= 32 6 160 20,clip]{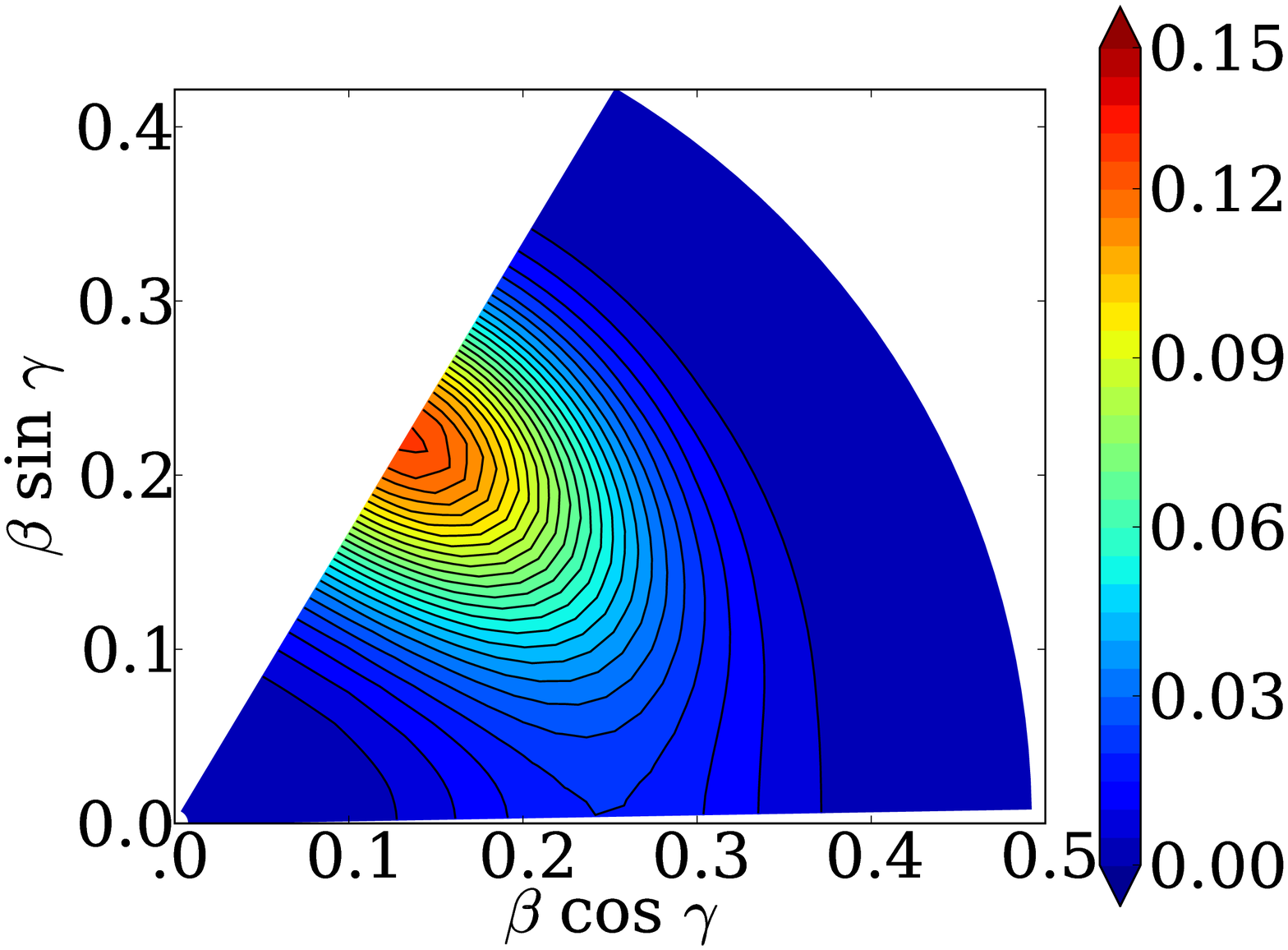}} 
\subfigure[$6_1$ state]{\includegraphics[height=0.2\textwidth,keepaspectratio,trim= 32 6 160 20,clip]{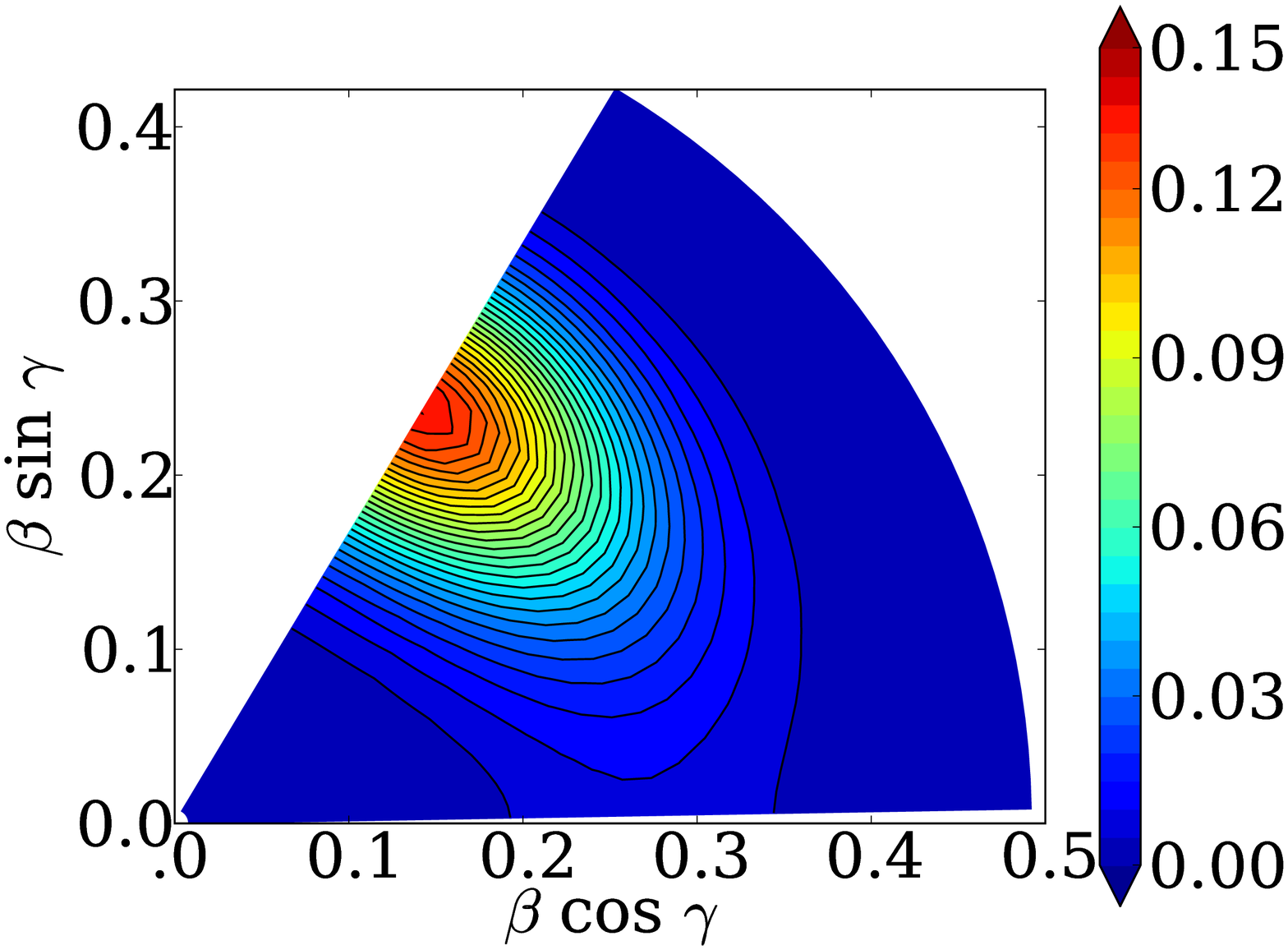}} 
\subfigure[$8_1$ state]{\includegraphics[height=0.2\textwidth,keepaspectratio,trim= 32 6 50 20,clip]{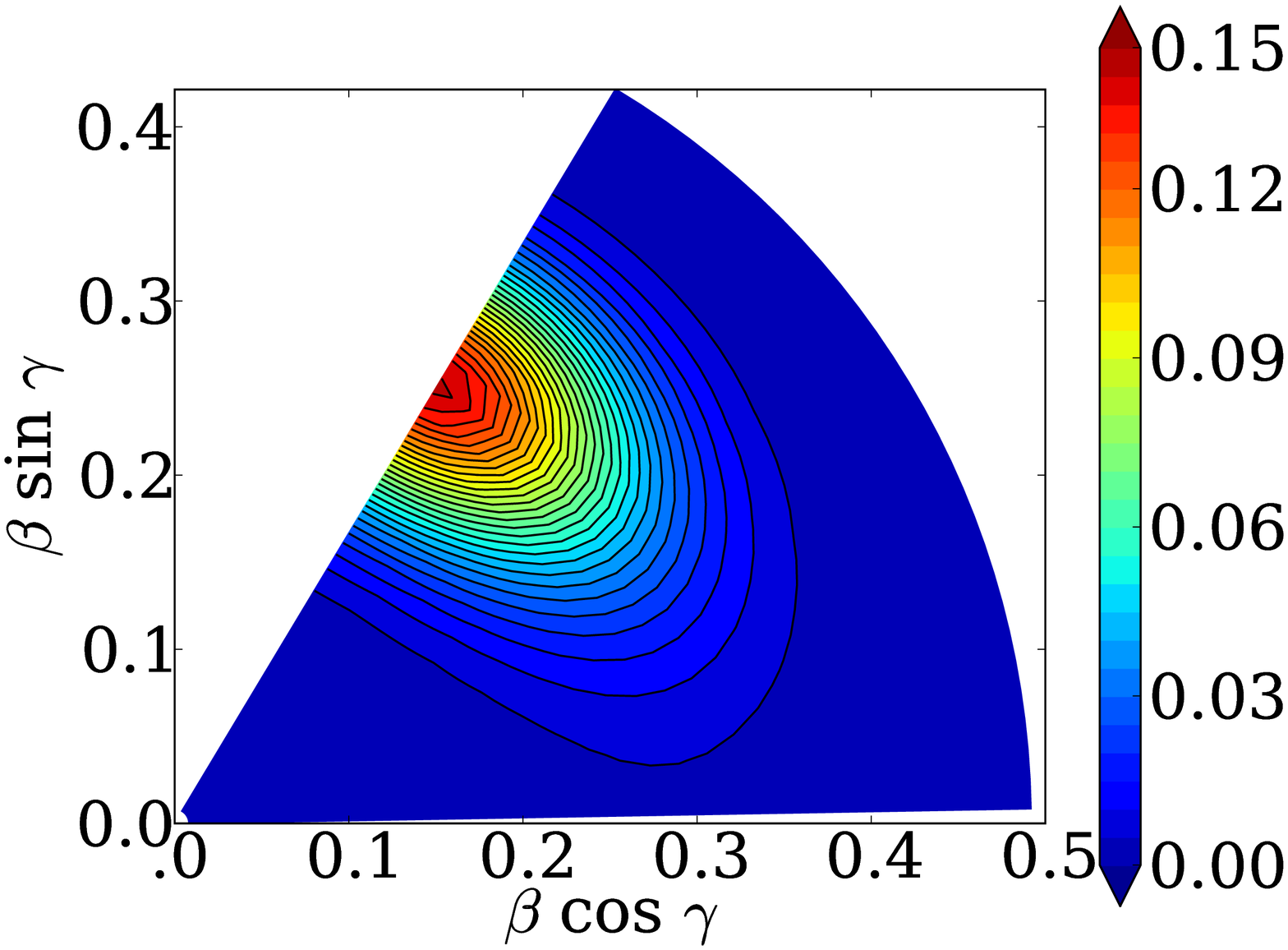}}   
\subfigure[$0_2$ state]{\hspace{-2.em}\includegraphics[height=0.2\textwidth,keepaspectratio,trim= 10 6 160 20,clip]{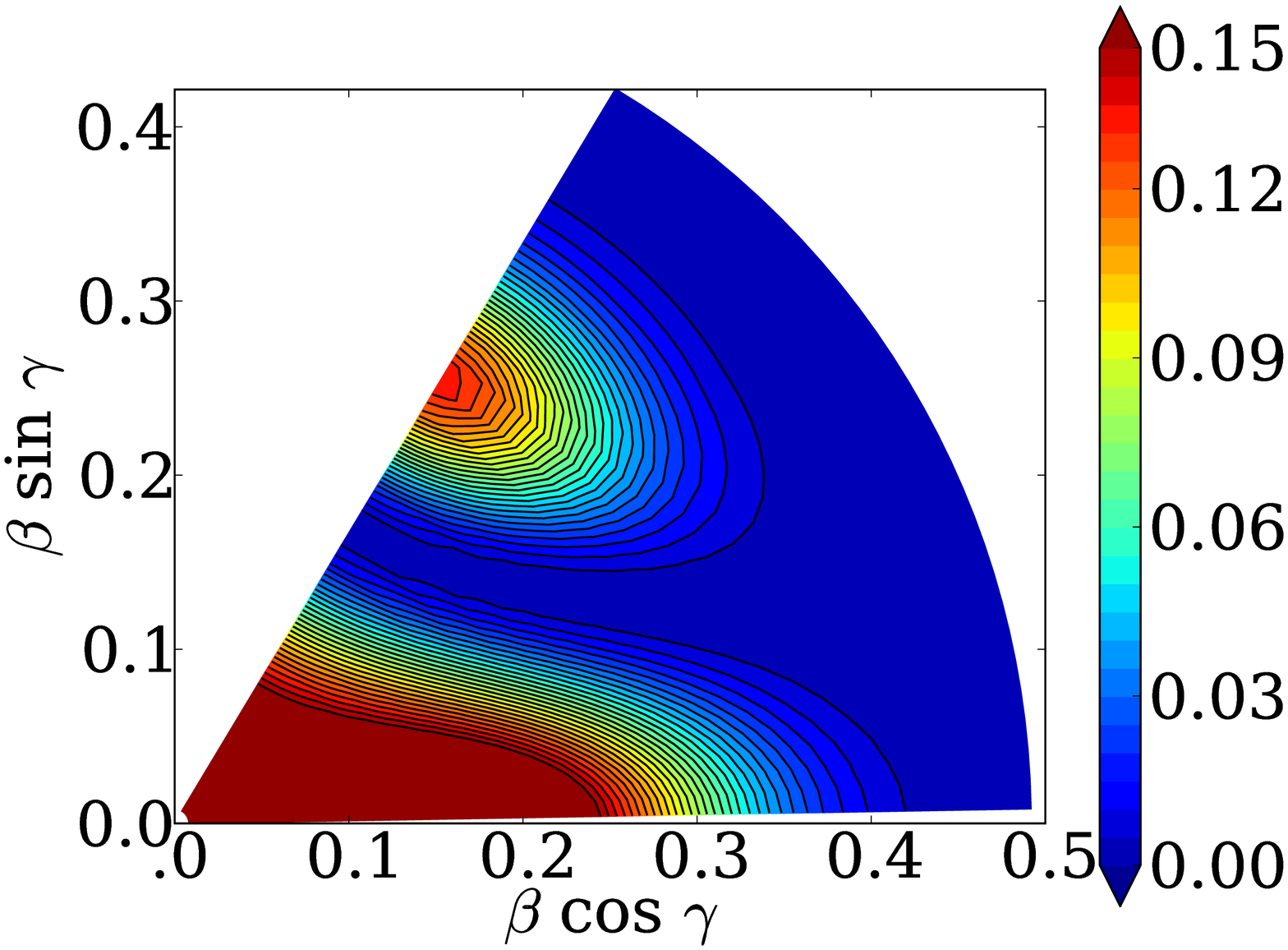}} 
\subfigure[$2_2$ state]{\includegraphics[height=0.2\textwidth,keepaspectratio,trim= 32 6 160 20,clip]{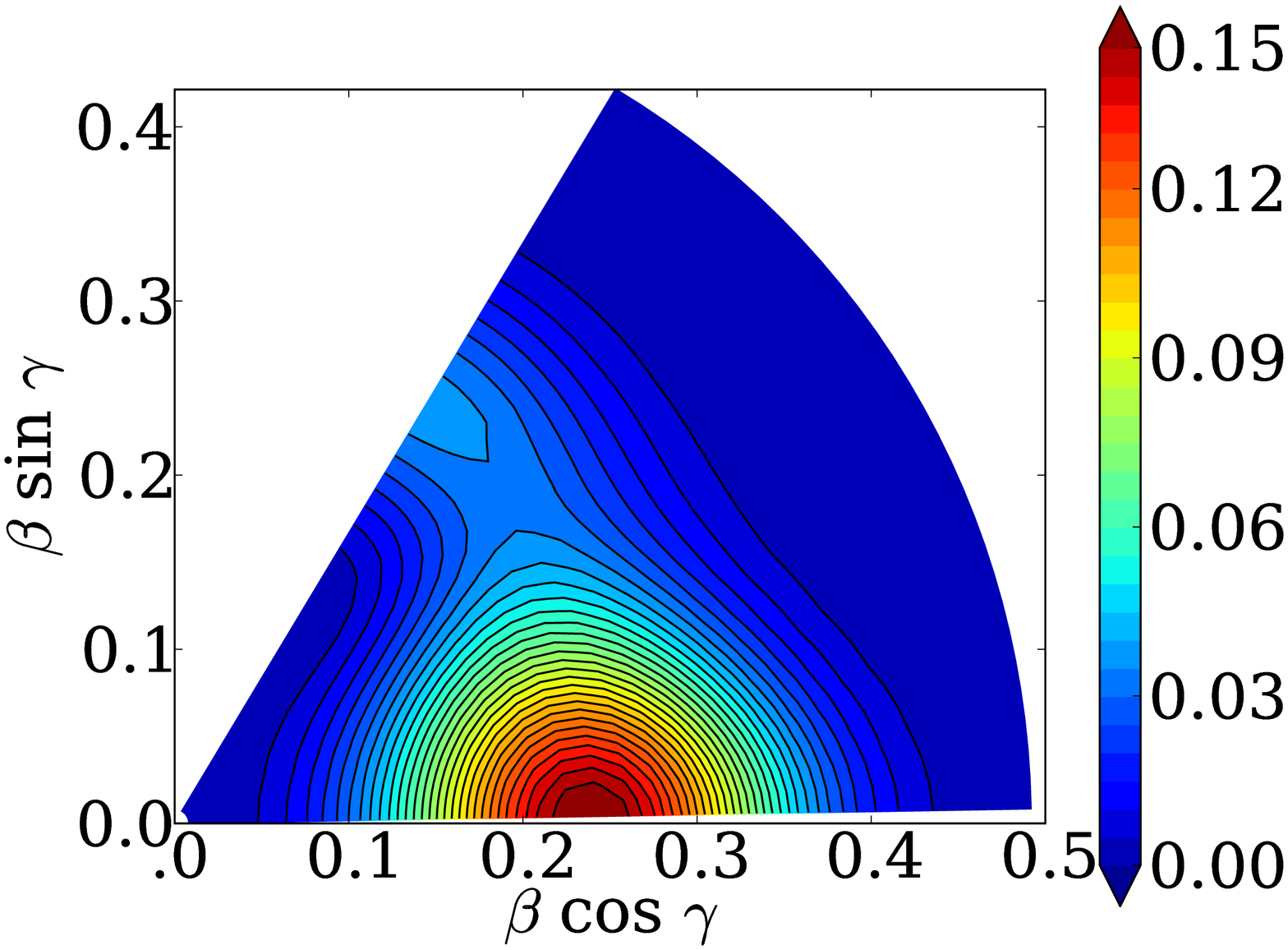}} 
\subfigure[$4_2$ state]{\includegraphics[height=0.2\textwidth,keepaspectratio,trim= 32 6 160 20,clip]{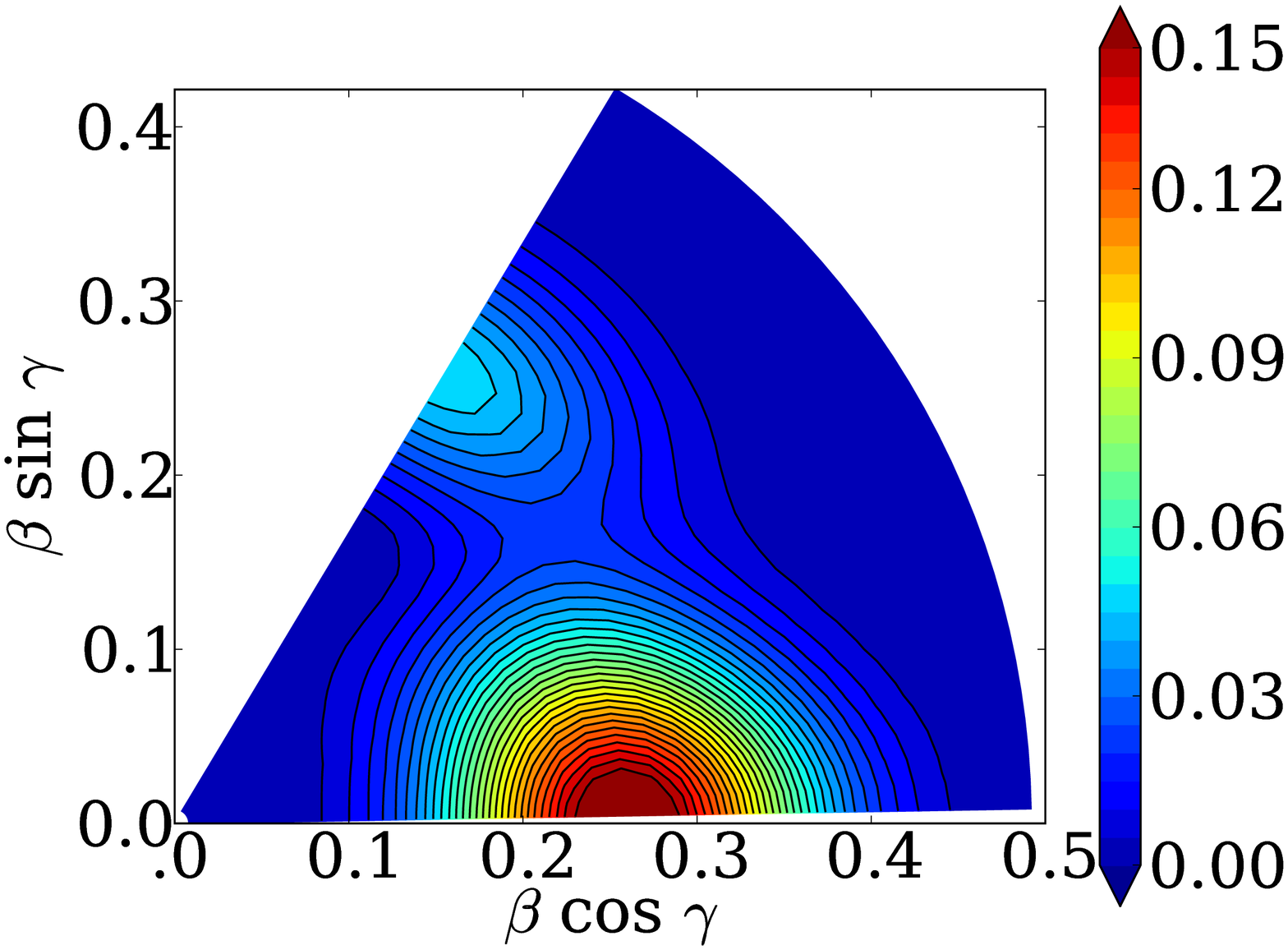}} 
\subfigure[$6_2$ state]{\includegraphics[height=0.2\textwidth,keepaspectratio,trim= 32 6 160 20,clip]{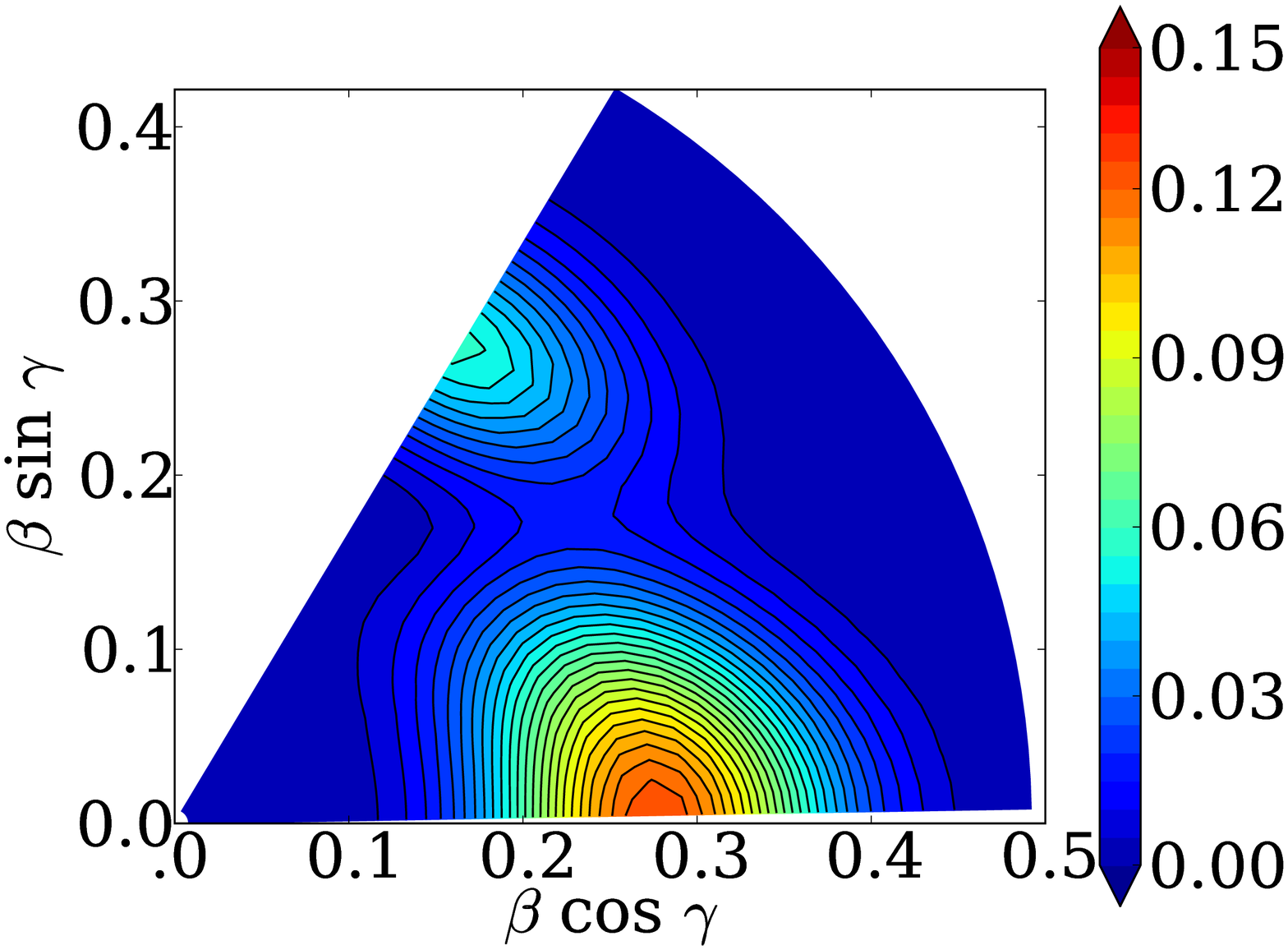}} 
\subfigure[$8_2$ state]{\includegraphics[height=0.2\textwidth,keepaspectratio,trim= 32 6 50 20,clip]{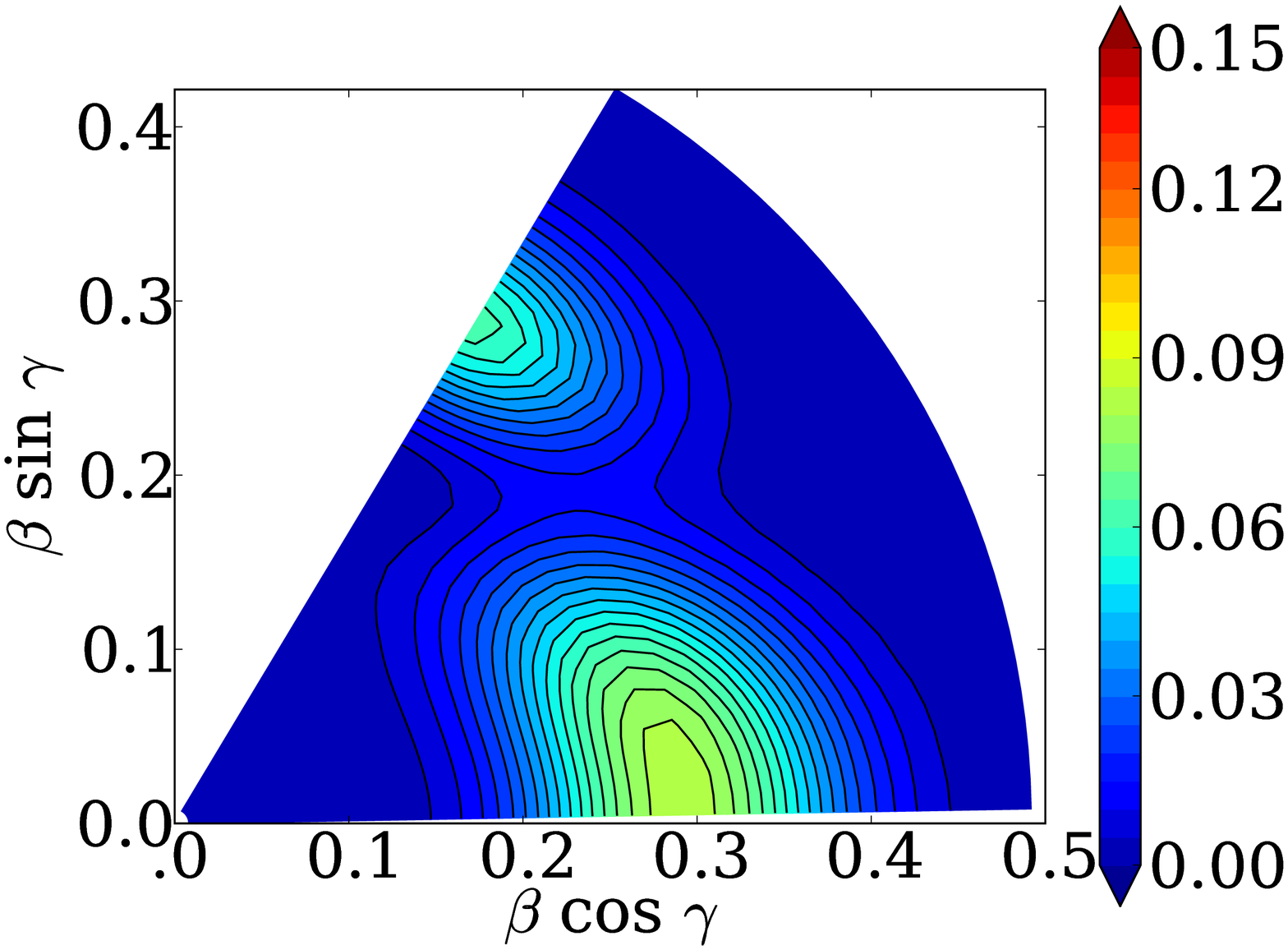}} 
\end{center}
\caption{Collective wave functions squared $\sum_K|\Phi_{IK\alpha}(\beta,\gamma)|^2$
  calculated for the mass-asymmetry parameter $\E=0.5$ and 
  the two-dimensional collective potential $V(\beta,\gamma)$ 
  with $C=800.0, V_0=1.0,~V_1=0.5,~C_6=1000.0$ MeV and $\beta_0^2=0.1$. 
  The upper and the lower panels display the results 
  for the yrast and the yrare states, respectively.}
\label{fig:2Dwfs_C+800_C6+1000}
\end{figure} 


\begin{figure}[htb]
\begin{center}
\subfigure[$0_1$ state]{\hspace{-2.em}\includegraphics[height=0.2\textwidth,keepaspectratio,trim= 10 6 160 20,clip]{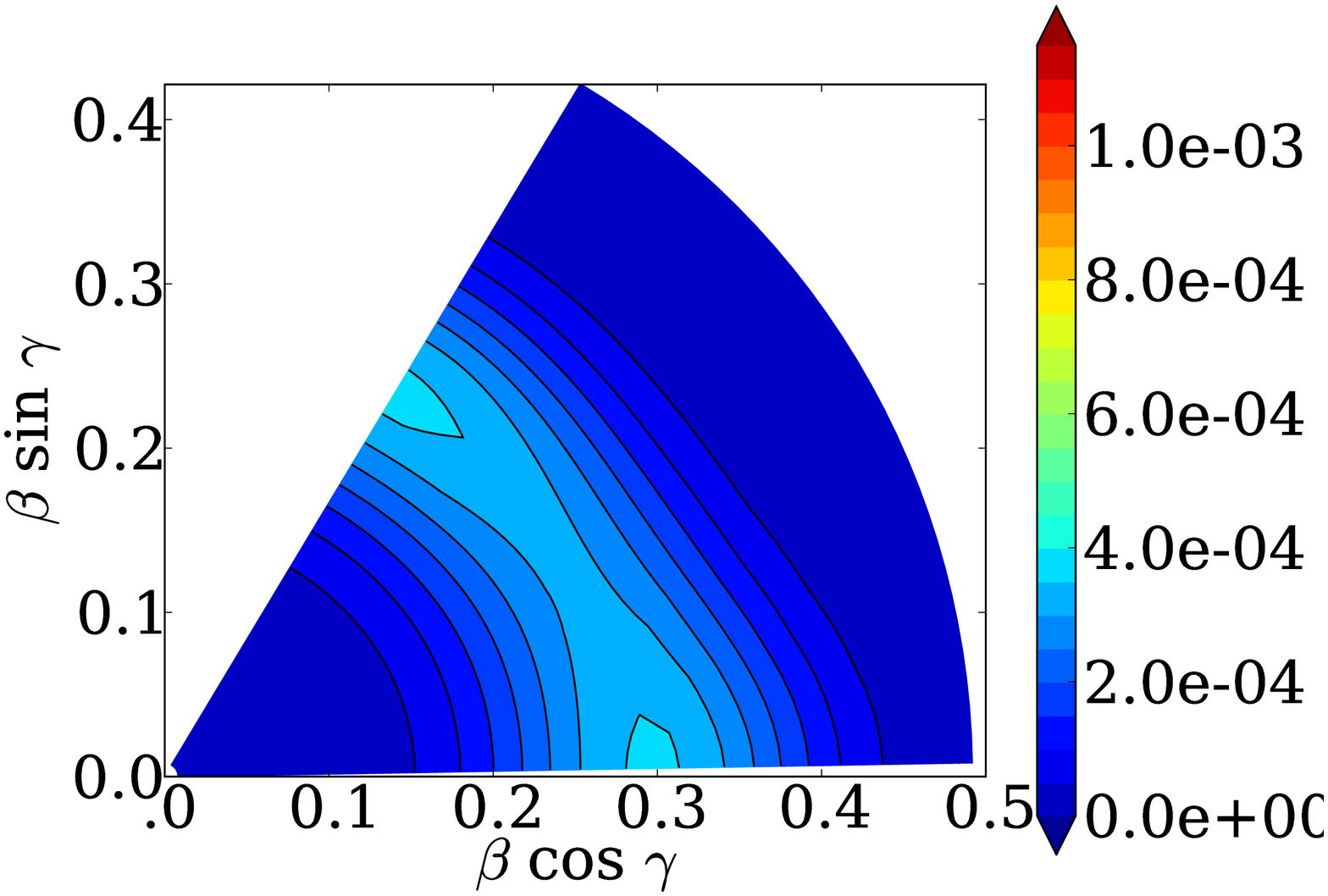}} 
\subfigure[$2_1$ state]{\includegraphics[height=0.2\textwidth,keepaspectratio,trim= 32 6 160 20,clip]{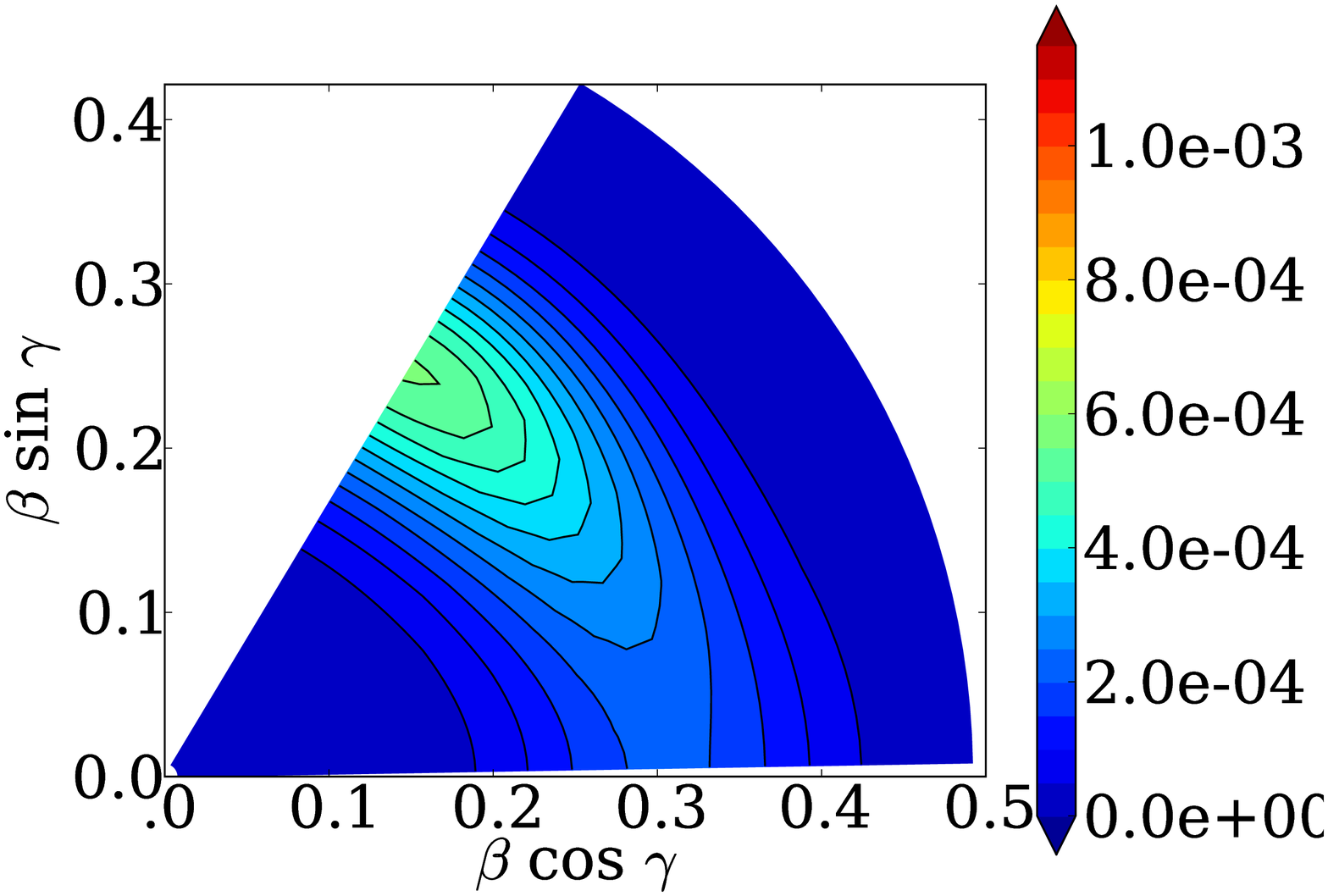}} 
\subfigure[$4_1$ state]{\includegraphics[height=0.2\textwidth,keepaspectratio,trim= 32 6 160 20,clip]{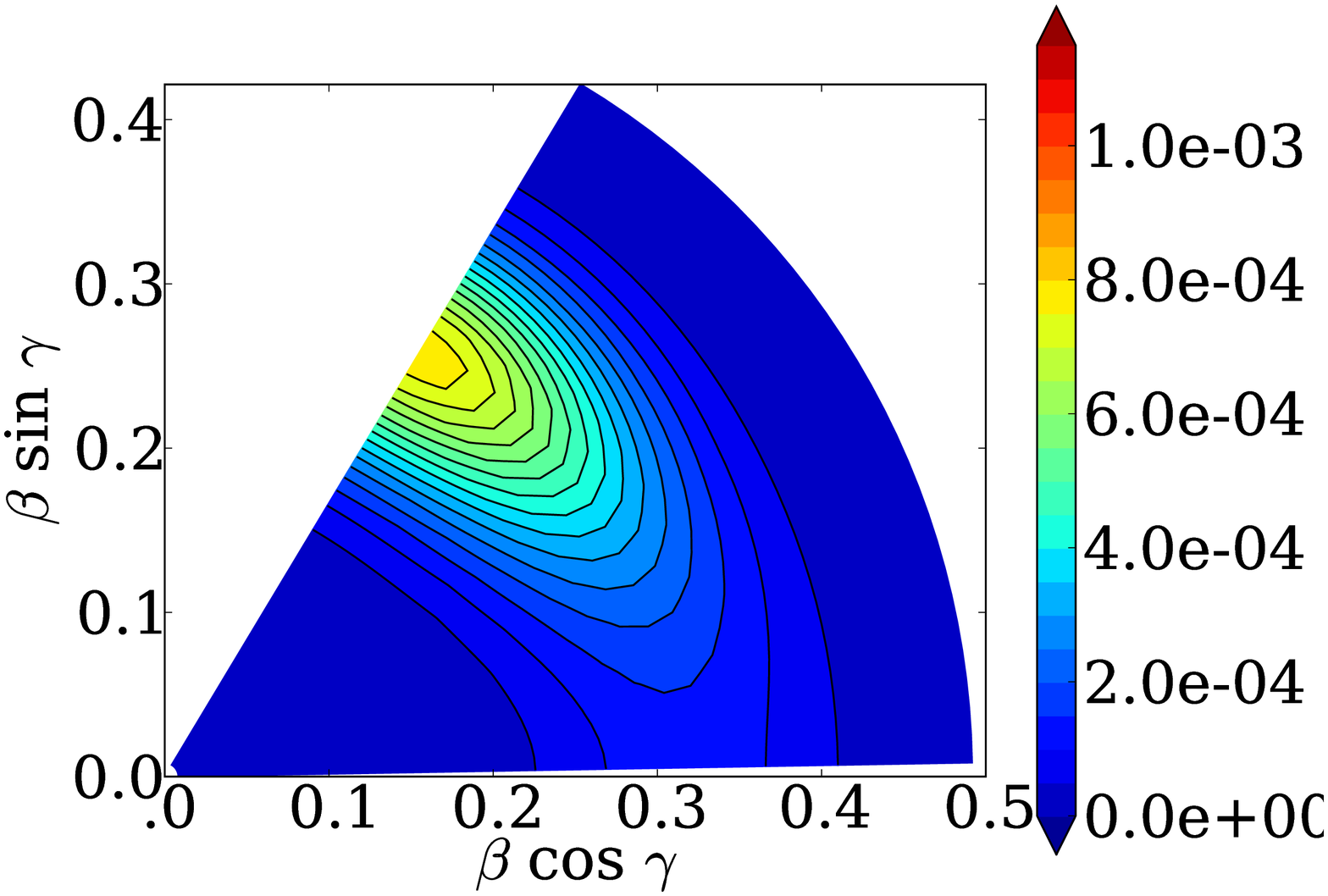}} 
\subfigure[$6_1$ state]{\includegraphics[height=0.2\textwidth,keepaspectratio,trim= 32 6 160 20,clip]{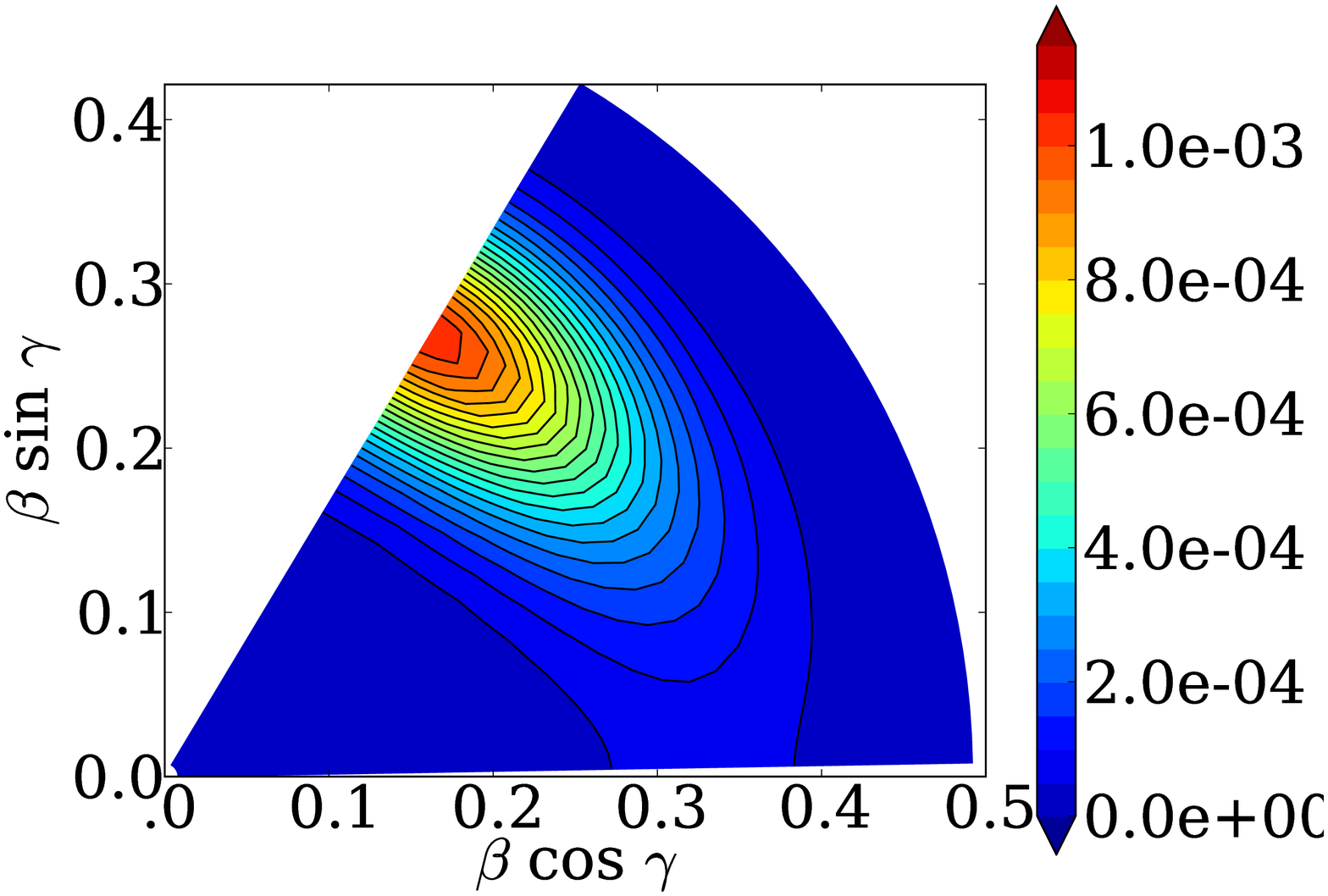}} 
\subfigure[$8_1$ state]{\includegraphics[height=0.2\textwidth,keepaspectratio,trim= 32 6 15 20,clip]{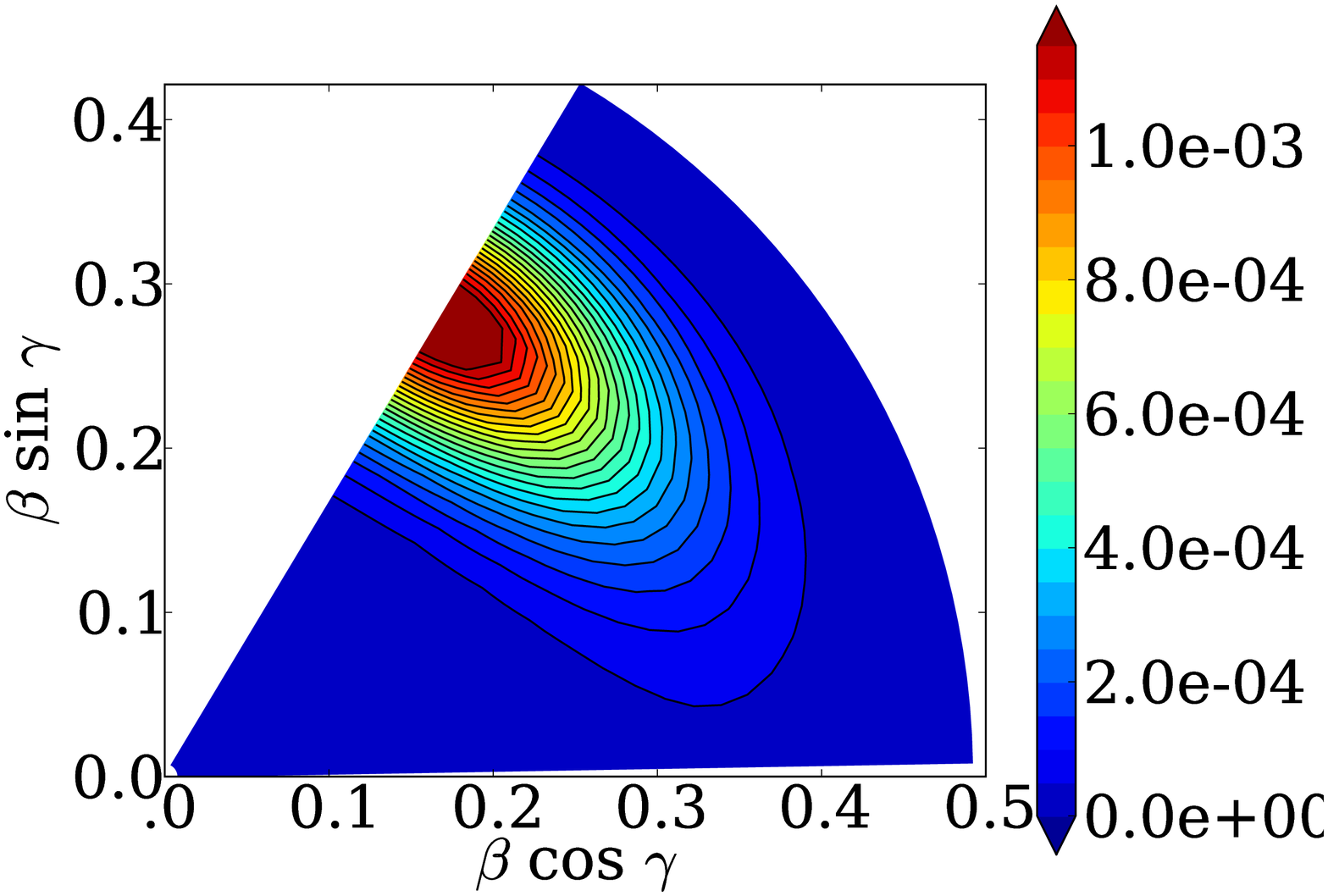}} 
\subfigure[$0_2$ state]{\hspace{-2.em}\includegraphics[height=0.2\textwidth,keepaspectratio,trim= 10 6 160 20,clip]{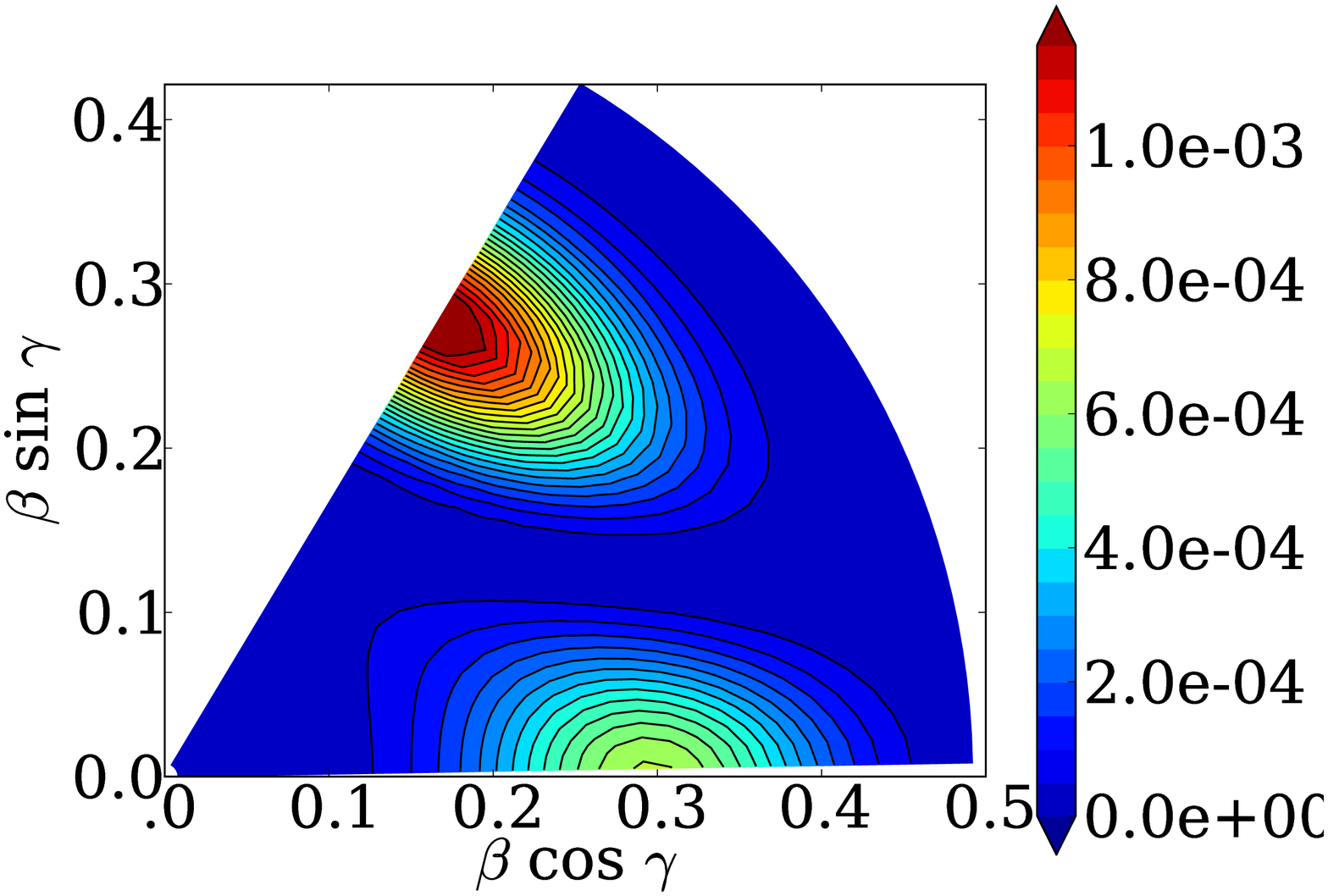}} 
\subfigure[$2_2$ state]{\includegraphics[height=0.2\textwidth,keepaspectratio,trim= 32 6 160 20,clip]{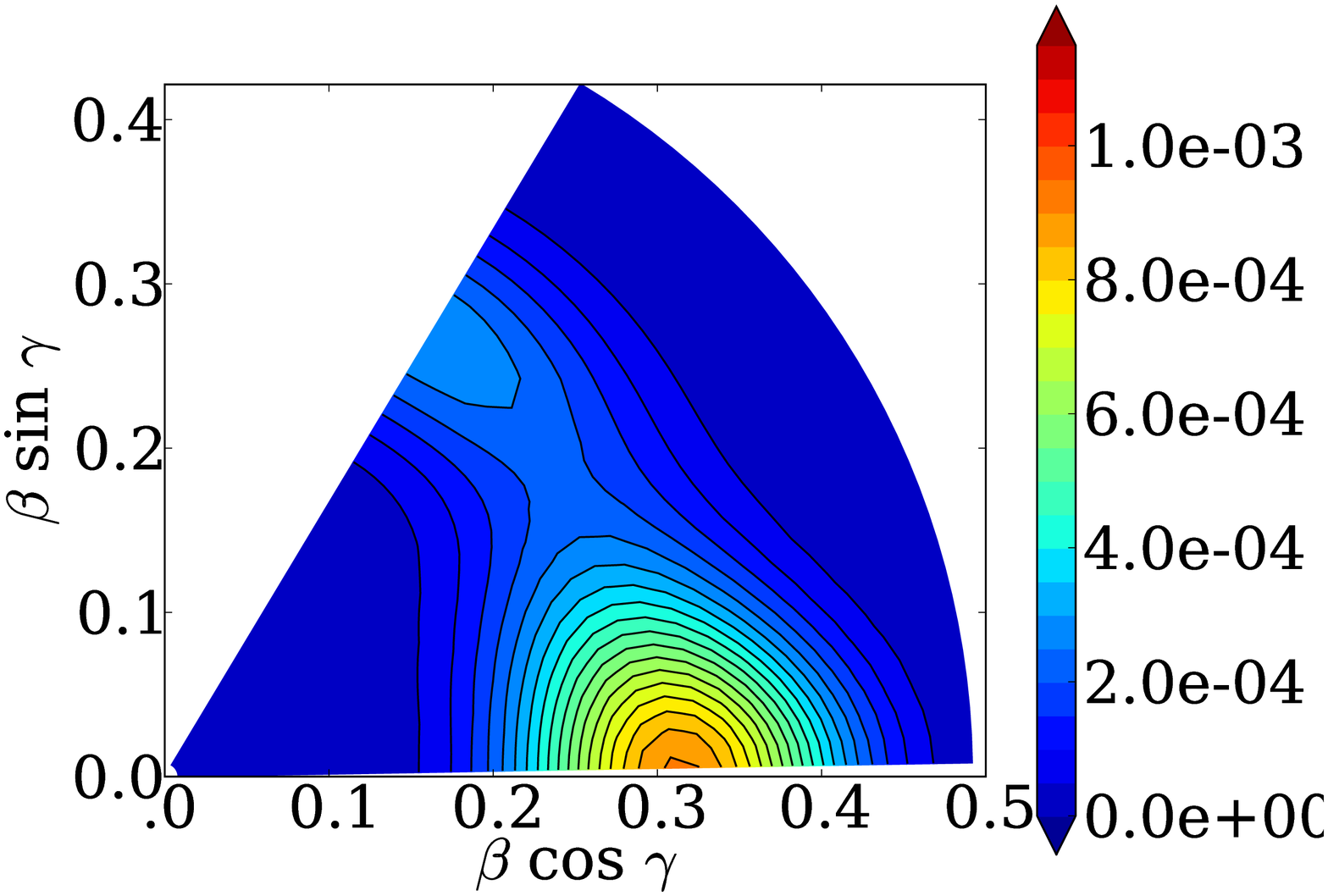}} 
\subfigure[$4_2$ state]{\includegraphics[height=0.2\textwidth,keepaspectratio,trim= 32 6 160 20,clip]{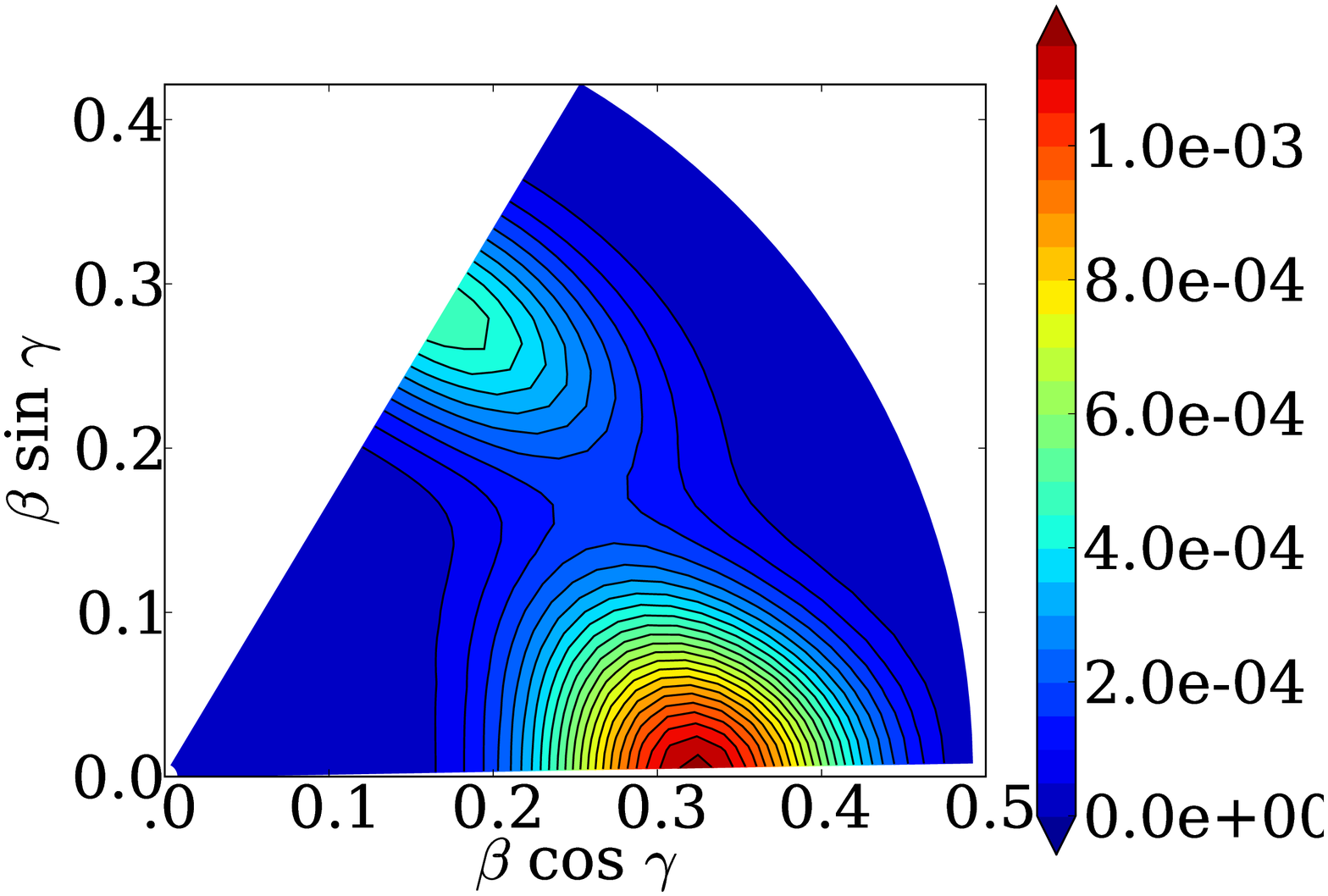}} 
\subfigure[$6_2$ state]{\includegraphics[height=0.2\textwidth,keepaspectratio,trim= 32 6 160 20,clip]{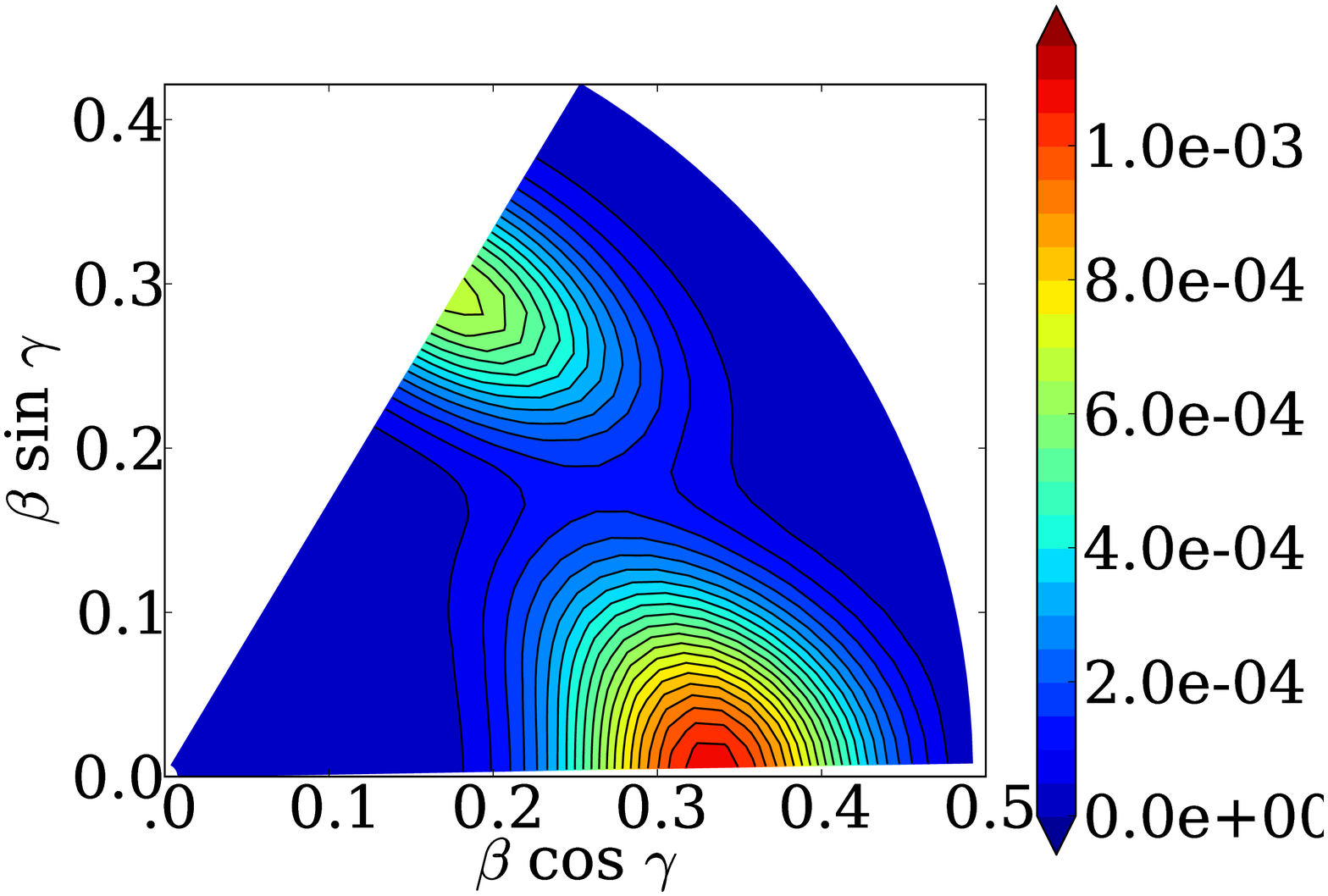}} 
\subfigure[$8_2$ state]{\includegraphics[height=0.2\textwidth,keepaspectratio,trim= 32 6 15 20,clip]{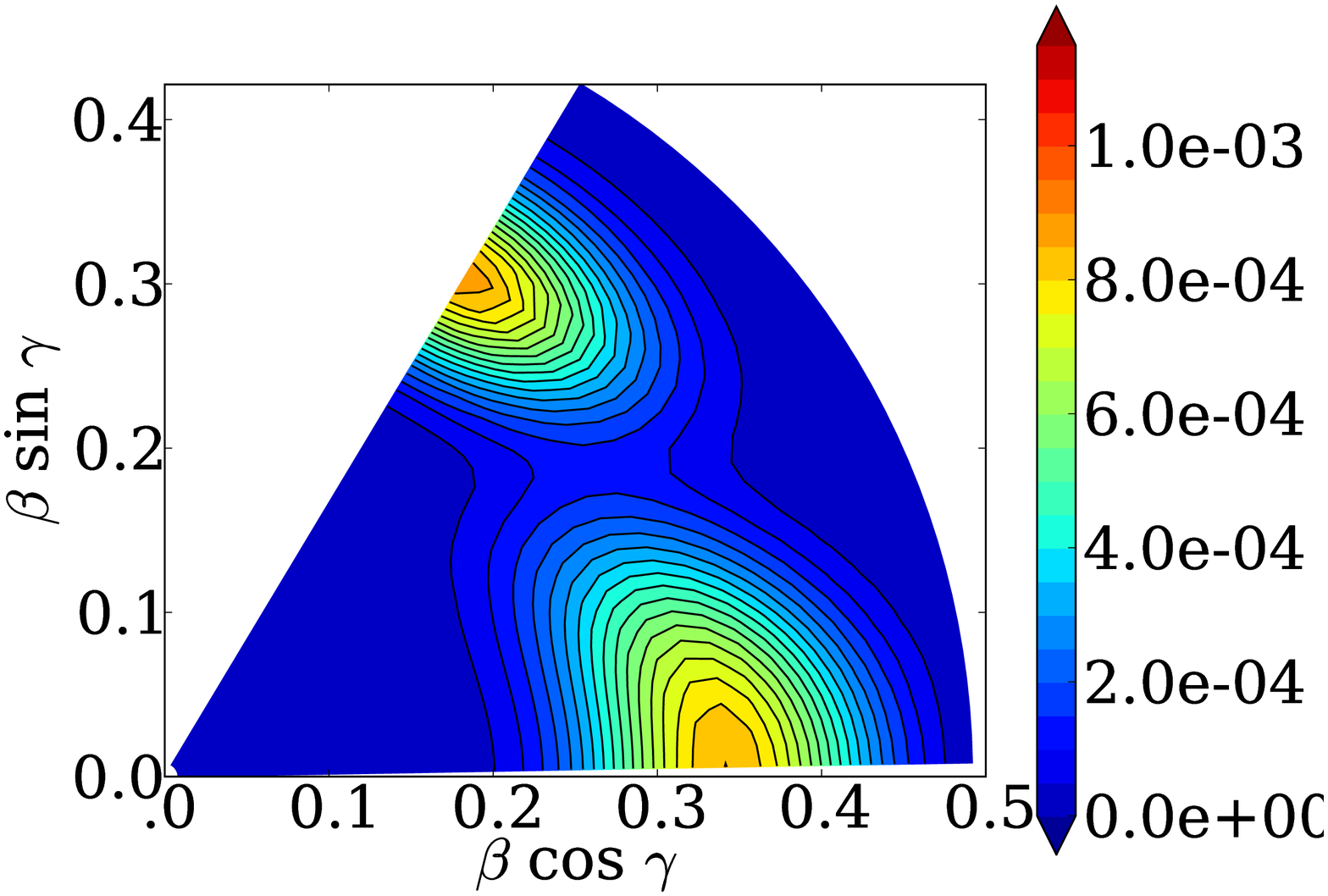}} 
\end{center}
\caption{ The same as Fig.~18 but for weighted collective wave functions squared 
  $\beta^4\sum_K|\Phi_{IK\alpha}(\beta,\gamma)|^2$.}
\label{fig:w2Dwfs_C+800_C6+1000}
\end{figure} 


Finally we compare in Fig.~20 the excitation spectrum obtained  
in the (2+3)D model with that in the (1+3)D model.  
It is evident that they agree very well. 
Aside from the quantitative difference that the excitation energies 
are a little higher and the $B(E2)$ values increase more rapidly 
with increase in the angular momentum in the (2+3)D model 
than in the (1+3)D model, 
the essential features of the excitation spectrum are the same. 
In particular, we find a beautiful agreement in the level sequences: 
the energy ordering of these eigenstates are exactly the same 
between the two calculations.    
This agreement implies that the $\beta$-$\gamma$ coupling plays 
only a secondary role here,  
and the major feature of the excitation spectrum is determined 
by triaxial deformation dynamics. 
In this dynamics, the $\gamma$ dependence of the collective mass functions 
plays an important role as well as that of the collective potential. 
 
The excitation spectra of Fig.~20  
 are quite different from any of the patterns known well 
in axially symmetric deformed nuclei,  
in the rigid triaxial rotor model and 
in the $\gamma$-unstable model. 
It also deviates considerably from the spectrum expected in 
an ideal situation of the oblate-prolate shape coexistence 
where two rotational bands keep their identities 
without strong mixing between them.  
Among a number of interesting features, we first notice the unique character of 
the $0_2^+$ state.  It is significantly shifted up in energy 
from its position expected when the yrare 
$0_2^+, 2_2^+, 4_2^+, 6_2^+$ and $8_2^+$ states 
form a regular rotational band.  
As we have discussed in connection with Fig.~3, 
the position of the $0_2^+$ state relative to the $2_2^+$ state 
serves as a sensitive measure indicating where the system locates  
between the $\gamma$-unstable situation ($V_0=V_1=0$) 
and the ideal oblate-prolate shape coexistence (large $V_0$ and small $V_1$). 
Thus, the results of our calculation suggest that 
experimental data for the excitation energy of the $0_2^+$ state 
provide  very valuable information on 
the barrier height between the oblate and prolate local minima. 
In this connection, we also note that the $3_1^+$  ($5_1^+$) state 
is situated slightly higher in energy than the $6_1^+$ and $4_2^+$  
($8_1^+$ and $6_2^+$) states. These are other indicators suggesting that 
the system is located in an intermediate situation between the two limits 
mentioned above.

\begin{figure}[t]
\begin{center}
  \includegraphics[width=\textwidth]{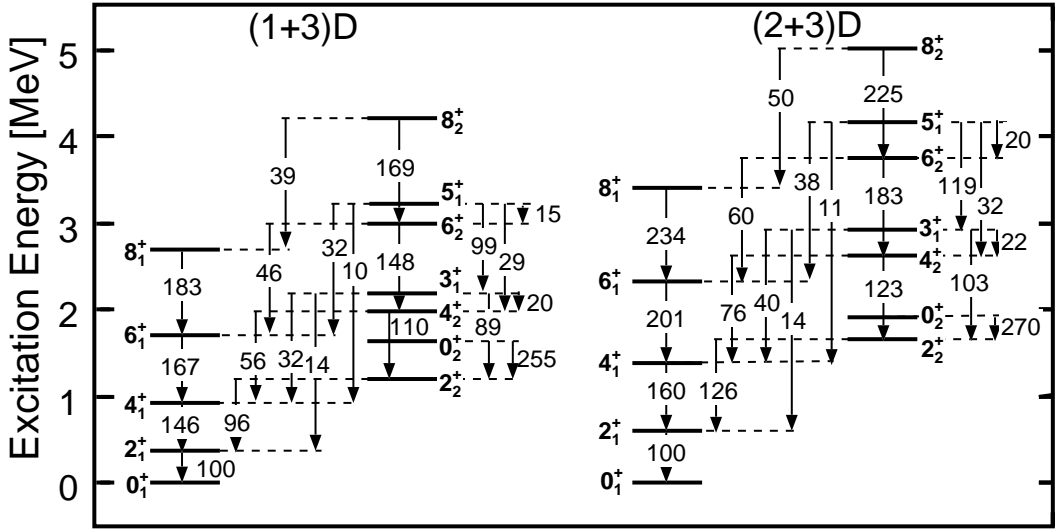}
  \caption{Comparison of the excitation spectra and the $E2$-transition properties 
  in the (1+3)D model (left side) and those in the (2+3)D model (right side), 
  calculated for the potential parameters $V_0=1.0, V_1=0.5$ MeV 
  and the mass-asymmetry parameter $\E=0.5$.  
  Additional parameters for the two-dimensional potential $V(\beta,\gamma)$ 
  in the (2+3)D model are $C=800.0$ and $C_6=1000.0$ MeV.
  The $B(E2)$ values are written on the transition arrows normalizing 
  $B(E2;2_1 \rightarrow 0_1)$ as 100. 
  Weak $E2$ transitions whose $B(E2)$ values smaller than 10 are not shown. } 
\label{fig:CompSpectra}
\end{center}
\end{figure}

  
In Fig.~20, we furthermore notice interesting properties of the 
$E2$ transitions from the yrare to the yrast states: 
for instance, the $E2$ transitions with $\Delta I= -2$, or $B(E2;I_{\rm yrare} \rightarrow (I-2)_{\rm yrast})$, and 
those with $\Delta I= +2$, or $B(E2;I_{\rm } \rightarrow (I+2)_{\rm yrast})$,
are much smaller than other yrare-to-yrast $E2$ transitions. 
These transitions are forbidden in the $\gamma$-unstable model 
\cite{wil56} because they transfer the boson seniority $v$ 
by $\Delta v=2$ and $\Delta v=0$, respectively.  
Thus, some features characteristic to the $\gamma$-unstable situation persist here. 
On the other hand, although $B(E2;3_1 \rightarrow 4_2)$ and $B(E2;5_1 \rightarrow 6_2)$ 
are also forbidden transitions with $\Delta v=0$ 
in the $\gamma$-unstable limit, 
they are not very small in Fig.~20 and indicate a significant deviation from the $\gamma$-unstable limit.

\begin{figure}[h]
\begin{tabular}{cc}
\subfigure[(1+3)D]
{\includegraphics[width=0.5\textwidth]{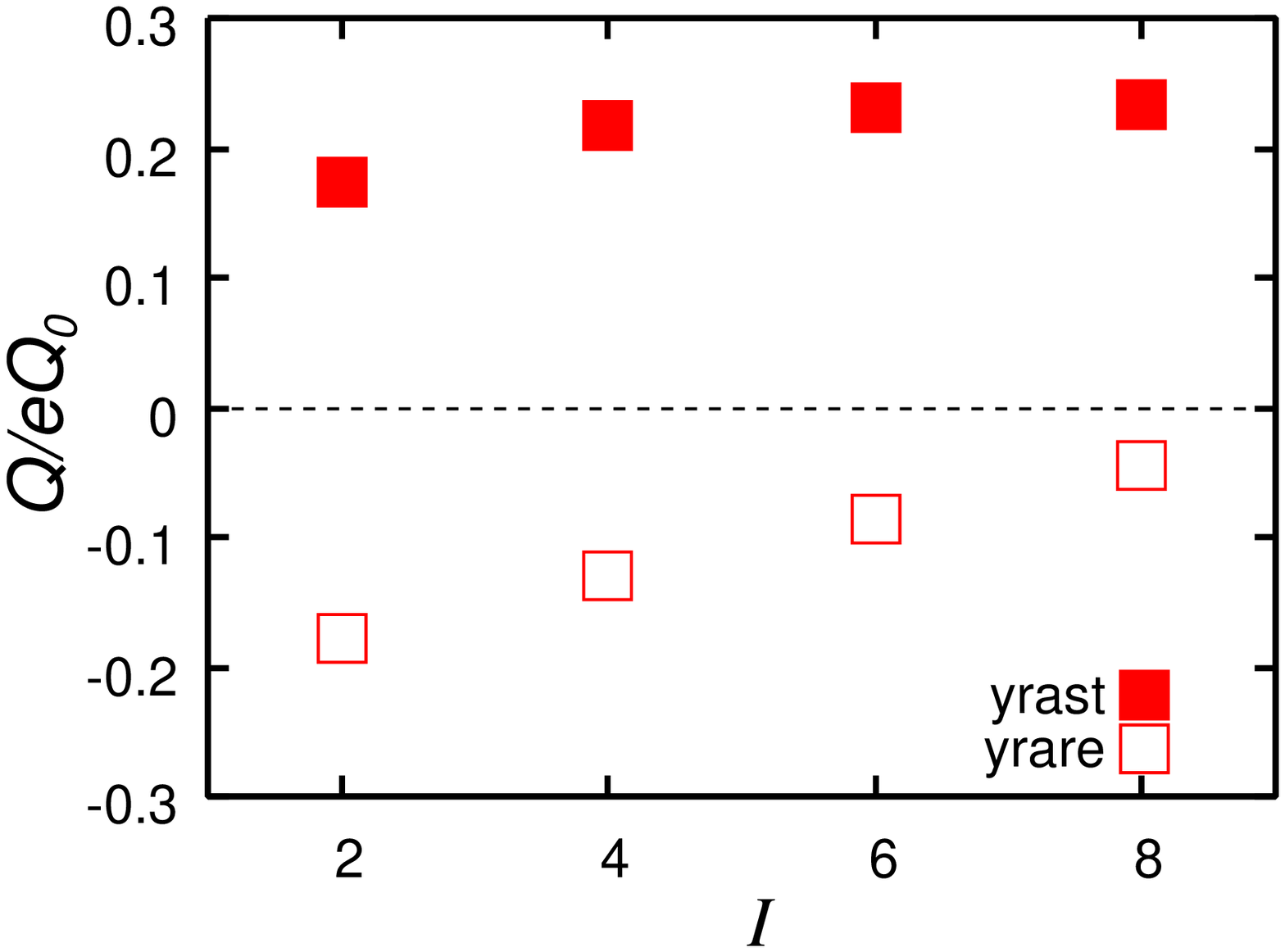} }
\subfigure[(2+3)D]
{\includegraphics[width=0.5\textwidth]{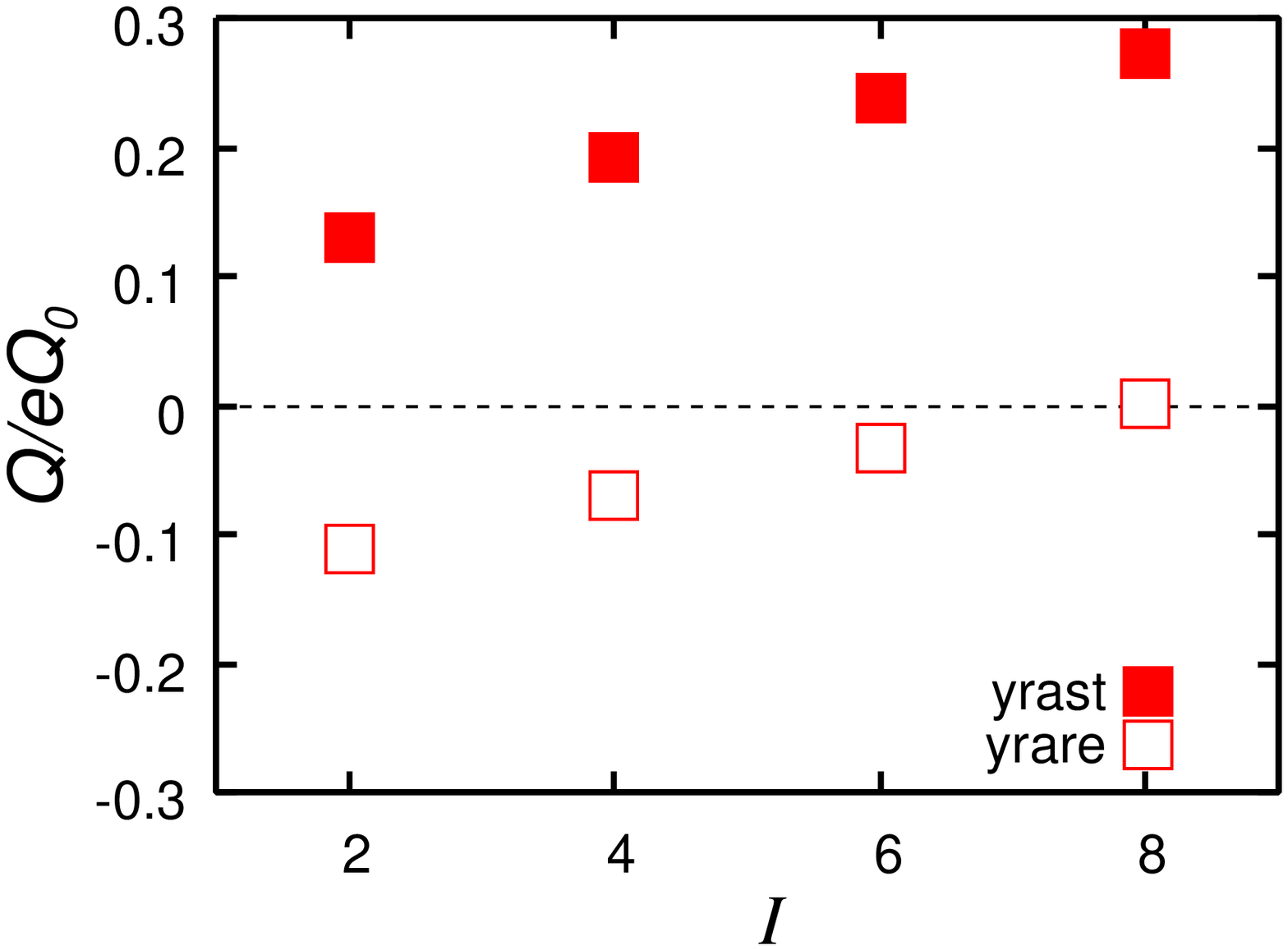}}
\end{tabular}
 \caption{The same as Fig.~20 but for the spectroscopic quadrupole moments  
  in units of the intrinsic quadrupole moment, $eQ_0=3e/\sqrt{5\pi} ZR_0^2\beta_0$. 
  The results of calculation in the (1+3)D model and in the (2+3)D model 
  are displayed as functions of angular momentum $I$ 
  in the left and right panels, respectively. 
  The values for the yrast (yrare) states are shown 
  by filled (open) squares.}
\label{figs:Qcomp}
\end{figure}

The spectroscopic quadrupole moments $Q$ calculated in the (1+3)D  and 
the (2+3)D models are compared in Fig.~21.  
It is seen again that the two calculations show 
the same qualitative features: 
both calculations yield the positive (negative) sign 
for the spectroscopic quadrupole moments of the yrast (yrare) states 
to indicate an oblatelike (prolatelike) character. 
Quantitatively, in the (2+3)D model, the yrast $Q$ value increases with angular momentum 
more significantly and the absolute values of 
the yrare $Q$ moments are slightly smaller than those of the (1+3)D model. 
As we discussed above, in both calculations, the  $Q$ value of the yrare states approaches zero 
with the angular momentum increasing due to the cancellation mechanism 
associated with the growth of the two peak structure 
in the collective wave functions. 
In spite of such deviations from a simple picture,
we can see in Fig.~21 some qualitative features characteristic to the 
oblate-prolate shape coexistence. 

We conclude that the excitation spectrum and the properties of the quadrupole 
transitions and moments exhibited in Figs.~20 and 21 can be regarded 
as those characteristic to an intermediate situation between 
the well-developed oblate-prolate shape coexistence and 
the $\gamma$-unstable limit.

\section{Concluding Remarks}

From a viewpoint of the oblate-prolate symmetry and its breaking, 
we have proposed a simple (1+3)D model capable of describing  
the coupled motion of the large-amplitude shape fluctuation in the 
$\gamma$-degree of freedom and the three-dimensional rotation.    
Using this model, we have made a systematic investigation of the 
oblate-prolate shape coexistence phenomena and their relationships 
to other classes of low-frequency quadrupole modes of excitation, 
including particular cases described by the $\gamma$-unstable model 
and the rigid triaxial rotor model.    
We have also adopted the (2+3)D model to check the validity of freezing the $\beta$ degree of freedom
in the (1+3)D model.
We have obtained a number of interesting suggestions 
for the properties of low-lying states 
that are characteristic to an intermediate situation between 
the well-developed oblate-prolate shape-coexistence and 
 $\gamma$-unstable limits. In particular,
1) the relative energies of the excited $0^+$ states can be indicators of the barrier height of the
collective potential.
2) Specific $E2$ transition probabilities are sensitive to the oblate-prolate symmetry breaking.
3) Nuclear rotation can assist the localization of the 
collective wave functions in the $(\beta,\gamma)$ deformation space.
However, even if the rotation-assisted localization is realized in the yrast band, it is not necessarily in the yrare band:
the two-peak structure may develop in the yrare band.
  

\section*{Acknowledgements}

One of the authors (N. H.) is supported by the Special Postdoctoral 
Researcher Program of RIKEN. 
This work is supported by Grants-in-Aid for Scientific Research
(Nos. 20540259, 21340073) from the Japan Society for the 
Promotion of Science and the JSPS Core-to-Core
Program ``International Research Network for Exotic Femto Systems''.



\end{document}